\documentclass[12pt,a4paper,dvips]{article}
\usepackage{graphicx,epsfig}
\usepackage{hhline}

\usepackage{amsmath,amssymb}
\usepackage[varg]{txfonts}
\DeclareMathAlphabet{\mathbold}{OML}{txr}{b}{it}

\usepackage{rotating}

\usepackage[numbers,square,comma,sort&compress]{natbib}
\usepackage{hypernat}
\usepackage{textcomp}
\usepackage[dvipsnames]{color}
\definecolor{rltred}{rgb}{0.75,0,0}
\definecolor{rltgreen}{rgb}{0,0.5,0}
\definecolor{rltblue}{rgb}{0,0,0.5}

\newcounter{pdfadd}    
\usepackage[hyperindex,bookmarks,bookmarksnumbered,breaklinks,a4paper,unicode]{hyperref}
\hypersetup{%
  pdftitle        = {A Precision Measurement of the Inclusive ep Scattering Cross Section at HERA},
  urlcolor        = rltblue,       
  urlbordercolor  = 0 0 0.5,
  filecolor       = rltblue,       
  filebordercolor = 0 0 0.5,
  linkcolor       = rltred,        
  linkbordercolor = 0.75 0 0,
  citecolor       = rltgreen,      
  citebordercolor = 0 0.5 0,
  pagecolor       = rltgreen,      
  pagebordercolor = 0 0.5 0,
  menucolor       = rltgreen,      
  menubordercolor = 0 0.5 0,
  colorlinks    = true,
  pdfauthor     = {H1 Collaboration},
  pdfsubject    = { },
  pdfkeywords   = {High-Energy Physics, Particle Physics, Proton Structure, DIS}
}

\newlength{\dinwidth}
\newlength{\dinmargin}
\newcommand{\empz}{\mbox{$E-P_z$}}

\newcommand{\ee}{\mbox{$E_e^{\prime}$}}
\newcommand{\thetae}{\mbox{$\theta_e$}}

\newcommand{\pt}{\mbox{$P_\perp$}}

\newcommand{\Eh}{E_{\rm h}}
\newcommand{\pth}{P_{\perp}^h}
\newcommand{\pte}{P_{\perp}^e}

\newcommand{\Fig}{\mbox{figure}}
\newcommand{\Tab}{\mbox{table}}

\newcommand{\Sec}{\mbox{section}}

\newcommand{\FFig}{\mbox{Figure}}

\newcommand{\Tabs}{\mbox{tables}}

\renewcommand{\perp}{{\rm T}}

\newcommand{\thetamax}{173^{\circ}}
\newcommand{\thetamin}{158^{\circ}}

\newcommand{\honepdf}{H1PDF~2009}

\def\dof{\mathop{n_{\rm dof}}\nolimits}

\newcommand{\gevsq}{\,\mathrm{GeV}^2}
\newcommand{\gev}{\,\mathrm{GeV}}
\newcommand{\pbi}{\,\mathrm{pb}^{-1}}

\begin{document}

\makeatletter \def\NAT@space{} \makeatother

\begin{titlepage}
 
\noindent
\begin{flushleft}
{\tt DESY 09-005    \hfill    ISSN 0418-9833} \\
{\tt September 2009}                  \\
\end{flushleft}

\noindent 

\vspace*{3.5cm}

\begin{center}
\begin{Large}

{\bfseries A Precision Measurement of the \\
 Inclusive $\mathbold{ep}$ Scattering Cross Section at HERA}

\vspace*{2cm}

H1 Collaboration

\end{Large}
\end{center}

\vspace*{2cm}

\begin{abstract} \noindent A measurement of the inclusive
  deep inelastic neutral current $e^+p$ scattering cross section is reported
  in the  region of four-momentum transfer squared, $12\gevsq\leq Q^2\leq 150\gevsq$, and Bjorken $x$, $2 \cdot 10^{-4} \leq x \leq
  0.1$. The results are based on data collected by the H1
  Collaboration at the $ep$ collider HERA at positron and proton beam
  energies of $E_e =27.6\gev$ and $E_p = 920\gev$, respectively. The
  data are combined with previously published data, taken at $E_p = 820\gev$.
  The accuracy of the combined measurement is
  typically in the range of $1.3 - 2\%$. A QCD analysis
  at next-to-leading order is performed to determine the parton
  distributions in the proton based on H1 data.
\end{abstract}

\vspace*{1.5cm}

\begin{center}
{Accepted by {\em Eur. Phys. J.} {\bf C}}
\end{center}

\end{titlepage}

\begin{flushleft}

F.D.~Aaron$^{5,49}$,           
C.~Alexa$^{5}$,                
K.~Alimujiang$^{11}$,          
V.~Andreev$^{25}$,             
B.~Antunovic$^{11}$,           
A.~Asmone$^{33}$,              
S.~Backovic$^{30}$,            
A.~Baghdasaryan$^{38}$,        
E.~Barrelet$^{29}$,            
W.~Bartel$^{11}$
% ...

\bigskip{\it
% ...
$ ^{5}$ National Institute for Physics and Nuclear Engineering (NIPNE) ,
          Bucharest, Romania \\
% ...
$ ^{11}$ DESY, Hamburg, Germany \\
% ...
$ ^{25}$ Lebedev Physical Institute, Moscow, Russia$^{ e}$ \\
% ...
$ ^{29}$ LPNHE, Universit\'{e}s Paris VI and VII, IN2P3-CNRS,
          Paris, France \\
 $ ^{30}$ Faculty of Science, University of Montenegro,
          Podgorica, Montenegro$^{ e}$ \\
% ...
$ ^{33}$ Dipartimento di Fisica Universit\`a di Roma Tre
          and INFN Roma~3, Roma, Italy \\
% ...
$ ^{38}$ Yerevan Physics Institute, Yerevan, Armenia \\
% ...
$ ^{49}$ Also at Faculty of Physics, University of Bucharest,
          Bucharest, Romania \\
% ...
$ ^e$ Supported by the Deutsche Forschungsgemeinschaft \\
% ...
}
\end{flushleft}

\newpage

\section{Introduction} \label{sec:introduction}

The electron-proton collider HERA extends the kinematic range
of deep inelastic lepton-nucleon scattering (DIS), determined
by the four-momentum transfer squared, $Q^2$, and Bjorken $x$,
by two orders of magnitude towards  high $Q^2$ and small $x$,
compared with fixed target experiments.
This allows proton structure to be thoroughly investigated
as is vital for the understanding of strong interactions,
described by Quantum Chromodynamics (QCD).

This paper presents the most accurate cross section data to date for
the inclusive neutral current process $e^+p \rightarrow e^+X$,
measured in the kinematic region $12 \gevsq \leq Q^2 \leq 150 \gevsq$
and $2 \cdot 10^{-4} \leq x \leq 0.1$. The data newly presented here
were taken in the year 2000 with positrons of energy $E_e = 27.6\gev$
and protons of energy $E_p = 920\gev$, corresponding to a centre of
mass energy $\sqrt{s} = 319\gev$. The luminosity amounts to $22\pbi$.
The measurements are combined with similar data taken in 1996/97 at
$E_p = 820\gev$~\cite{h1alphas}. The present paper refers to methods
and detailed explanations given in a recent publication~\cite{h1lowq2} of
lower $Q^2$ data, $0.2 \gevsq \leq Q^2 \leq 12 \gevsq$.
The results can be compared with data obtained by
the ZEUS Collaboration~\cite{Chekanov:2001qu,Chekanov:2009na}.

The double differential neutral current DIS cross section
in its reduced form and neglecting
contributions from $Z$ boson exchange, is given by
\begin{equation} \label{sigred0}
\sigma_r = \frac { Q^4 x} { 2\pi \alpha^2 [1+(1-y)^2]} \cdot
\frac{{\rm d}^2\sigma }{{\rm d}x dQ^2} = 
F_2(x,Q^2) -  f(y) \cdot  F_L(x,Q^2)
\end{equation}
with $\alpha$ denoting the fine structure constant and
$f(y)=y^2/[1+(1-y)^2]$. The inelasticity $y$ is related to $Q^2$, $x$
and the centre-of-mass energy squared, $s=4 E_e E_p$, by $y=Q^2/sx$. A
first measurement of the longitudinal structure function $F_L$ at low
$x$ was recently performed by H1~\cite{h1fl}. The measurement
presented here is restricted to the region of inelasticity $y \leq
0.6$ where the contribution of $F_L$ is small and thus it focuses on
the structure function $F_2$. The data are used to determine the
derivative $(\partial F_2 / \partial \ln Q^2)_x$ which provides a
sensitive test of the evolution dynamics of partons, and the
derivative $(\partial \ln F_2 / \partial \ln x)_{Q^2}$, which
quantifies the rise of $F_2(x,Q^2)$ at fixed $Q^2$ towards low $x$
\cite{Adloff:2001rw}.

The new data cover
the $Q^2$ region of deep inelastic scattering,
from a few $\mathrm{GeV}^2$ to about $150 \gevsq$,
with unprecedented accuracy at low $x$.
A QCD analysis at next-to-leading order (NLO) is performed to obtain a new
set of parton distribution functions (PDFs) from the inclusive DIS
cross section measurements of the H1 experiment alone. The QCD analysis is
based on the results presented here, the low $Q^2$ data~\cite{h1lowq2},
and the neutral and charged current (NC and CC) data sets at high
$Q^2$~\cite{Adloff:1999ah,Adloff:2000qj,Adloff:2003uh}.

\section{Measurement Technique}
\label{sec:xsection}

The analysis techniques are similar to those applied at lower
$Q^2$~\cite{h1lowq2}, more details of the present analysis can also be found in~\cite{kretzschmarth}.

\subsection{Detector}

The H1 detector~\cite{h1det,h1det2} was built and upgraded for the
accurate measurement of inelastic $ep$ interactions at HERA. The
detector components most relevant to this measurement are the central
tracker, the backward\footnote{The backward direction is determined by
  the outgoing positron beam direction. H1 uses a coordinate system
  with the positive $z$ axis given by the outgoing proton beam
  direction and the nominal interaction point at $z=0$.}
lead-scintillator calorimeter (SpaCal)~\cite{spacalc} and the liquid
argon calorimeter (LAr)~\cite{Andrieu:1993kh}. The central tracker
consists of the central jet drift chamber, two complementary $z$ drift
chambers, the central inner
(CIP)~\cite{Muller:1992jk,Eichenberger:1992km} and outer
proportional chambers, and the central silicon tracker
(CST)~\cite{cst}. The drift chambers and the CST are used for the
measurement of tracks from the hadronic final state. The momenta of
the tracks are determined and the event vertex is reconstructed. The
polar angle of the scattered positron is determined by the planar
backward drift chamber (BDC) and the vertex position. Complementary
tracking information is obtained from the backward silicon tracker
(BST)~\cite{Eick:1996gv}. The SpaCal contains electromagnetic and
hadronic sections. Its energy resolution for electromagnetic energy
depositions is $\delta E/E \approx 0.07/\sqrt{E/\mathrm{GeV}} \oplus
0.01$. It also provides a trigger based on the scattered positron
energy. The LAr allows the hadronic final state to be reconstructed.
Its energy resolution was determined to be $\delta E/E \approx
0.50/\sqrt{E/\mathrm{GeV}} \oplus 0.02$ with pion test beam
data~\cite{Andrieu:1993tz}.
Two electromagnetic crystal calorimeters, a photon tagger and an
electron tagger, located close to the beam pipe 
at $z = -103.1$\,m and $z = -33$\,m,
respectively, are used to monitor the luminosity via the measurement
of the Bethe-Heitler process $ep\to \gamma ep$.

\subsection{Online event selection} \label{sec:online}

The online trigger conditions used in this analysis are based on an
energy deposition in the electromagnetic section of the SpaCal. Three
trigger conditions with different energy and radius thresholds are
used, which largely overlap in phase space. The dominant source of
trigger inefficiency are the veto conditions against beam related
background. For these a global inefficiency of $(0.5 \pm 0.3)\%$ is
determined and corrected for. The inefficiency of the online software filter
is determined to be $0.2\%$, which is applied as a global correction
with a systematic uncertainty of half that size. The residual trigger
inefficiencies from other sources are smaller than $0.1\%$.

\subsection{Kinematics}
\label{sec:kinematics}

The DIS event kinematics are reconstructed from scattered positron and
hadronic final state information. The positron energy $\ee$ and
scattering angle $\theta_e$ are used. For the hadronic final state a
sum over the particles' energies $E_i$ and longitudinal momenta
$P_{z,i}$ is formed, $\Sigma_h = \sum_i (E_i - P_{z,i})$. The total
difference between energy and longitudinal momentum $\empz$ is
obtained by adding to $\Sigma_h$ the positron contribution, $\empz =
\Sigma_h + \ee(1-\cos \theta_e)$. Based on these variables,
measurements of $Q^2$ and $y$ are obtained using the electron and the
$\Sigma$ methods as explained in~\cite{h1lowq2}. In order to optimise
the measurement accuracy, the electron method is used at larger $y \gtrsim 0.1$,
while the $\Sigma$ method is used at lower $y$. For this analysis,
positrons with scattering angles between $\theta_e \approx \thetamin$
and $\theta_e \approx \thetamax$, energies $\ee > 11\gev$ and $Q^2_e>10\,\text{GeV}^2$ are included. The cross section measurement is performed in bins of $x$
and $Q^2$ chosen similarly to the previous
measurement~\cite{h1alphas}, with small modifications due to the
different centre of mass energy. For each bin and reconstruction
method the purity $P=N_{\rm rec,gen}/N_{\rm rec}$ and the stability
$S=N_{\rm rec,gen}/N_{\rm gen}$ are calculated. Here $N_{\rm rec}$
($N_{\rm gen}$) is the total number of reconstructed (generated) Monte
Carlo events in the bin and $N_{\rm rec,gen}$ is the number of events
which are both generated and reconstructed in the same bin. The values
for purity and stability exceed $40\%$ in all
analysis bins and are typically well above $50\%$ for the chosen
reconstruction method~\cite{kretzschmarth}.

\subsection{Positron and hadronic final state reconstruction}
\label{sec:eventkinerec}

The reconstruction of the scattered positron is based on the
measurement of a deposition of energy, termed a cluster, with a
limited transverse size characteristic of an electromagnetic shower.
The energy of the cluster $\ee$ is obtained by summing over all cells of the
cluster in the electromagnetic section of the SpaCal. Its transverse
size is characterised by $R_{log}$, which is obtained from the
positions of all SpaCal cells belonging to a cluster using a
logarithmic energy weighting~\cite{Awes:1992yp}. A cut $R_{log}<4\,\text{cm}$ is applied. For additional
background suppression, the energy deposition in the hadronic section
of the SpaCal behind the electromagnetic cluster, $\Eh$, is required
to be less than $15\%$ of $\ee$. The positron candidate cluster is
further required to be associated to a track in the BDC, formed by at
least $4$ hits from the $8$ layers. This ensures an accurate
measurement of the polar angle \thetae\ in combination with the $z$
position of the interaction vertex, $z_{vtx}$, determined with the
central track detectors.

The reconstruction of the hadronic final
state uses information from the
central tracker and the calorimeters LAr and SpaCal~\cite{h1lowq2}.
The determination of $\Sigma_h$ is affected by the presence of 
noise in the calorimeters. The resulting bias is particularly
strong for small $y_h=\Sigma_h/2E_e$.
Contributions of noise from the SpaCal and the LAr are suppressed as 
described in \cite{h1lowq2}.

\subsection{Monte Carlo event simulations}
\label{sec:mc}

Monte Carlo (MC) simulations are used to correct for detector
acceptance and resolution effects, and for the background subtraction.
The DIS signal events are generated using the DJANGOH~\cite{Schuler:1991yg}
event generator, while the PHOJET~\cite{Engel:1995yda} program is used
for the photoproduction background.
Elastic QED Compton events are generated using the COMPTON event
generator~\cite{Courau:1992ht}. The cross section measurement is
corrected for QED radiation up to order $\alpha$ using
HERACLES~\cite{heracles}. The radiative corrections are cross checked
with HECTOR~\cite{hector}. An agreement to better
than $0.3\%$ is found in the kinematic range of this measurement.

All generated events are passed through the full
GEANT~\cite{geant} based simulation of the H1 apparatus and are
reconstructed using the same program chain as for the data. For
consistency, the calibrations of the SpaCal and the LAr, as well as
the alignment, are performed for the reconstructed MC events in the
same way as for the data. The calorimeter noise is determined using
events from random triggers and is overlaid on the simulated events.
The simulated events are reweighted to match the cross section derived
from the QCD fit which is described in section~\ref{sec:qcdfit}.

\section{Data Analysis}
\label{sec:dataanal}

The data for the present measurement were recorded in the year 2000
and correspond to an integrated luminosity of $22\pbi$. In the
following a description of the analysis is given. Further information
can be found in~\cite{h1lowq2, kretzschmarth}.

\subsection{Event selection}

An overview of the selection criteria is given in \Tab~\ref{tab:cuts}.
Detection of the scattered positron is required within the
backward calorimeter SpaCal, as discussed in section
\ref{sec:kinematics} and \ref{sec:eventkinerec}. In order to ensure
that the cluster is well contained in the SpaCal, the extrapolation of
the associated BDC track segment to the SpaCal plane is required to be within a distance from the beam line of $r_{\rm Spac} < 73\,\text{cm}$.
The event vertex is reconstructed either by the central drift chambers and the CST or by using the CIP and the positron cluster position. Its $z$-position is required to be within $\pm 35\,\text{cm}$ of the centre of the detector.

\begin{table}[tb]
\centerline{%
\begin{tabular}{ll}
\hline
\multicolumn{1}{l}{\bfseries Description} & \multicolumn{1}{l}{\bfseries Requirement}\\[3pt]
\hline
Kinematic Range  & $Q^{2}_{e} > 10\gevsq$\\[3pt]
Scattered positron energy & $\ee > 11\gev$\\[3pt]
SpaCal cluster radius     & $R_{\rm log}<4$\,cm \\[3pt]
Energy in hadronic SpaCal section & $\Eh/\ee < 0.15$  \\[3pt]
BDC validation         & $\ge 4$ linked hits, BDC-SpaCal radial
match $< 2.5$\,cm\\[3pt]
Radial cluster position   & $r_{\rm Spac} < 73$\,cm\\[3pt]
Vertex $z$ position       & $|z_{\rm vtx}| < 35$\,cm \\[3pt]
Transverse momentum balance         & $\pth/\pte > 0.3$\\[3pt]
Longitudinal momentum balance    & $ \empz > 35\gev$ \\[3pt]
QED Compton Rejection  & Topological veto\\[3pt]
\hline
\end{tabular}
}
\caption{\label{tab:cuts} Event selection criteria.}
\end{table}

Events for which the hadronic final state is poorly reconstructed are
rejected by demanding that the total measured hadronic transverse
momentum $\pth$ be at least $30\%$ of the positron transverse momentum
$\pte$. This efficiently removes migrations from very low $y$. Events
with high energy initial state photon radiation are excluded
by requiring $\empz > 35\gev$. The QED Compton process is suppressed
using a topological cut against events with two back-to-back
electromagnetic clusters reconstructed in the SpaCal.

\subsection{Efficiency determination}

The efficiencies of the positron identification requirements
(transverse cluster size, hadronic energy fraction, BDC validation)
exceed $99\%$ in most of the phase space. They are well
described by the simulation. The only significant local inefficiency, of
about $5\%$, is
observed for the BDC validation requirement at a radial distance from
the beam line of $r_{\rm BDC} \sim 25\,$cm where the geometry of the BDC
drift cells changes. A detailed map of this inefficiency as a function
of $r_{\rm BDC}$ and the azimuthal angle of the scattered positron is obtained
using positron candidates validated by the BST. The
simulation is adjusted accordingly and an additional $0.5\%$
systematic uncertainty is added to the cross section measurement
error, uncorrelated from bin-to-bin.
 
The efficiency of the vertex reconstruction is determined using events
independently reconstructed by the BST. This efficiency is
  determined to be close to $100\%$ for all
but a few bins at $Q^{2} < 20 \gevsq$ and $y < 0.03$, where it drops
to about $75 - 95\%$. It is described by the
simulation to an accuracy of $0.3\%$ for $y > 0.01$. For lower $y$ the
description is accurate to about $1\%$, which is accounted for by an
additional uncorrelated uncertainty in the corresponding bins.

\subsection{Alignment and calibration}
\label{sec:alignment}

\begin{figure}
  \epsfig{file=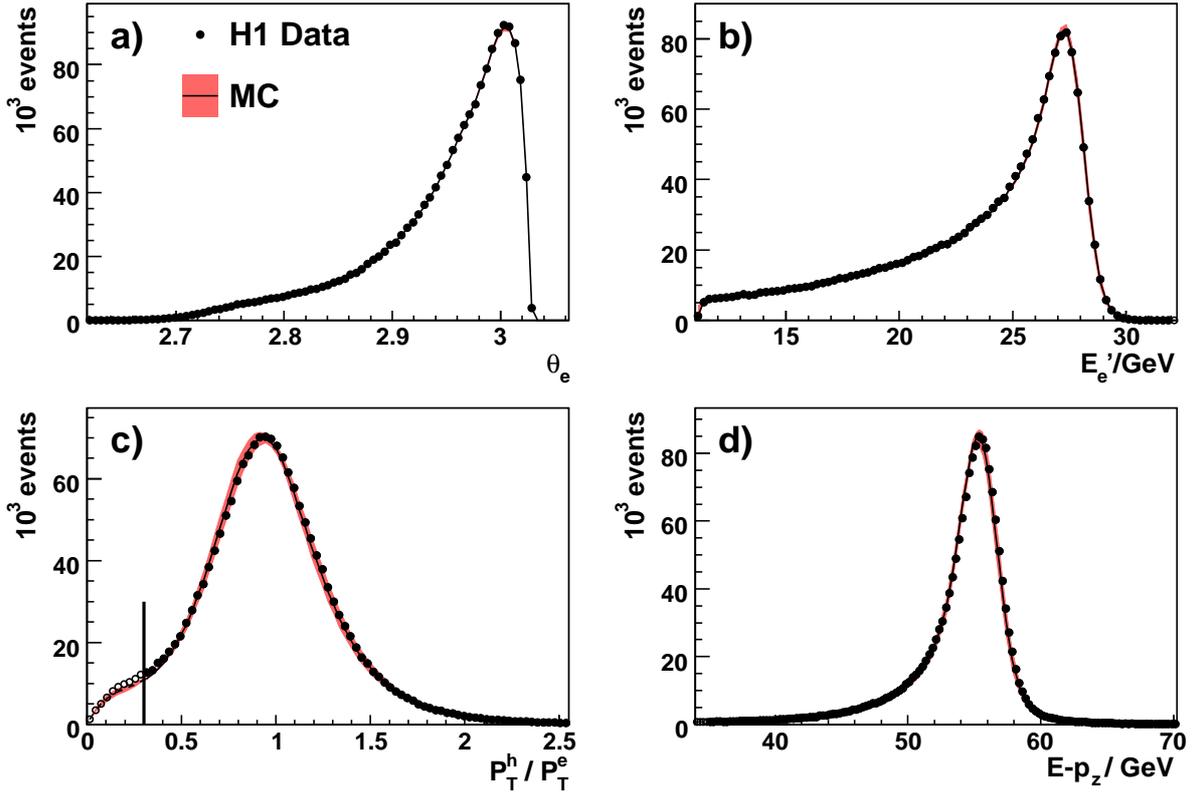,width=\linewidth}
  \caption{\label{fig:controltech} Distribution of events
    requiring $y_e < 0.6$ and $y_\Sigma > 0.005$:
    the polar angle a) and the energy b) of the scattered positron,
    the transverse momentum ratio $\pth/\pte$ (data outside the analysis
    selection indicated by the vertical line shown as open symbols)
    c) and $\empz$ d).
    The curves represent the MC simulation
    normalised to the luminosity, including a very small contribution
    from photoproduction.
    The narrow bands illustrate the correlated systematic
    uncertainties of the measurement.}
\end{figure}

The alignment of the H1 detector starts from the internal adjustment
of the central tracker and proceeds with the backward detectors, BDC,
SpaCal and BST. The alignment of the BDC is performed using the tracks
of positron candidates that are reconstructed in the central tracker
with high accuracy. The SpaCal position is adjusted based on the
positron tracks measured in the BDC. Finally the BST is aligned using
events with a well reconstructed central vertex and a positron track
measured in the BDC. The resulting agreement of the polar angle
measurements is better than $0.2$\,mrad, which defines the associated
systematic uncertainty. The measured and
simulated scattered positron polar angle distributions are shown in
\Fig~\ref{fig:controltech}a). The data are well described by the
simulation.

The calibration of the electromagnetic scale of the SpaCal
corrects for differences in the gain factors of individual SpaCal
cells, for local non-uniformities at the sub-cell level and for global
non-linearity in the energy response. The calibration is based mainly on the positron
candidates at low $y$. For these the kinematics can be reliably
reconstructed using the double angle reconstruction method as described
in~\cite{h1lowq2}, which employs only the polar angle information of the
hadronic final state and the scattered positron, in addition to the positron beam
energy. The non-linearity of the energy response is determined and
corrected for using a
sample of $\pi^0 \to \gamma\gamma$ events. The energy scale is then
checked using elastically produced $J/\psi$ particles decaying
to $e^+e^-$ and QED Compton events, $ep \to ep\gamma$, where the
scattered positron and photon are both reconstructed in the SpaCal.
The relative data-to-simulation
scale uncertainty is derived from two contributions: a
global $0.2\%$ arising from the double angle calibration and a part resulting
from the $\pi^0$ studies, which is $1\%$ at
$E=2\gev$ linearly decreasing to zero at $E=27.6\gev$. The measured and
simulated scattered positron energy distributions are in very good
agreement, as shown in \Fig~\ref{fig:controltech}b).

The calibration of the calorimeters employed for the hadronic final
state energy measurement is based on kinematic constraints relating
the scattered positron to the hadronic final state. For the
calibration of the LAr calorimeter, the conservation of the total
transverse momentum \pt\ is used. The  hadronic 
SpaCal calibration utilises the
conservation of \empz.

The transverse momentum balance between the scattered positron and the
hadronic final state is studied as a function of various
variables, such as $P^e_\perp$, the polar angle of the hadronic final
state, and $y_\Sigma$. \FFig~\ref{fig:controltech}c) shows the overall
$\pth/\pte$ distribution with a vertical line at $0.3$ indicating the
analysis cut value. For $\pth/\pte$ values larger than this cut, the data
distribution lies inside a band given by varying the LAr hadronic
energy scale in the simulation by $2\%$ for $y > 10^{-2}$. At the
lowest $y \simeq 0.005$ considered in the measurement, the hadronic
final state is produced at small polar angles, and a significant part
escapes detection. The systematic
uncertainty on the hadronic energy scale is increased from $2\%$ at $y
= 10^{-2}$ linearly in $\log y$ to $10\%$ at $y = 10^{-3}$.
Furthermore, for bins with $y < 10^{-2}$, an additional uncorrelated
cross section uncertainty of $2\%$ is estimated by varying the
$\pth/\pte$ cut between $0.25$ and $0.35$.

A topological finder is used to identify and subtract LAr noise.
The fraction of hadronic energy attributed to noise
is described by the simulation to within $15\%$,
which is taken as a systematic uncertainty.

For large values of $y$, the contribution of the SpaCal to $\Sigma_h$
becomes larger than the combined contribution of the LAr calorimeter
and the tracks, and thus the total \empz{}, expected to be $2E_e$,
provides the calibration for the hadronic final state measurement in
the SpaCal. A study at $y \simeq 0.5$ shows that the hadronic energy
measurement in the SpaCal is described by the simulation to
$0.3\gev$. \FFig~\ref{fig:controltech}d) shows the \empz\
distribution for the data and the simulation. The simulation
reproduces the data within the combined calibration uncertainties.

\subsection{Background}
\label{sec:bg}

A small source of background for this analysis arises from
photoproduction events, in which the scattered positron escapes
detection in the backward beam pipe while a particle from the hadronic
final state mimics the positron. For a fraction of photoproduction
events the scattered positron is detected by the electron tagger of
the luminosity system.
The photoproduction MC (PHOJET) is normalised globally
based on a comparison with tagged events applying all selection
criteria apart from the \empz\
cut. The systematic uncertainty on the background normalisation is
taken to be $15\%$, based on the studies described
in~\cite{h1lowq2}. The background contribution at the highest $y \sim
0.5$ considered here typically amounts to $5\%$, hence yielding an
uncertainty of less than $1\%$ on the cross section. Potential
background from non-$ep$ 
interactions is studied using non-colliding bunches and found to be
negligible.

\subsection{Summary of systematic uncertainties}
\label{sec:sysuncert}

\begin{table}
\centerline{%
\begin{tabular}{l c}
\hline\hline
\multicolumn{2}{c}{{\bfseries Correlated systematic errors}} \\[3pt]
{\bfseries Source} &  {\bfseries Uncertainty}\\[3pt]
\hline
\ee\ scale   & $0.2\%$ \\[3pt]
\ee\ linearity  & $1\%$ at $2\gev$ to $0\%$ at
$27.6\gev$  \\[3pt]
Polar angle \thetae  & $0.2\,$mrad  \\[3pt]
LAr hadronic scale  & $2\%$ for $y > 0.01$\\
 & increasing linearly in $\log y$ to $10\%$ at $y=0.001$ \\[3pt]
LAr noise contribution to \empz   & $15\%$ \\[3pt]
SpaCal hadronic scale & $ 0.3\gev$ \\[3pt]
$\gamma p$ background normalisation & $15\%$ \\[3pt]
Luminosity and other global uncertainties & $1.2\%$ \\[3pt]
\hline\hline
\multicolumn{2}{c}{{\bfseries Uncorrelated systematic errors}} \\[3pt]
{\bfseries Source} &  {\bfseries Uncertainty} \\[3pt]
\hline
BDC efficiency  & $0.5\%$\\[3pt]
Vertex efficiency  & $0.3\%$\\[3pt]
Radiative corrections & $0.3\%$  \\[3pt]
Additional uncertainty for bins with $y \leq 0.01$ & $2.0\%$  \\[3pt]
\hline\hline
\end{tabular}
}
\caption{\label{tab:syssum}
  Summary of the systematic uncertainties. For the correlated error
  sources, the uncertainties are given in terms on the uncertainty
  on the corresponding source.
  For the uncorrelated error sources, the uncertainties
  are quoted directly in terms of the effect on the measured cross section.}
\end{table}

The systematic uncertainties are classified into two groups,
bin-to-bin correlated and uncorrelated errors. They are
summarised in \Tab~\ref{tab:syssum}. The electromagnetic and hadronic
energy scales, the positron scattering angle, LAr noise,
background subtraction and normalisation are all considered to be
correlated sources of uncertainty. The uncorrelated errors arise from
various efficiencies and the radiative corrections. For most of the
analysis phase space none of the sources of systematic uncertainty
dominates the result. For very low $y < 0.01$, many uncertainties
increase strongly, as the reconstruction efficiency drops rapidly and
the modelling of the hadronic final state suffers from increased
losses in the forward direction.

The uncertainty on the global normalisation of the measurement is
determined mostly from the luminosity measurement, which is accurate
to within $1.1\%$. The additional
corrections discussed in \Sec~\ref{sec:online} increase the overall
normalisation uncertainty to $1.2\%$.

\subsection{Data - Monte Carlo Comparison}

\begin{figure}
  \epsfig{file=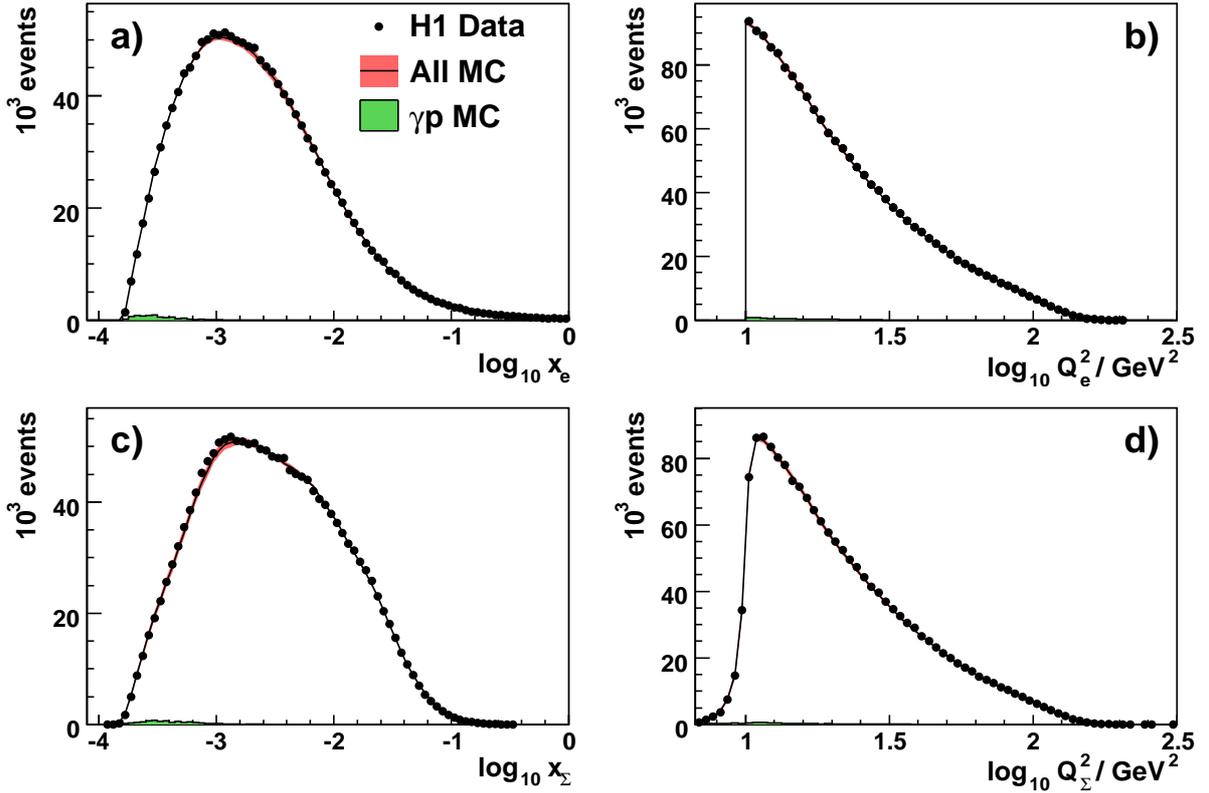,width=\linewidth}
  \caption{\label{fig:controlplots00} Bjorken-$x$ and $Q^2$ distributions of events
    requiring $y_e < 0.6$ and $y_\Sigma > 0.005$:
    $x_e$ and $Q^2_e$ reconstructed using the electron method a)-b),
    and $x_\Sigma$ and $Q^2_\Sigma$ reconstructed using the $\Sigma$
    method c)-d).
    The curves represent the MC simulation
    normalised to the luminosity. 
    The narrow bands illustrate the correlated systematic
    uncertainties of the measurement.
    The small photoproduction background is drawn shaded.}
\end{figure}

The quality of the MC description of the measurement can be verified
by comparing experimental and simulated distributions in the kinematic
range of the measurement, corresponding approximately to 
$y_\Sigma > 0.005$ and $y_e < 0.6$. The DIS MC cross section prediction is
reweighted to the QCD fit discussed in \Sec~\ref{sec:qcdfit}. The
distributions are normalised to the luminosity.
\FFig~\ref{fig:controlplots00} a)-d) shows the $x$ and $Q^2$
distributions of the selected events for the two kinematic
reconstruction methods, electron and $\Sigma$. A good overall agreement is
obtained in the description of the data by the simulation.

\section{DIS Cross Section Results}
\label{sec:results}

The measurement of the inclusive double differential cross section
for deep inelastic positron-proton scattering, $e^+p \rightarrow
e^+X$ with the $E_p = 920\gev$ data, is reported in the tables~\ref{tab:table00} and
\ref{tab:table00_2}. The tables show the statistical, uncorrelated and 
the various correlated error contributions. The region of small Bjorken $x$, $2
\cdot 10^{-4} \leq x \leq 0.1$, and four-momentum transfer squared,
$12 \gevsq \leq Q^2 \leq 150 \gevsq$, is covered.

\begin{figure}
\centerline{\epsfig{file=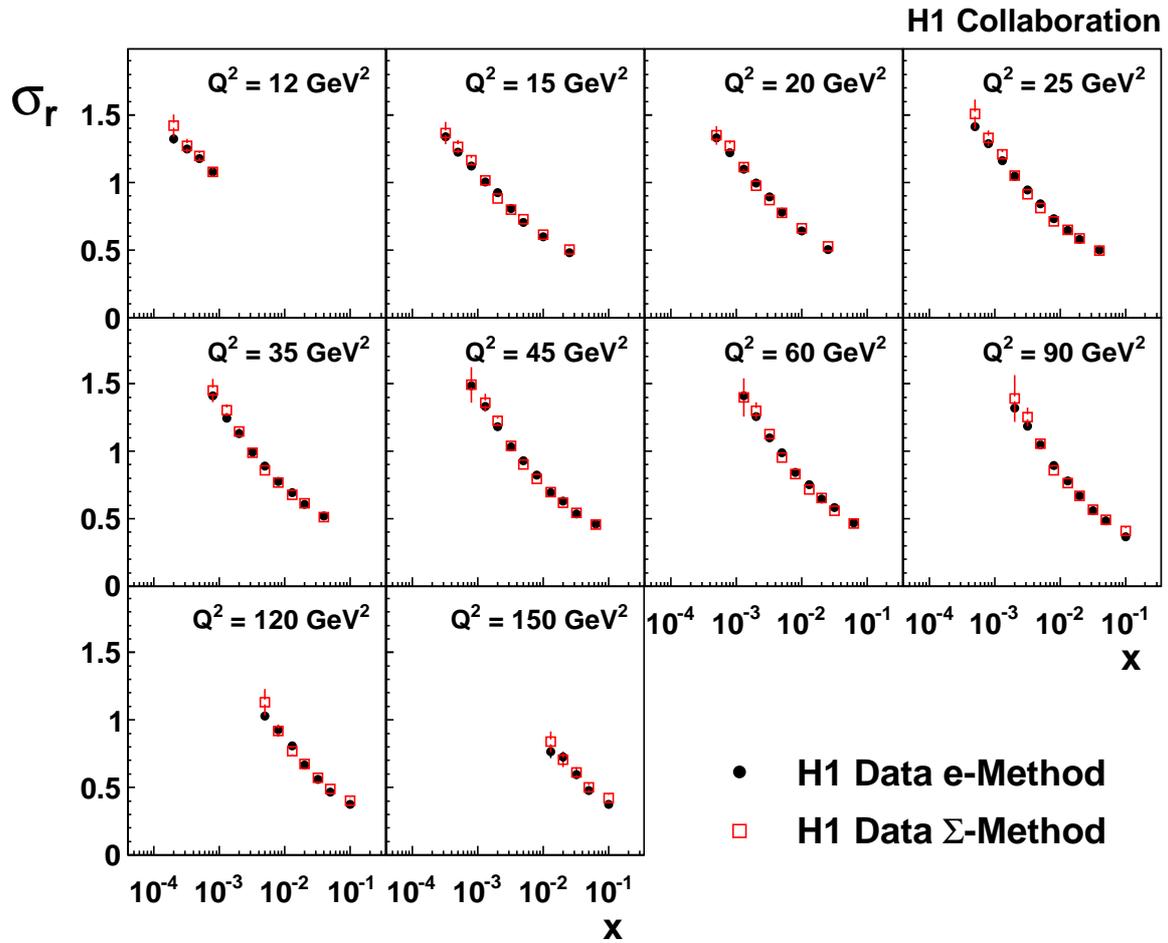,width=\linewidth}}
\caption{\label{fig:elsigma}
Comparison of reduced cross sections as obtained from the $E_p =
920\gev$ data with the electron (closed circles) 
and $\Sigma$ (open squares) reconstruction methods. The error bars
represent the total measurement uncertainties.}
\end{figure}

The stability of the cross section measurement is tested with a set of
dedicated cross checks. This is done by splitting the data into two
approximately equal subsamples and comparing the results obtained
using the same reconstruction method on each. For example, the data
are compared as measured with the upper and the lower half of the
SpaCal, for negative and positive $z_{vtx}$ positions, and dividing
the sample into early and late data taking periods. These tests are
sensitive to local efficiency problems, energy miscalibrations, or the
stability of the luminosity measurement. A particularly interesting
test is the comparison of the cross section measurements performed
with the electron and $\Sigma$ methods, which have different
sensitivities to systematic error sources. The results shown in
\Fig~\ref{fig:elsigma} demonstrate very good agreement taking the
correlated uncertainties into account. For the final result, the
method with the smaller total uncertainty is chosen in each bin, which
results in a transition near $y = 0.1$.

This analysis improves the uncertainties by up to a factor of two
with respect to previously published results on the
inclusive DIS cross section in this kinematic range by
H1~\cite{h1alphas} and by ZEUS~\cite{Chekanov:2001qu,Chekanov:2009na}. 

\section{Combination of 820 and 920\,GeV Data}

The present measurement is based on data taken with $920\gev$ proton
beam energy. A similar data set was obtained in 1996/97 at $820\gev$
and published in~\cite{h1alphas}. The results were obtained from
two different samples, called \textit{A} and \textit{B}, covering higher $Q^2 \geq 12
\gevsq$ and lower $Q^2 \leq 12 \gevsq$ regions, respectively. A
comparison of the cross section measurement for sample \textit{A} with the
present analysis revealed a significant deviation, which required a
dedicated study as described below.

\subsection{Correction of the 820\,GeV data}
\label{sec:oldpubcorr}

For the analysis of the $820\gev$ data an older DIS event simulation
program version was used.
If that program version is used without event weighting, the MC cross
section agrees to better than $0.3\%$ with the HECTOR calculation.
However, in order to improve the statistical accuracy of the DIS MC
event sample, the simulated
events in~\cite{h1alphas} were generated with $Q^2$ dependent weights for $Q^2 <
50\gevsq$. This introduced a bias which depends only on $Q^2$. The
cross section as obtained in~\cite{h1alphas} can be corrected by a
factor defined as
\begin{equation}
  c(Q^2) = \frac{N_{MC}^{weighted}(Q^2)}{N_{MC}(Q^2)} \,.
\end{equation}
Here $N_{MC}(Q^2)$ and $N_{MC}^{weighted}(Q^2)$ are the sums of weights of
events generated for a given $Q^2$ bin in unweighted and weighted
mode, respectively. 
The dependence of the correction factor on $Q^2$ is
empirically parameterised by
\begin{equation}
c(Q^2) =
\begin{cases}
  c_0 & \mbox{for $Q^2 > 50\gevsq$} \\
  c_0 + c_1 \cdot \log_{10} (Q^2/50\gevsq) 
     & \mbox{for $Q^2 \leq 50\gevsq$}
\end{cases}
\label{e:cdjango}
\end{equation}
with $c_0=1.027$ and $c_1=0.0352$.
This correction procedure introduces an additional uncorrelated
cross section uncertainty of $0.5\%$ on the corrected published data.

For a further cross check the 1996/97 data are reanalysed within the
framework of the present analysis. The luminosity determination is
improved, resulting in an additional normalisation shift by $+0.5\%$,
corresponding to one third of a standard deviation of the quoted
normalisation uncertainty. No further significant deviation is
observed between the published results and the reanalysis.

Tables\,\ref{tab:table97} and \ref{tab:table97_2} present the 1996/97
data from \cite{h1alphas}
corrected for the $Q^2$ weighting bias and the small luminosity shift.
These tables therefore replace the previous data for $Q^2 \geq
12\gevsq$, sample \textit{A}.
The measurements of $F_2$ for $y < 0.6$ are extracted from the
corrected cross sections accounting 
for the $F_L$ influence, as described in \Sec~\ref{sec:f2}.
The sample \textit{B} in \cite{h1alphas}, extending to lower
values of $Q^2 \leq 12\gevsq$, was not affected by the MC weighting
problem and  has been combined with further H1
data as described in \cite{h1lowq2}.

\subsection{Combined cross sections}
\label{sec:averaging}
\begin{figure}
\centerline{\epsfig{file=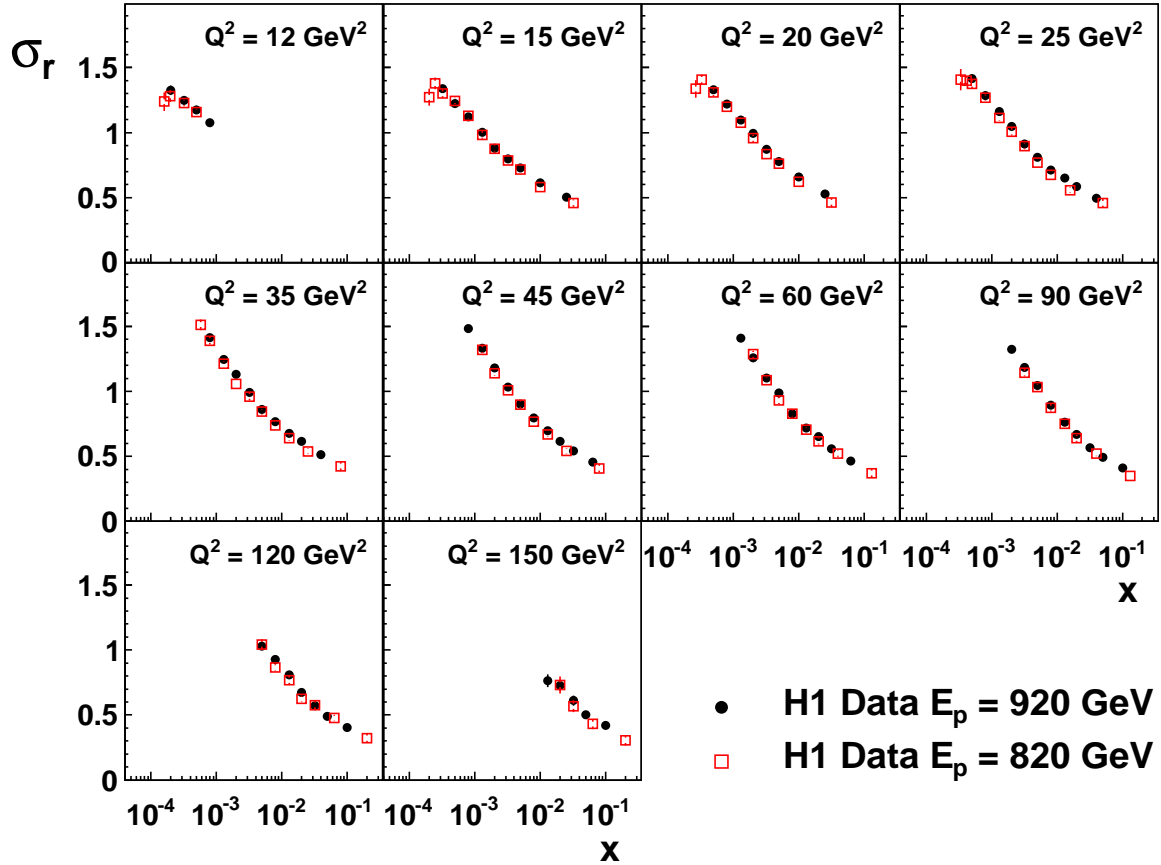,width=\linewidth}}
\caption{\label{fig:xsec_9700}
Comparison of reduced cross sections as obtained from the $E_p =
920\gev$ (closed circles) and the corrected results from $E_p =
820\gev$ (open squares). The error bars
represent the total measurement uncertainties.}
\end{figure}

The corrected $820\gev$ data and the present $920\gev$ data are
shown in figure~\ref{fig:xsec_9700} and combined to determine a new H1
measurement at $12 \gevsq \leq Q^2 \leq 150 \gevsq$. The combination
of the data sets is based on the prescription introduced
in~\cite{glazov} and developed further in~\cite{h1lowq2}. The
combination uses the reduced cross section data and consistently
treats the correlated and uncorrelated uncertainty information, which
is given in \Tabs~\ref{tab:table00}-\ref{tab:table97_2}. The data are
averaged for the region of large $x$, defined as in~\cite{h1lowq2} by
$y < 0.35$, but kept separate at higher $y$ where the
reduced cross section depends significantly on the centre of mass
energy via the $F_L$ term. Small differences in the measurement grid
are corrected for by adjusting cross section values for points close
in $x$ such that they correspond to a common $x$ value, prior to the
averaging.

In the combination of the two data sets,
assumptions have to be made about the relationship between systematic
uncertainty sources, which may not be fully independent between the
different analyses of the 1996/97 and 2000 data.
Reasons for correlations between data sets are the
similarity in the calibration procedure and the detector setup,
whereas uncorrelated effects include variations with
time, beam conditions and changes in the analysis procedure. To estimate the
sensitivity of the averaged result to the correlation assumptions,
different averages of the 1996/97 and 2000 data are compared, 
obtained considering all possible assumptions as to which systematic 
error sources are taken to be correlated or uncorrelated between the data sets.
The observed variation 
of the averaged cross sections is typically at the per-mille level, which is 
negligible compared with the total uncertainty. The variations of the total 
uncertainties never exceed $10\%$ of the final uncertainty. 
The assumption that the systematic uncertainties on the two data sets
are uncorrelated yields the largest uncertainty. Thus no correlations between
the error sources of the 1996/97 and 2000 data are considered
in the combination procedure.

The 1996/97 and 2000 data sets are fully consistent, as determined in the
averaging procedure, with $\chi^2_{\rm tot}/\dof = 51.6/61$. As a result
of the averaging, 
the central values of the correlated systematic uncertainties are
modified. The shifts of the central values in units of the original
uncertainties are given in
\Tab~\ref{tab:avg9700}. It is observed that none
of the absolute values of the shifts exceeds one standard deviation.

\begin{table}[tb]
  \centerline{%
  \begin{tabular}{l|rr}
    \hline\hline
    {\bfseries Systematic Source} & \multicolumn{2}{c}{\bfseries Shift}\\
    &   1996/97 &  2000 \\
    \hline
    \ee ~scale            & $0.72$    & $0.50$  \\
    \ee ~linearity        & ---       & $-0.39$  \\
    Polar angle \thetae   & $-0.46$   & $0.09$  \\
    LAr hadronic scale     & $-0.86$   & $-0.13$  \\
    LAr noise              & $-0.22$   & $0.04$  \\
    SpaCal hadronic scale   & ---      & $0.35$  \\
    $\gamma p$ background  & $0.11$    & $-0.11$  \\
    Luminosity             & $0.64$    & $-0.46$  \\
    \hline\hline
  \end{tabular}%
}
  \caption{\label{tab:avg9700} 
    Shifts in the central values of the systematic uncertainties 
    determined for the combination of the 1996/97 data at $E_p=820\gev$
    and the 2000 data at $E_p=920\gev$. The value 
    for each shift is given
    in units of the original uncertainty. Note that for
    the two data sets some of the correlated systematic error sources
    were defined differently.}
\end{table}

The combined measurement of the reduced $ep$ scattering cross section
and its uncertainties are listed in \Tabs~\ref{tab:table9700comb} and
\ref{tab:table9700comb2}. The correlated uncertainties are given as 14
new sources, $\delta_i$, after diagonalisation of the error matrix as
explained in~\cite{h1lowq2}. The combined H1 data on the inclusive
cross section measurement have total uncertainties of $1.3-2\%$ in most of the
phase space.

\subsection{$\mathbold{F_2}$ and its derivatives}
\label{sec:f2}

\begin{figure}
\centerline{\epsfig{file=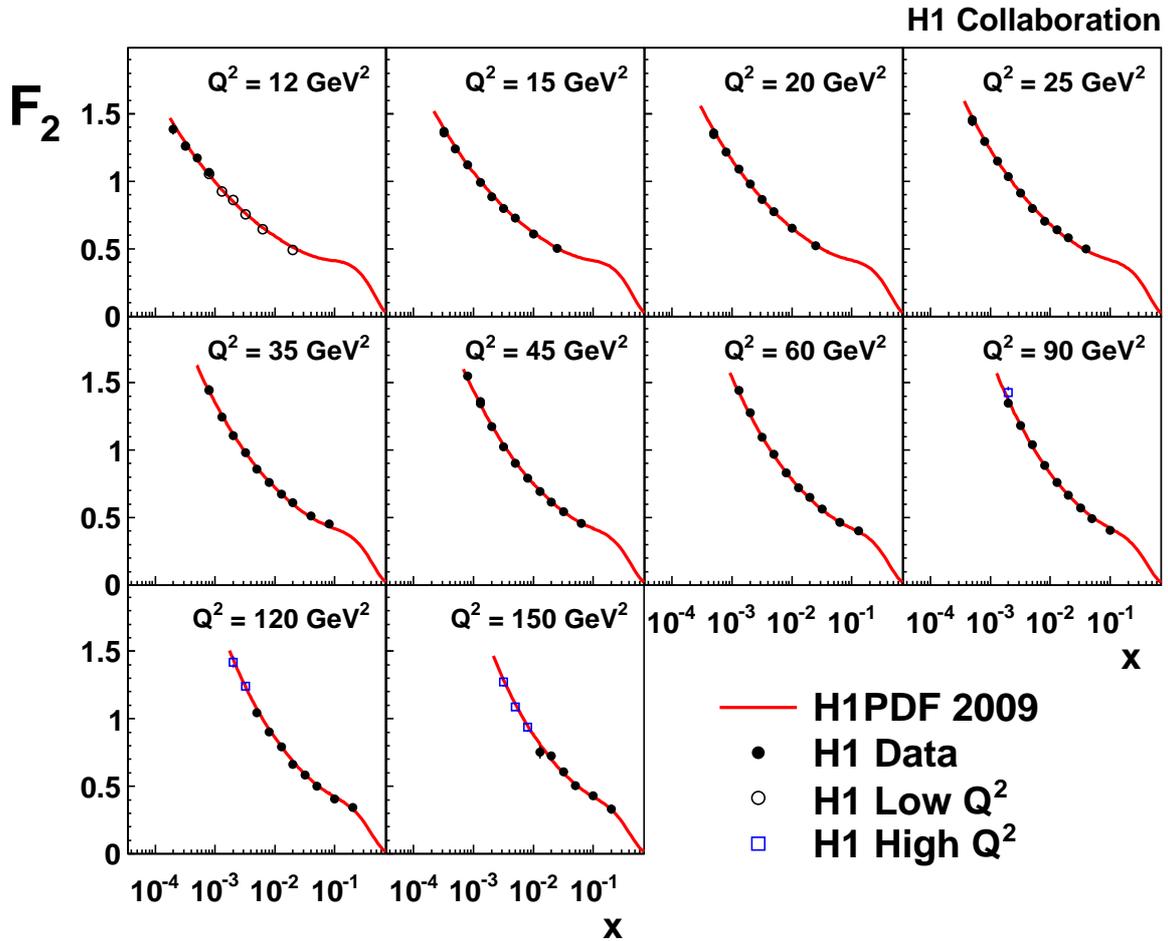,width=\linewidth}}
\caption{\label{fig:xseccomb} Measurement of the structure function
  $F_2$ at fixed $Q^2$ as a function of $x$. The data of this
  measurement (closed circles) are complemented by the previously
  published data at low $Q^2$ (open circles)~\cite{h1lowq2} and high
  $Q^2$ (open boxes)~\cite{Adloff:2003uh}. The error bars represent
  the total measurement uncertainties. The curve represents the QCD
  fit described in this paper.}
\end{figure}

The cross section data are used to extract $F_2$ 
for $y < 0.6$, correcting for the small
influence of $F_L$ using
\begin{equation}
  F_2 = \frac{\sigma_r}{1-f(y)\frac{R}{1+R}} \,.
\label{e:rflf2}
\end{equation}
The values used for $R=F_L/(F_2-F_L)$  
are quoted in the cross section tables and
are taken from the NLO QCD fit introduced below.
\FFig~\ref{fig:xseccomb} shows the structure function
$F_2$  at fixed $Q^2$ as a function of $x$ together with H1 data from
lower~\cite{h1lowq2} and higher $Q^2$~\cite{Adloff:2003uh}.
The data are well described by the NLO QCD fit.
The rise of $F_2$ towards low $x$ is thus established 
 at a much improved accuracy  compared
with the first observations \cite{h1firstf2,zeusfirstf2}. 
There is no indication for a saturation of this
behaviour in the $Q^2, x$ region of study.

\begin{figure}
\centerline{%
\epsfig{file=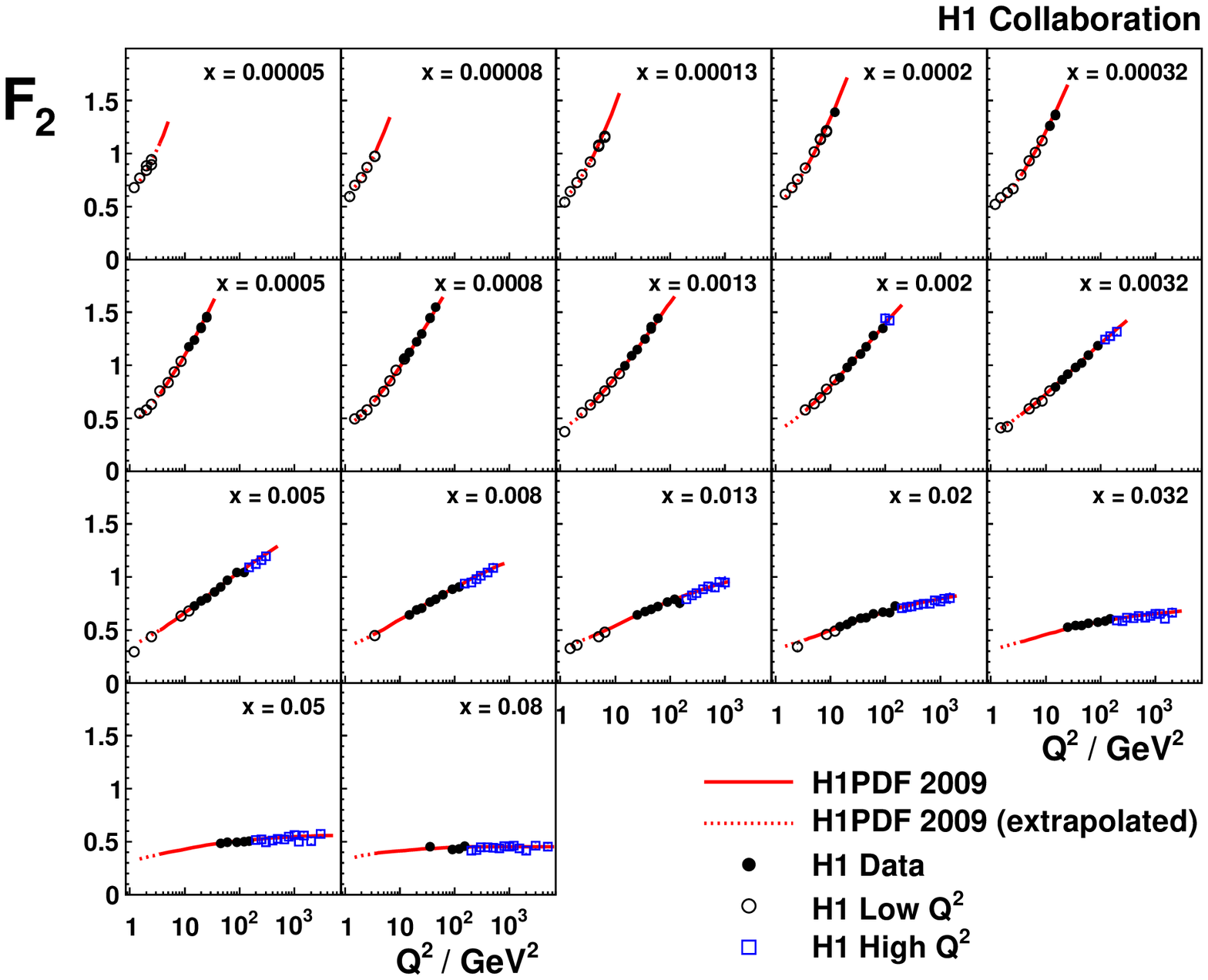,width=\linewidth}
}
\caption{\label{fig:f2_q2dep} Measurement of the structure function
  $F_2$ as a function of $Q^2$ at various values of $x$. The data of
  this measurement (closed circles) are complemented by the previously
  published data at low $Q^2$ (open circles)~\cite{h1lowq2} and high
  $Q^2$ (open boxes)~\cite{Adloff:2003uh}. The error bars represent
  the total measurement uncertainties. The solid curve represents the
  QCD fit described in this paper for $Q^2 \geq 3.5\gevsq$, which is
  also shown extrapolated down to $Q^2 = 1.5\gevsq$ (dashed).}
\end{figure}

\FFig~\ref{fig:f2_q2dep} shows the measurement of the 
structure function $F_2$ at fixed  values of $x$ 
 as a function of $Q^2$, compared with the QCD fit described below.
The strong  rise with $Q^2$
at low $x$ is a consequence of the large gluon density in this
region. The data are well described by the QCD fit for
$Q^2>3.5\gevsq$. Also the extrapolation of the fit down to $Q^2=1.5 \gevsq$
gives a reasonable description of the data.
At the largest $x$ value covered by the data presented
here, $x \simeq 0.1$, the structure function becomes almost independent of $Q^2$
as the result of a compensation of quark and gluon
contributions to the $\ln Q^2$ derivative of $F_2$.

\begin{figure}
\centerline{%
\epsfig{file=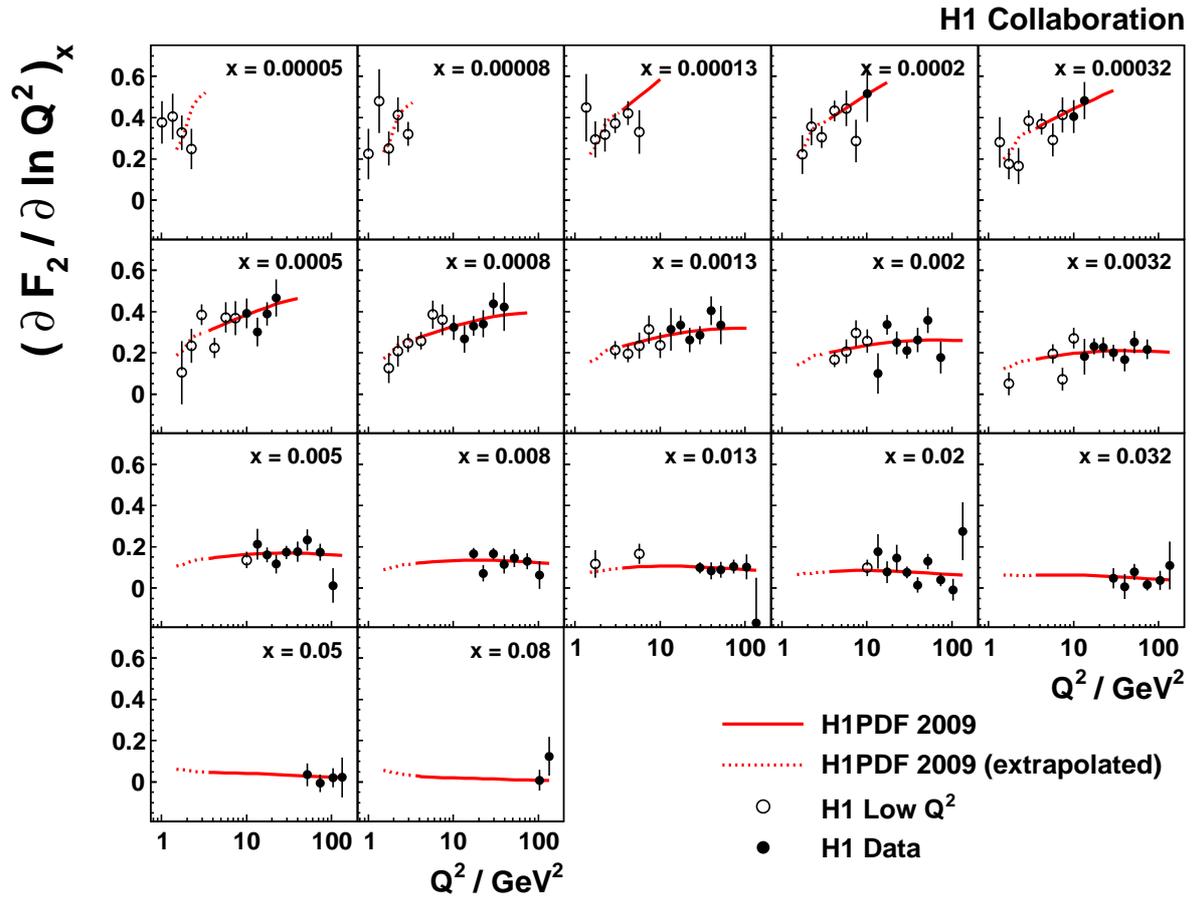,width=\linewidth}
}
\caption{\label{fig:df2dlnq2} Logarithmic $Q^2$ derivative of the structure
  function $F_2$ as a function of $Q^2$ at various values of $x$.
  The data of this measurement
  (closed circles) are complemented with the published data at
  lower $Q^2$ (open circles)~\cite{h1lowq2}.
  The error bars
  represent the total measurement uncertainties.
  The solid curve represents the prediction of the QCD fit for
  $Q^2 \geq 3.5\gevsq$, which is also shown extrapolated down to $Q^2 =
  1.5\gevsq$ (dashed).}
\end{figure}

The DGLAP evolution
equations~\cite{Gribov:1972ri,Gribov:1972rt,Lipatov:1974qm,Dokshitzer:1977sg,Altarelli:1977zs}
 determine the derivative
$(\partial F_2 / \partial \ln Q^2)_x$ taken at fixed $x$,
with the dominant low $x$ contribution arising 
from gluon splitting into a quark-anti-quark pair.
 The measurement of this derivative is a powerful constraint on 
the gluon distribution $xg$ and the strong coupling constant
$\alpha_s$~\cite{Johnson:1977dy}. 
A study of $(\partial F_2 / \partial \ln Q^2)_x$ at
low $x$ has been presented previously by the H1
Collaboration~\cite{h1alphas}. The method described there
is used to determine this derivative for the new,
combined $F_2$ data set, including the low $Q^2$
data\,\cite{h1lowq2}. The results are shown in 
\Fig~\ref{fig:df2dlnq2} for
different $x$ values as a function of $Q^2$.
The dependence of the derivative on $Q^2$ is well
reproduced by the QCD fit.

\begin{figure}
\centerline{%
\epsfig{file=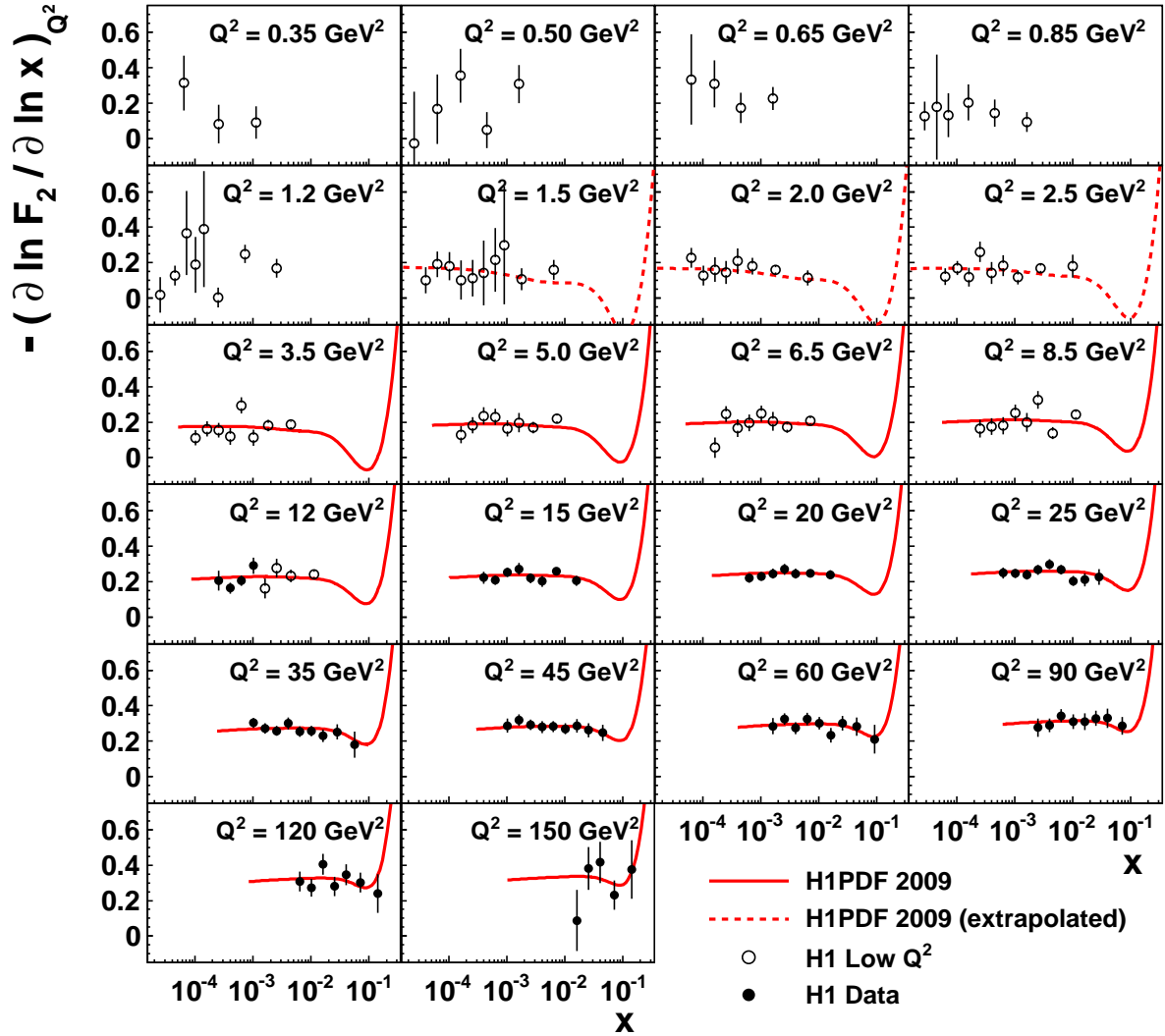,width=\linewidth}
}
\caption{\label{fig:df2dlnx} Measurement of the function $\lambda(x, Q^2)$, defined
  as the negative logarithmic $x$ derivative of $\ln F_2$
  as a function of $x$ at various values of $Q^2$.
  The data of this measurement
  (closed circles) are complemented with the published data at
  lower $Q^2$ (open circles)~\cite{h1lowq2}.
  The error bars
  represent the total measurement uncertainties.
  The solid curve represents the prediction of the QCD fit for
  $Q^2 \geq 3.5\gevsq$, which is also shown extrapolated down to $Q^2 =
  1.5\gevsq$ (dashed).}
\end{figure}

The rise of the structure function $F_2(x,Q^2)$
towards low $x$ may be quantified
by the derivative $\lambda = -(\partial \ln F_2 / \partial \ln x)_{Q^2}$
which is shown in \Fig~\ref{fig:df2dlnx}. The result
is more accurate than the previous measurement \cite{Adloff:2001rw}
and extends to lower $Q^2$. Within the uncertainty
of the data, the derivative is constant at small $x < 0.01$,
i.e. $F_2$ for fixed $Q^2$ is consistent with a power law
$F_2 \propto x^{-\lambda}$. Small departures from this behaviour,
as are inherent in the QCD fit, cannot be excluded either. The value of 
$\lambda$ increases from about $0.1$ to $0.3$ in the
$Q^2$ region covered, from about $1$ to $100\gevsq$.
Data from \cite{h1lowq2} allow 
this measurement to be extended to $Q^2$ values below the region of
validity of the DGLAP evolution.

\section{QCD Analysis}
\label{sec:qcdfit}

The neutral current cross section measurements presented here,
together with the measurements at lower $Q^2$~\cite{h1lowq2} and the
previously published NC and CC data at higher $Q^2$ 
\cite{Adloff:1999ah,Adloff:2000qj,Adloff:2003uh}, provide an
accurate H1 data set for the determination of the parton density
functions (PDFs) of the proton. A new QCD analysis, referred to as \honepdf,
is performed, which supersedes the previous H1PDF~2000
fit~\cite{Adloff:2003uh}, as it relies on the more accurate new data. It
also uses a general variable flavour number scheme (VFNS)
treatment~\cite{Thorne:2006qt} of the heavy quarks, unlike the former
fit, which used a zero mass variable flavour number scheme (``massless"
scheme).

\subsection{Framework and settings}

The QCD analysis decomposes the structure functions into a set
of parton densities: the gluon $xg$, the valence quarks, $x u_v$ and
$x d_v$, and the combined anti-up type and anti-down type quarks, $x
\bar{U} = x \bar{u} + x \bar{c}$ and $x \bar{D} = x \bar{d} + x
\bar{s} + x \bar{b}$. These are parameterised\footnote{
  The previous H1PDF~2000 fit used a very similar decomposition
  of the quark flavours, but instead of the
  valence quark distributions $x u_v$ and $x d_v$, the combined up
  and down quark distributions $x U$ and $x D$ were used. These
  determine $x u_v =x U - x \bar{U}$ and $x d_v=x D - x \bar{D}$
  assuming symmetry between the sea-quarks and anti-quarks for each
  flavour.} at a starting scale $Q^2_0$ and are evolved
using the DGLAP evolution equations. An adjustment of the parton
distribution parameters is performed to best fit the measured cross
sections.

The analysis is performed at NLO within the $\overline{MS}$ renormalisation
scheme. The program QCDNUM~\cite{QCDNUM} is used to solve the evolution equations.
From the evolved parton distributions, the structure functions
are calculated in the VFNS scheme~\cite{Thorne:2006qt, ThorneCode},
using $O(\alpha_s^2)$ coefficient functions for the calculation of $F_L$.
The factorisation and renormalisation scales are both set to $Q^2$.
The $\chi^2$ function is defined as for the cross section averaging
discussed in section~\ref{sec:averaging} and~\cite{h1lowq2}. It is minimised using the
MINUIT package.
The correlations between data points caused by systematic uncertainties
are taken into account, following the numerical
method presented in~\cite{Pascaud:1995qs, Pumplin:2002vw}. The correlations
between the systematic error sources of the different high $Q^2$ data sets are treated
as described in table 2 in~\cite{Adloff:2003uh}.
The error sources of the data presented
here and of the lower $Q^2$ data~\cite{h1lowq2} are all taken to be
uncorrelated with any other source. The only exception is a $0.5\%$
luminosity uncertainty, which is associated with the theoretical calculation
of the Bethe-Heitler cross section and which
is common to all H1 cross section measurements.

Following~\cite{Martin:2009iq}, the masses of the charm and beauty
quarks are set to $m_c = 1.4\gev$ and $m_b = 4.75\gev$.
The value of the strong coupling constant is taken to be
$\alpha_s (M^2_Z) = 0.1176$~\cite{PDG-08}. The starting scale $Q^2_0$
is chosen to be slightly below the charm threshold, $Q^2_0 =
1.9\gevsq$. Hence at the starting scale the anti-quark densities
simplify to $x \bar{U}(x) = x \bar{u}(x)$ and 
$x \bar{D}(x) = x (\bar{d}(x) +  \bar{s}(x))$. 
The anti-strange quark density, 
$x \bar{s}(x) = f_s x \bar{D} (x)$, 
is taken to be a constant fraction, $f_s = 0.31$,
of $x \bar{D}$ at the starting scale \cite{Adloff:2000qj}. 
A cut $Q^2 > Q^2_{min} = 3.5\gevsq$ 
is applied in order to ensure that the data used in the fit
correspond to a kinematic domain where leading twist perturbative QCD
can be used to predict the cross sections. Variations around these
central values are taken into account as model uncertainties, as
described below.

\subsection{Parameterisation}
\label{sec:fitpar}

The parton distributions $xP$
are parameterised at $Q^2_0$ using the general form
\begin{equation}
 xP(x) = A_P x^{B_P} (1-x)^{C_P} \left[ 1 + D_P x + E_P x^2 + \ldots \right ]\,.
 \label{eq:FitGeneral}
\end{equation}
The specific choice of the parameterisations is obtained as follows:
the parameters $D$, $E$, $\ldots$ are added one-by-one in an iterative
procedure and kept only, if they reduce the $\chi^2$ by more than $3$
units. Parameterisations obtained are discarded if the structure
functions $F_2$ and $F_L$ are negative anywhere in the range $10^{-5}
< x < 1$ for $Q^2 > Q_0^2$. Parameterisations leading to very low
valence quark distributions as compared to the total sea-quark
density $xS(x) = 2x(\bar{U}(x) +\bar{D}(x))$ at large $x$, which
dramatically fail to describe $\nu p$ and $\nu d$ fixed target
measurements of the structure function $xF_3$~\cite{Allasia:1985hw},
are also not considered. Some parameterisations of the $xu_v$ density
involving a quadratic $E$ or cubic $F$ term in $x$ in the expansion
(\ref{eq:FitGeneral}) are examples of these. They are, however,
included in the parameterisation uncertainty discussed below. The same
applies to fits where the parton distributions are negative at very
high $x$, as happens for some parameterisations of the gluon density
involving an $E$ or $F$ term. Among the possible parameterisations the
one with the lowest $\chi^2$ is selected. This procedure, at
$Q_0^2=1.9\gevsq$, leads to the following choice:
\begin{eqnarray}
\label{eq:xgpar}
xg(x) &=   & A_g x^{B_g} (1-x)^{C_g} \left[ 1  + D_{g} x \right] ,  \\
xu_v(x) &=  & A_{u_v} x^{B_{u_v}}  (1-x)^{C_{u_v}} , \\
\label{eq:xuvpar}
xd_v(x) &=  & A_{d_v} x^{B_{d_v}}  (1-x)^{C_{d_v}} , \\
\label{eq:xdvpar}
x\bar{U}(x) &=  & A_{\bar{U}} x^{B_{\bar{U}}} (1-x)^{C_{\bar{U}}} , \\
\label{eq:xubarpar}
x\bar{D}(x) &= & A_{\bar{D}} x^{B_{\bar{D}}} (1-x)^{C_{\bar{D}}} .
\label{eq:xdbarpar}
\end{eqnarray}
The normalisation parameters $A_{u_v}$ and $A_{d_v}$ are not fitted,
but are obtained from the other parameters via the quark counting rules.
Since the existing data have a limited sensitivity to the behaviour of
the valence quark distributions at low $x$, it is assumed that
$B_{u_v} = B_{d_v}$. Similarly, the behaviour of the up and down
anti-quarks at low $x$ is assumed to be governed by the same power,
$B_{\bar{U}} = B_{\bar{D}}$. As in~\cite{Adloff:2003uh}, the
normalisations of the $\bar{U}$ and $\bar{D}$ distributions are
related by $ A_{\bar{U}} = A_{\bar{D}} ( 1 - f_s )$ which corresponds
to the usual assumption that $\bar{d} / \bar{u} \rightarrow 1$ as $x
\rightarrow 0$. Finally, the normalisation $A_g$ of the gluon
distribution is derived from the momentum sum rule.
The total number of free parameters is thus equal to ten.

\subsection{Fit results}

The central fit of this analysis as specified above
has a $\chi^2$ of $587$ for $644$ degrees of freedom.
The $\chi^2$ value for each data set is given in
table~\ref{tab:chi2values}.
No significant tension is observed
between the fit results and the systematic uncertainties on the low,
medium, and high $Q^2$ data sets.
\begin{table}
\begin{center}
 \begin{tabular}{l r|c|r}
 \hline
 \hline
 \textbf{Data Set}   &    &  \textbf{Data Points}    &
 \multicolumn{1}{c}{\textbf{$\chi^2_{\rm unc}$}}    \\ \hline
 low $Q^2$  & \cite{h1lowq2}      &   $58$    &   $55.9$ \\ 
 medium $Q^2$   &  this measurement  & $99$   &   $81.5$ \\
 $e^+p$ NC high $Q^2$, $94-97$   &  \cite{Adloff:1999ah}  & $130$~~   &  $92.6$ \\
 $e^+ p$ CC high $Q^2$, $94-97$   &  \cite{Adloff:1999ah}  & $25$    &  $21.2$  \\
 $e^- p$ NC high $Q^2$, $98-99$   &  \cite{Adloff:2000qj}  & $139$~~   &  $112.1$ \\
 $e^- p$ CC high $Q^2$, $98-99$   &  \cite{Adloff:2000qj}  & $28$    &  $17.3$    \\
 $e^+p$ NC high $Q^2$, $99-00$  &  \cite{Adloff:2003uh}  & $147$~~   & $137.4$    \\
 $e^+p$ CC high $Q^2$, $99-00$  &  \cite{Adloff:2003uh}  & $28$    &  $31.1$     \\
 \hline
 \hline
\end{tabular}
\end{center}
\caption{\label{tab:chi2values}
    For each data set used in the \honepdf\ 
    fit, the number of data points is shown, along with
    the $\chi^2$ contribution determined using the uncorrelated
    errors only ($\chi^2_{\rm unc}$).
  }
\end{table}

\begin{table}
\begin{center}
    \begin{tabular}{c|c|c|c|c}
      \hline
      \hline
      $xP$      & $A_P$       & $B_P$           & $C_P$    & $D_P$    \\\hline
      $xg$      & $5.66^*$~~  & $0.243$~         & $18.76$~~  &  $34.0$  \\
      $xu_v$    & $5.15^*$~~  & $0.784$~         & $3.25$   &  ---  \\
      $xd_v$    & $3.29^*$~~  & $0.784^*$        & $4.77$  &  ---    \\
      $x\bar{U}$ & $0.105^*$  & $-0.177$~~~      & $2.42$  &  ---    \\
      $x\bar{D}$ & $0.152$~   & $-0.177^*$~~     & $3.42$  &  ---   \\
      \hline
      \hline
    \end{tabular}
\end{center}
  \caption{\label{tab:fit_parameters}
    Fitted parameters corresponding to the distributions $xg(x)$, $xu_v(x)$,
    $xd_v(x)$, $x \bar{U}(x)$ and $x \bar{D}(x)$ at the starting scale $Q^2_0 = 1.9\gevsq$
    (see section~\ref{sec:fitpar}). The symbol $^*$ indicates that the corresponding
    parameter is not a free parameter of the fit, but is derived from the other parameters.
  }
\end{table}

\begin{figure}
\centerline{%
\epsfig{file=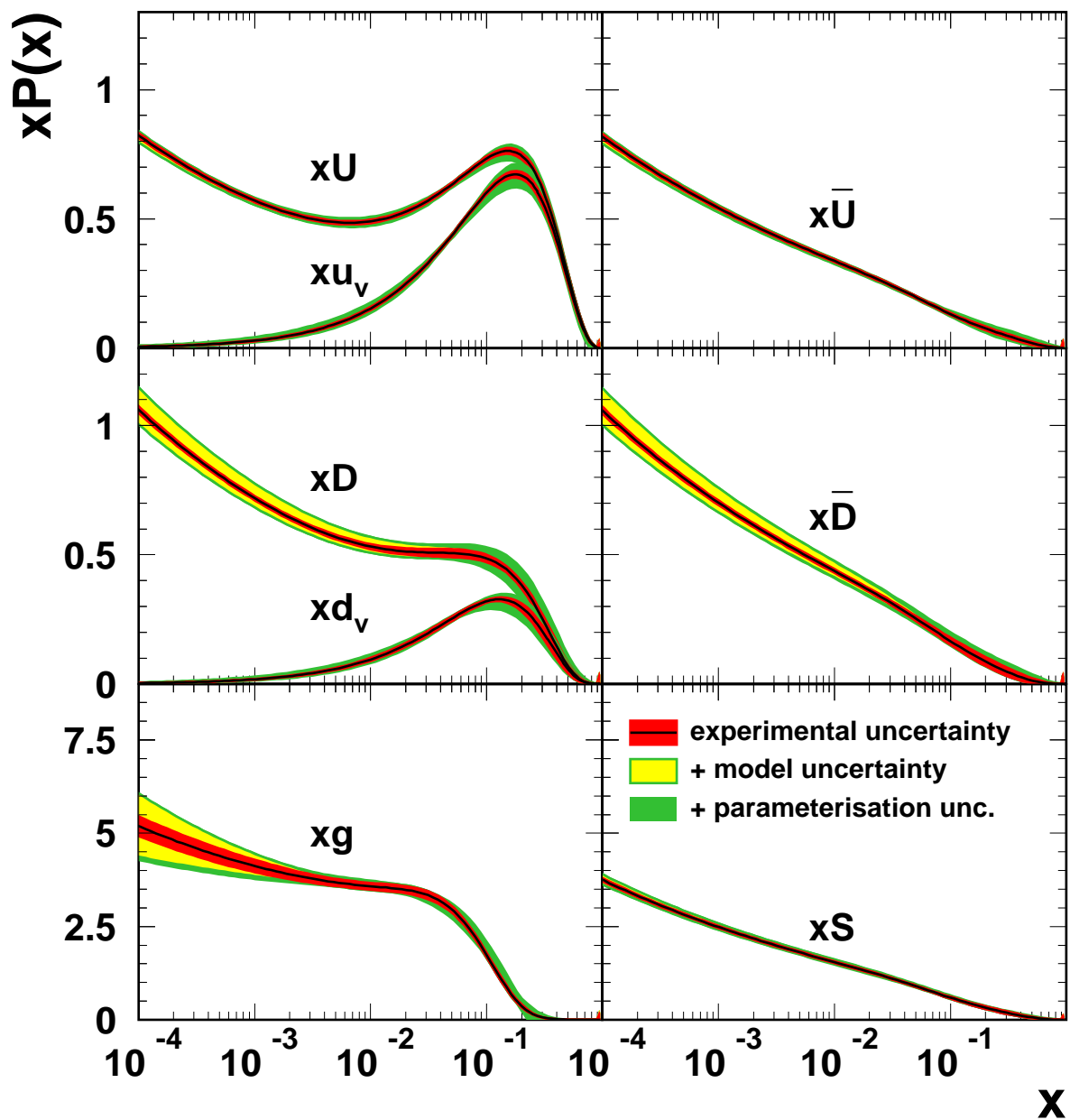,width=\linewidth}
}
\caption{\label{fig:pdf4} Parton distributions 
 as determined by the \honepdf\ QCD fit at $Q^2 = 4\gevsq$. Shown are the
 combined up and down quark distributions, $xU=x(u+c)$ and
 $xD=x(d+s)$, their anti-quark counter parts, $x\bar{U}$ and
 $x\bar{D}$, the valence quark distributions, $xu_v$ and $xd_v$,
 the total sea distribution, $xS=2x(\bar{U}+\bar{D})$, and the gluon
 distribution, $xg$.
 The inner error bands show the experimental uncertainty, the middle
 error bands include the theoretical model uncertainties of the fit
 assumptions, and the outer error band represents the total
 uncertainty including the parameterisation uncertainty.
}
\end{figure}

The fitted parameters of the distributions at the starting scale are given in
table~\ref{tab:fit_parameters}. The resulting
parton distributions, including the total sea-quark density, are shown
at $Q^2 = 4\gevsq$ in
figure~\ref{fig:pdf4}. The inner error band describes the experimental
uncertainty, obtained from the criterion $\Delta \chi^2 = 1$ using the Hessian
method as described in~\cite{Pumplin:2002vw} and the numerical algorithm
presented in~\cite{Pumplin:2000vx}. The middle error band
represents the experimental and model uncertainties added in quadrature.
The model uncertainties are obtained by varying:
\begin{itemize}
 \item the charm mass $m_c$ between $1.38\gev$ and
   $1.47\gev$;
 \item the bottom mass $m_b$ between $4.3\gev$ and $5.0\gev$;
 \item the strange fraction $f_s$ from $0.25$ to $0.40$;
 \item the value of $Q^2_{\min}$, from $2.25\gevsq$ to $5.0\gevsq$;
 \item the starting scale $Q^2_0$ down to $1.5\gevsq$.
\end{itemize}

The resulting model uncertainty at low $x$ is dominated by the
sensitivity of the fit to the $Q_0^2$ variation~\footnote{ An
  uncertainty contribution due to the $Q_0^2$ variation is calculated
  by allowing $Q_0^2$ to be as low as $1.5\gevsq$. This causes an
  increase of the $\chi^2$, which is comparable to the $\chi^2$ change
  in a massless fit, when $Q_0^2$ is varied from $4\gevsq$ to $2\gevsq$, as
  has conventionally been done in the past. Since the VFNS scheme
  implementation requires $Q_0^2 \leq m_c^2$, a default of $Q_0^2 =
  1.9\gevsq$ is chosen and the calculated uncertainty is symmetrised.
}.

In addition to the model uncertainty a new error contribution is
introduced resulting from the parameterisation choice. As explained in
section~\ref{sec:fitpar}, alternative parameterisations leading to
good fit quality but peculiar behaviour at large $x$ are used to
estimate the parametrisation uncertainties. An envelope of these fit
solutions is built which is added in quadrature to the contributions of the
experimental and model uncertainties.

\begin{figure}
\centerline{
\begin{tabular}{cc}
\epsfig{file=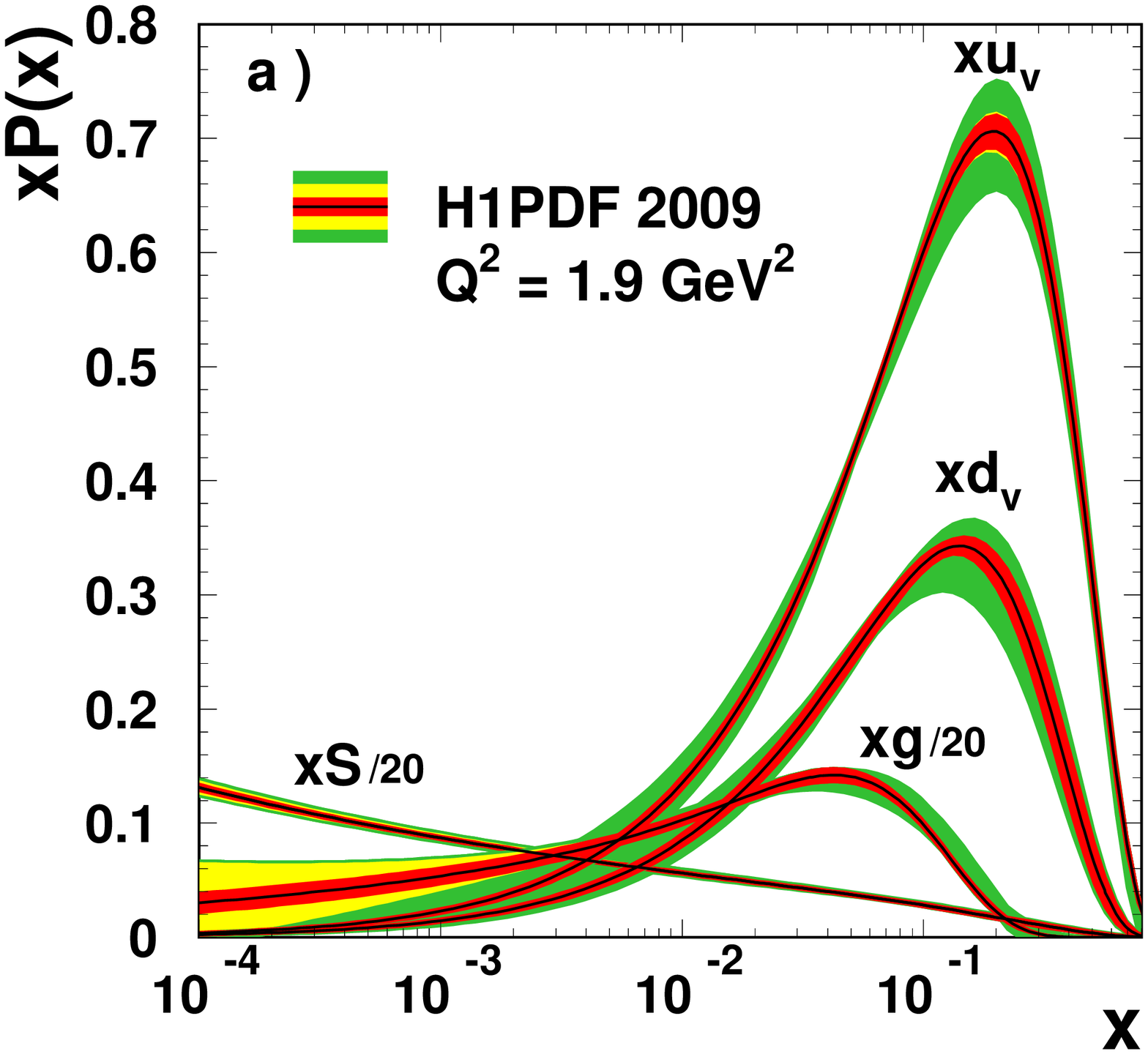,width=0.5\linewidth} &
\epsfig{file=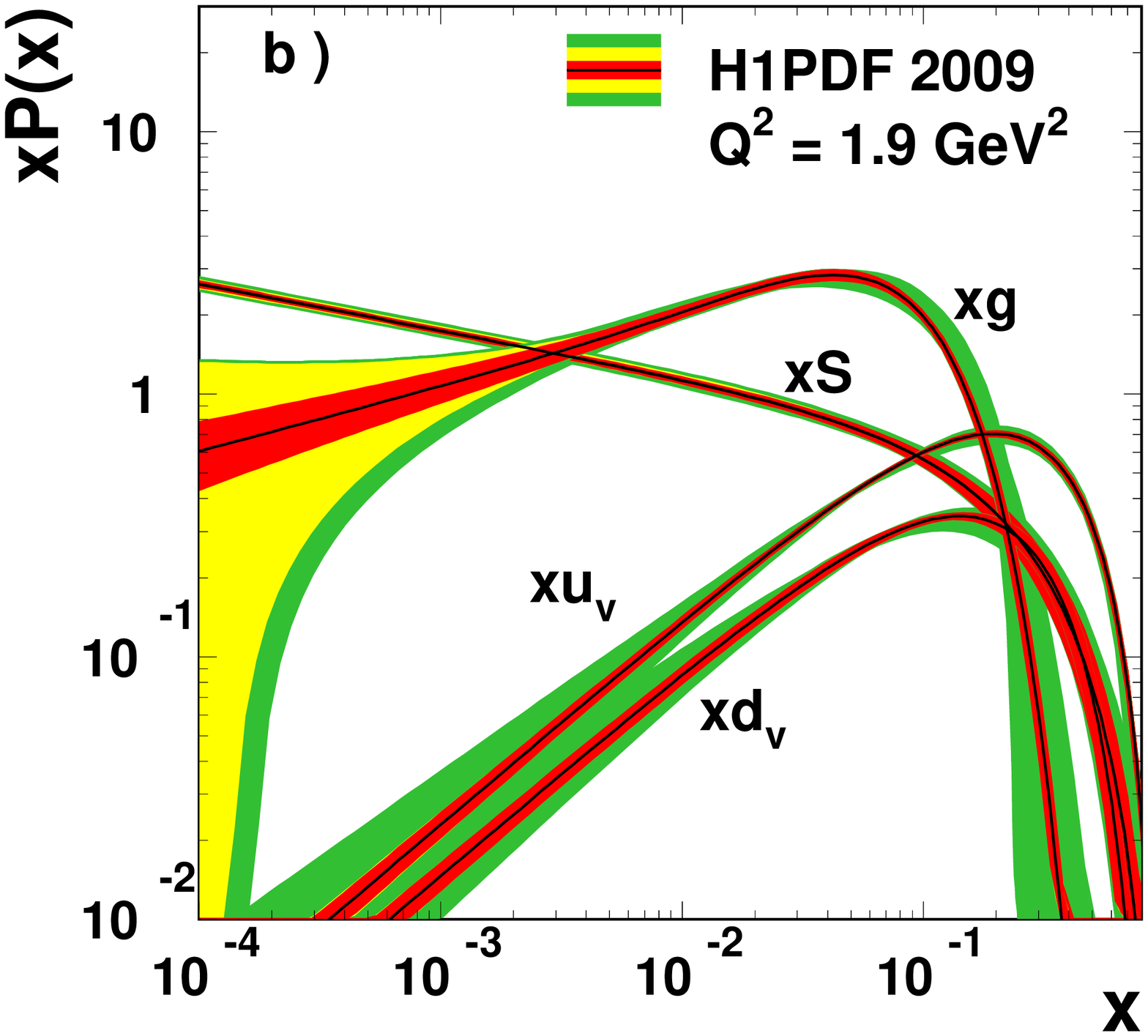,width=0.5\linewidth} \\
\epsfig{file=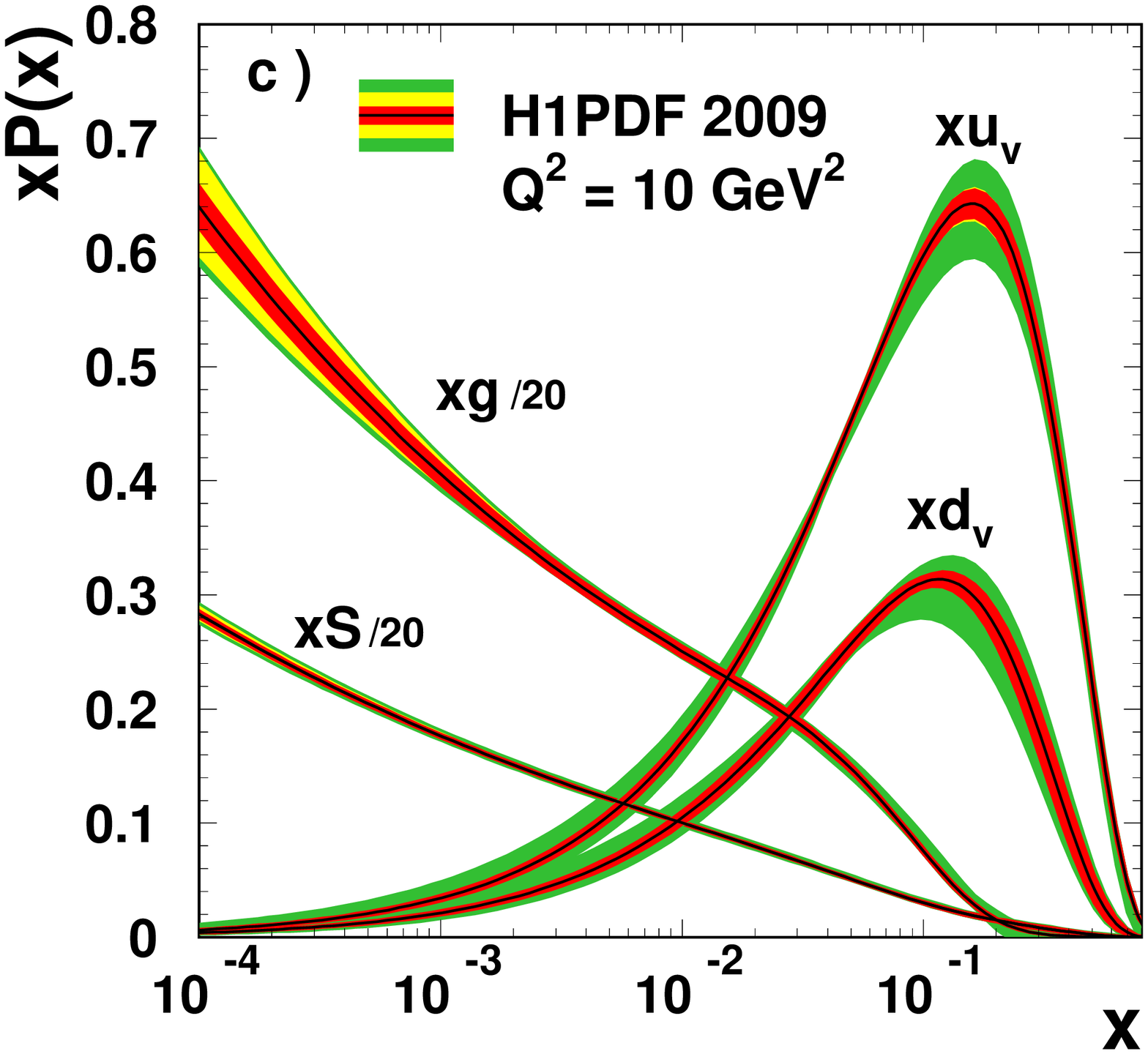,width=0.5\linewidth} &
\epsfig{file=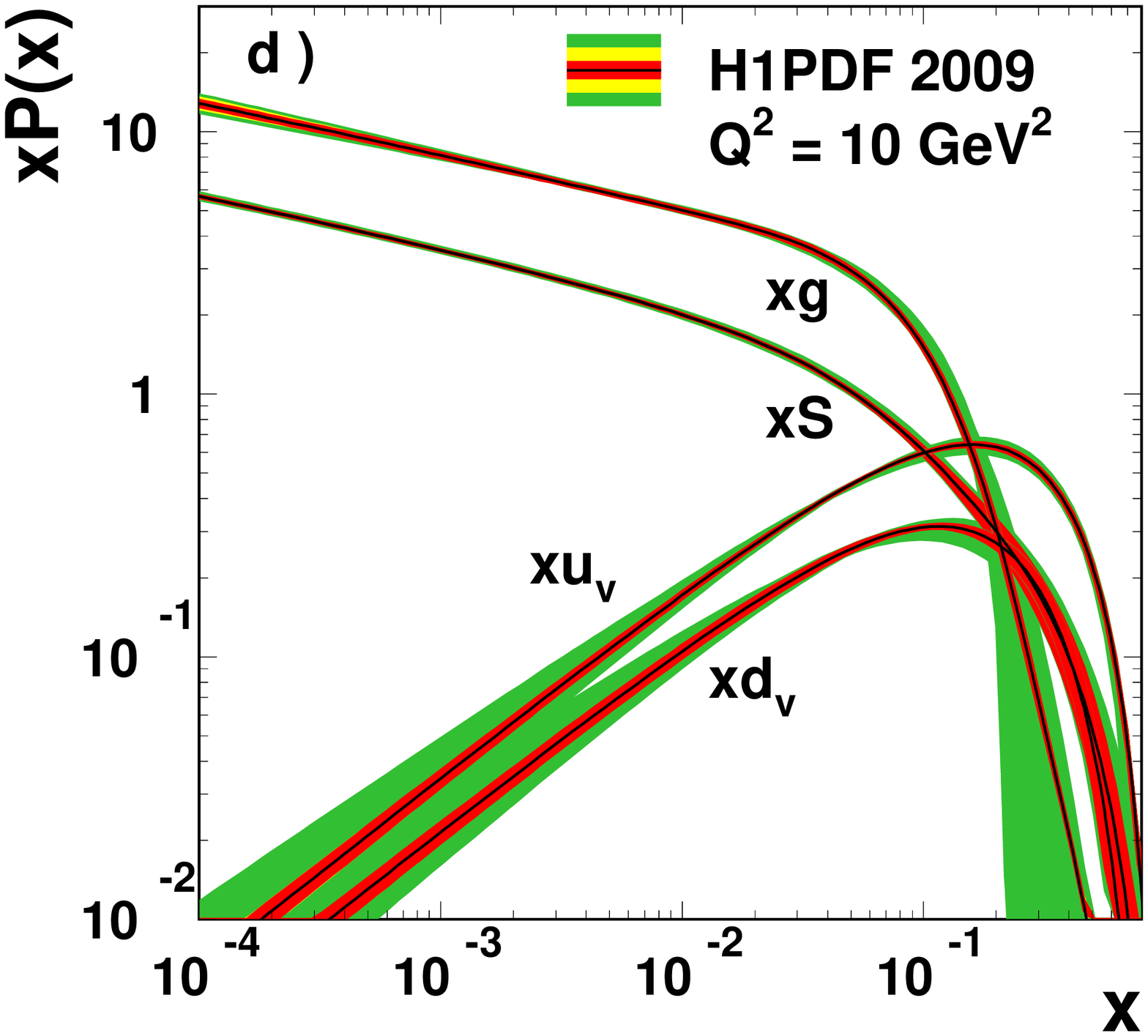,width=0.5\linewidth}
\end{tabular}
}
\caption{\label{fig:partons}Parton distributions as determined by
 the \honepdf\ QCD fit at $Q^2 = 1.9\gevsq$ a)-b) and at $Q^2 = 10\gevsq$
 c)-d). In a) and c) (linear vertical scale), the gluon and sea-quark
 densities are downscaled by a factor $0.05$.
 The inner error bands show the experimental uncertainty, the middle
 error bands include the theoretical model uncertainties of the fit
 assumptions, and the outer error band represents the total
 uncertainty including the parameterisation uncertainty, see text.
}
\end{figure}

The distributions of $x u_v(x)$, $x d_v(x)$, $x g(x)$ and $x S(x)$ are
shown at the starting scale $Q^2 = 1.9\gevsq$ and at $Q^2 = 10\gevsq$
in figures~\ref{fig:partons}~a)-b) and figures~\ref{fig:partons}~c)-d),
respectively, with both linear and logarithmic scales. A comparison of
the PDFs at the starting scale with their behaviour in the DIS region,
here represented by $Q^2 = 10\gevsq$, illustrates the rather dramatic
influence of the DGLAP evolution on the sea-quark and gluon densities.
At $Q^2 = 1.9\gevsq$ the sea-quark density rises towards low $x$, in contrast
to the gluon distribution which has a valence quark-like shape.
The $Q^2$ evolution rapidly changes the low-$x$ behaviour of the gluon
distribution, which starts to rise similarly  to the sea-quark distribution
towards low $x$. In contrast, the non-singlet valence quark
distributions evolve very slowly, as expected. As is shown in
figures~\ref{fig:partons}~c)-d), $xg$ is the dominating parton
distribution at low $x$ and higher $Q^2$, here displayed for $Q^2 = 10\gevsq$.

\section{Summary}
\label{sec:summary}

A new measurement is presented of the inclusive double differential
cross section for deep inelastic positron-proton scattering, $e^+p
\rightarrow e^+X$, in the region of small Bjorken $x$,  $2 \cdot 10^{-4}
\leq x \leq 0.1$, and four-momentum transfer squared,
$12 \gevsq \leq Q^2 \leq 150 \gevsq$. The data, corresponding to an
integrated luminosity of about $22\pbi$, were obtained with the H1
detector at the $ep$ collider HERA at beam energies $E_e=27.6\gev$
and $E_p=920\gev$.
A small bias in a similar previously published data set, taken at
$E_p=820\gev$, is found and corrected. The two data sets are then
combined and represent the most precise measurement in this kinematic
region to date, with typical total uncertainties in the range of $1.3 - 2\%$.

The kinematic range of the measurement corresponds to a wide range of
inelasticity $y$, from $0.005$ to $0.6$. The data are used to
determine the structure function $F_2(x,Q^2)$, which is observed to
rise continuously towards low $x$ at fixed $Q^2$.

The high precision of the data allows new measurements  to be
presented of the partial
derivatives of $F_2(x, Q^2)$ with respect to $Q^2$ and $x$.
For $x<0.01$, the derivative of $\ln{F_2}$ with respect to $\ln{x}$
confirms a power law dependence of $F_2 \propto x^{-\lambda}$ with
$\lambda$ depending primarily on $Q^2$.
The derivative of $F_2$ with respect to $\ln{Q^2}$ is a measure for the
gluon distribution $xg$ at low $x$.

An NLO QCD fit to the H1 data alone, including the new 
measurements presented in this paper, is performed.
The fit implements a variable flavour treatment of heavy quark
threshold effects. This new \honepdf\ fit supersedes the H1PDF 2000
previously obtained 
and provides a new determination of the gluon and quark
densities of the proton including experimental, model, and
parameterisation uncertainties.

\section*{Acknowledgements}
\refstepcounter{pdfadd} \pdfbookmark[0]{Acknowledgements}{s:acknowledge}

We are grateful to the HERA machine group whose outstanding
efforts have made this experiment possible.
We thank the engineers and technicians for their work in constructing
and maintaining the H1 detector, our funding agencies for
financial support, the DESY technical staff for continual assistance
and the DESY directorate for support and for the hospitality
which they extend to the non DESY members of the collaboration.
We thank R.~Thorne for useful discussions and for providing us with the code to calculate
the structure functions in the variable flavour number scheme.

\clearpage
\appendix

\clearpage
\begin{flushleft}
\bibliography{desy09-005}
\end{flushleft}
\clearpage
\begin{table}
\begin{tiny}
\begin{center}
\begin{tabular}{cccccccccrrrrrrr}
\hline\hline\\[-1.5ex]
$Q^2/\gevsq$ & $x$ & $y$ & $\sigma_{r}$ & $R$ & $F_2$ & $\delta_{\rm tot}$ &
 $\delta_{\rm stat}$ & $\delta_{\rm uncor}$ 
 & \multicolumn{1}{c}{$\gamma_{E_e^{\prime}}$}
 & \multicolumn{1}{c}{$\gamma_{E_{e, lin}^{\prime}}$}
 & \multicolumn{1}{c}{$\gamma_{\theta_e}$}
 & \multicolumn{1}{c}{$\gamma_{E_{\rm had}}$ }
 & \multicolumn{1}{c}{$\gamma_{\rm noise}$}
 & \multicolumn{1}{c}{$\gamma_{E_{\rm hadS}}$}
 & \multicolumn{1}{c}{$\gamma_{\gamma p}$} \\[1.5ex]
\hline
    $12$ & $2.00 \cdot 10^{-4}$ & $0.591$ & $1.324$ & $0.202$ & $1.394$ & $2.79$ & $1.21$ & $0.71$ & $-0.28$ & $-0.26$ & $-0.33$ & $0.19$ & $-0.22$ & $-0.25$ & $-2.33$\\ 
    $12$ & $3.20 \cdot 10^{-4}$ & $0.369$ & $1.247$ & $0.205$ & $1.268$ & $1.32$ & $0.86$ & $0.68$ & $-0.41$ & $-0.11$ & $-0.35$ & $0.19$ & $-0.15$ & $-0.02$ & $-0.40$\\ 
    $12$ & $5.00 \cdot 10^{-4}$ & $0.236$ & $1.175$ & $0.209$ & $1.182$ & $1.40$ & $0.89$ & $0.68$ & $-0.72$ & $0.00$ & $-0.37$ & $0.17$ & $-0.14$ & $0.00$ & $-0.04$\\ 
    $12$ & $8.00 \cdot 10^{-4}$ & $0.148$ & $1.077$ & $0.213$ & $1.079$ & $1.60$ & $0.93$ & $0.69$ & $-0.98$ & $0.19$ & $-0.43$ & $0.16$ & $-0.13$ & $0.01$ & $-0.01$\\ [1ex]
    $15$ & $3.20 \cdot 10^{-4}$ & $0.462$ & $1.338$ & $0.214$ & $1.378$ & $1.71$ & $0.84$ & $0.70$ & $-0.40$ & $-0.31$ & $-0.21$ & $0.20$ & $-0.22$ & $-0.14$ & $-1.14$\\ 
    $15$ & $5.00 \cdot 10^{-4}$ & $0.295$ & $1.225$ & $0.217$ & $1.238$ & $1.18$ & $0.69$ & $0.69$ & $-0.50$ & $-0.08$ & $-0.32$ & $0.20$ & $-0.14$ & $0.00$ & $-0.12$\\ 
    $15$ & $8.00 \cdot 10^{-4}$ & $0.185$ & $1.122$ & $0.220$ & $1.127$ & $1.33$ & $0.72$ & $0.69$ & $-0.76$ & $0.13$ & $-0.38$ & $0.18$ & $-0.11$ & $0.00$ & $-0.02$\\ 
    $15$ & $1.30 \cdot 10^{-3}$ & $0.114$ & $1.003$ & $0.224$ & $1.005$ & $1.65$ & $0.80$ & $0.70$ & $-1.17$ & $0.22$ & $-0.42$ & $0.15$ & $-0.09$ & $0.01$ & $0.00$\\ 
    $15$ & $2.00 \cdot 10^{-3}$ & $0.074$ & $0.881$ & $0.228$ & $0.881$ & $1.67$ & $0.83$ & $0.70$ & $0.44$ & $0.26$ & $-0.44$ & $-0.07$ & $-0.93$ & $0.53$ & $-0.01$\\ 
    $15$ & $3.20 \cdot 10^{-3}$ & $0.046$ & $0.798$ & $0.233$ & $0.798$ & $1.53$ & $0.91$ & $0.70$ & $0.42$ & $0.12$ & $-0.44$ & $0.06$ & $-0.76$ & $0.23$ & $-0.01$\\ 
    $15$ & $5.00 \cdot 10^{-3}$ & $0.030$ & $0.727$ & $0.237$ & $0.727$ & $1.52$ & $0.97$ & $0.71$ & $0.37$ & $0.08$ & $-0.44$ & $-0.10$ & $-0.61$ & $0.37$ & $0.00$\\ 
    $15$ & $1.00 \cdot 10^{-2}$ & $0.015$ & $0.614$ & $0.242$ & $0.614$ & $1.19$ & $0.78$ & $0.69$ & $0.27$ & $0.13$ & $-0.33$ & $-0.02$ & $-0.37$ & $0.08$ & $0.00$\\ 
    $15$ & $2.51 \cdot 10^{-2}$ & $0.006$ & $0.504$ & $0.224$ & $0.504$ & $4.31$ & $0.95$ & $2.12$ & $-0.12$ & $-0.02$ & $-0.02$ & $0.35$ & $3.59$ & $0.28$ & $0.00$\\ [1ex]
    $20$ & $5.00 \cdot 10^{-4}$ & $0.394$ & $1.329$ & $0.225$ & $1.357$ & $1.21$ & $0.73$ & $0.70$ & $-0.35$ & $-0.09$ & $-0.26$ & $0.18$ & $-0.15$ & $-0.05$ & $-0.43$\\ 
    $20$ & $8.00 \cdot 10^{-4}$ & $0.246$ & $1.219$ & $0.227$ & $1.228$ & $1.15$ & $0.63$ & $0.70$ & $-0.56$ & $0.06$ & $-0.28$ & $0.15$ & $-0.09$ & $0.01$ & $-0.05$\\ 
    $20$ & $1.30 \cdot 10^{-3}$ & $0.151$ & $1.099$ & $0.229$ & $1.102$ & $1.36$ & $0.65$ & $0.70$ & $-0.89$ & $0.16$ & $-0.29$ & $0.13$ & $-0.08$ & $0.01$ & $-0.01$\\ 
    $20$ & $2.00 \cdot 10^{-3}$ & $0.098$ & $0.996$ & $0.231$ & $0.997$ & $1.79$ & $0.70$ & $0.71$ & $-1.42$ & $0.22$ & $-0.34$ & $0.15$ & $-0.09$ & $0.00$ & $0.00$\\ 
    $20$ & $3.20 \cdot 10^{-3}$ & $0.062$ & $0.870$ & $0.233$ & $0.870$ & $1.32$ & $0.74$ & $0.71$ & $0.33$ & $0.10$ & $-0.25$ & $-0.06$ & $-0.64$ & $0.31$ & $0.00$\\ 
    $20$ & $5.00 \cdot 10^{-3}$ & $0.039$ & $0.776$ & $0.235$ & $0.776$ & $1.27$ & $0.78$ & $0.72$ & $0.32$ & $0.09$ & $-0.35$ & $-0.16$ & $-0.48$ & $0.13$ & $0.00$\\ 
    $20$ & $1.00 \cdot 10^{-2}$ & $0.020$ & $0.659$ & $0.235$ & $0.659$ & $1.49$ & $0.61$ & $0.69$ & $0.42$ & $0.11$ & $-0.33$ & $0.18$ & $-1.00$ & $0.20$ & $0.00$\\ 
    $20$ & $2.51 \cdot 10^{-2}$ & $0.008$ & $0.528$ & $0.207$ & $0.528$ & $2.67$ & $0.70$ & $2.12$ & $0.30$ & $0.10$ & $-0.31$ & $1.02$ & $0.85$ & $0.41$ & $0.00$\\ [1ex]
    $25$ & $5.00 \cdot 10^{-4}$ & $0.492$ & $1.414$ & $0.229$ & $1.467$ & $1.54$ & $0.93$ & $0.75$ & $-0.21$ & $-0.13$ & $-0.24$ & $0.21$ & $-0.20$ & $-0.19$ & $-0.84$\\ 
    $25$ & $8.00 \cdot 10^{-4}$ & $0.308$ & $1.287$ & $0.230$ & $1.302$ & $1.14$ & $0.67$ & $0.71$ & $-0.47$ & $0.00$ & $-0.26$ & $0.15$ & $-0.12$ & $-0.01$ & $-0.07$\\ 
    $25$ & $1.30 \cdot 10^{-3}$ & $0.189$ & $1.161$ & $0.231$ & $1.166$ & $1.24$ & $0.67$ & $0.71$ & $-0.69$ & $0.11$ & $-0.27$ & $0.12$ & $-0.06$ & $0.00$ & $-0.02$\\ 
    $25$ & $2.00 \cdot 10^{-3}$ & $0.123$ & $1.048$ & $0.232$ & $1.050$ & $1.46$ & $0.69$ & $0.72$ & $-1.00$ & $0.20$ & $-0.25$ & $0.12$ & $-0.05$ & $0.01$ & $-0.01$\\ 
    $25$ & $3.20 \cdot 10^{-3}$ & $0.077$ & $0.914$ & $0.232$ & $0.915$ & $1.32$ & $0.70$ & $0.73$ & $0.37$ & $0.23$ & $-0.32$ & $-0.01$ & $-0.63$ & $0.19$ & $0.00$\\ 
    $25$ & $5.00 \cdot 10^{-3}$ & $0.049$ & $0.809$ & $0.232$ & $0.809$ & $1.39$ & $0.71$ & $0.73$ & $0.35$ & $0.25$ & $-0.28$ & $0.01$ & $-0.75$ & $0.24$ & $0.00$\\ 
    $25$ & $8.00 \cdot 10^{-3}$ & $0.031$ & $0.712$ & $0.231$ & $0.712$ & $1.23$ & $0.74$ & $0.74$ & $0.29$ & $-0.04$ & $-0.27$ & $-0.14$ & $-0.46$ & $0.11$ & $0.00$\\ 
    $25$ & $1.30 \cdot 10^{-2}$ & $0.019$ & $0.649$ & $0.224$ & $0.649$ & $1.65$ & $0.78$ & $0.75$ & $0.55$ & $0.05$ & $-0.32$ & $0.07$ & $-1.04$ & $0.22$ & $0.00$\\ 
    $25$ & $2.00 \cdot 10^{-2}$ & $0.012$ & $0.587$ & $0.211$ & $0.587$ & $2.11$ & $0.83$ & $0.76$ & $0.32$ & $0.26$ & $-0.37$ & $0.62$ & $-1.54$ & $0.34$ & $0.00$\\ 
    $25$ & $3.98 \cdot 10^{-2}$ & $0.006$ & $0.496$ & $0.167$ & $0.496$ & $4.22$ & $0.73$ & $2.13$ & $0.54$ & $0.00$ & $-0.32$ & $0.14$ & $3.50$ & $0.25$ & $0.00$\\ [1ex]
    $35$ & $8.00 \cdot 10^{-4}$ & $0.431$ & $1.412$ & $0.231$ & $1.450$ & $1.28$ & $0.83$ & $0.74$ & $-0.38$ & $-0.23$ & $-0.18$ & $0.18$ & $-0.18$ & $-0.10$ & $-0.29$\\ 
    $35$ & $1.30 \cdot 10^{-3}$ & $0.265$ & $1.245$ & $0.231$ & $1.256$ & $1.19$ & $0.75$ & $0.73$ & $-0.49$ & $0.07$ & $-0.25$ & $0.11$ & $-0.05$ & $-0.01$ & $-0.03$\\ 
    $35$ & $2.00 \cdot 10^{-3}$ & $0.172$ & $1.129$ & $0.230$ & $1.133$ & $1.33$ & $0.76$ & $0.74$ & $-0.68$ & $0.28$ & $-0.31$ & $0.10$ & $-0.06$ & $0.00$ & $-0.02$\\ 
    $35$ & $3.20 \cdot 10^{-3}$ & $0.108$ & $0.991$ & $0.229$ & $0.992$ & $1.65$ & $0.79$ & $0.75$ & $-1.13$ & $0.40$ & $-0.30$ & $0.09$ & $-0.03$ & $-0.01$ & $0.00$\\ 
    $35$ & $5.00 \cdot 10^{-3}$ & $0.069$ & $0.860$ & $0.227$ & $0.860$ & $1.37$ & $0.78$ & $0.75$ & $0.44$ & $0.25$ & $-0.34$ & $0.07$ & $-0.56$ & $0.10$ & $0.00$\\ 
    $35$ & $8.00 \cdot 10^{-3}$ & $0.043$ & $0.766$ & $0.223$ & $0.766$ & $1.38$ & $0.80$ & $0.76$ & $0.45$ & $0.12$ & $-0.28$ & $-0.13$ & $-0.52$ & $0.31$ & $0.00$\\ 
    $35$ & $1.30 \cdot 10^{-2}$ & $0.027$ & $0.676$ & $0.214$ & $0.676$ & $1.47$ & $0.84$ & $0.78$ & $0.40$ & $-0.06$ & $-0.33$ & $-0.08$ & $-0.76$ & $-0.06$ & $0.00$\\ 
    $35$ & $2.00 \cdot 10^{-2}$ & $0.017$ & $0.614$ & $0.199$ & $0.614$ & $2.02$ & $0.88$ & $0.79$ & $0.45$ & $0.26$ & $-0.32$ & $0.25$ & $-1.42$ & $0.50$ & $0.00$\\ 
    $35$ & $3.98 \cdot 10^{-2}$ & $0.009$ & $0.514$ & $0.156$ & $0.514$ & $3.01$ & $0.76$ & $2.14$ & $0.41$ & $0.00$ & $-0.25$ & $0.13$ & $1.90$ & $0.31$ & $0.00$\\ [1ex]
    $45$ & $8.00 \cdot 10^{-4}$ & $0.554$ & $1.484$ & $0.231$ & $1.559$ & $1.74$ & $1.37$ & $0.89$ & $-0.16$ & $0.01$ & $-0.20$ & $0.23$ & $-0.18$ & $-0.23$ & $-0.41$\\ 
    $45$ & $1.30 \cdot 10^{-3}$ & $0.341$ & $1.332$ & $0.230$ & $1.352$ & $1.23$ & $0.86$ & $0.76$ & $-0.37$ & $0.02$ & $-0.18$ & $0.11$ & $-0.11$ & $-0.03$ & $-0.08$\\ 
    $45$ & $2.00 \cdot 10^{-3}$ & $0.222$ & $1.182$ & $0.228$ & $1.189$ & $1.30$ & $0.85$ & $0.76$ & $-0.53$ & $0.17$ & $-0.26$ & $0.07$ & $-0.03$ & $0.00$ & $0.00$\\ 
    $45$ & $3.20 \cdot 10^{-3}$ & $0.138$ & $1.032$ & $0.226$ & $1.034$ & $1.51$ & $0.88$ & $0.78$ & $-0.85$ & $0.33$ & $-0.24$ & $0.08$ & $-0.02$ & $-0.01$ & $0.00$\\ 
    $45$ & $5.00 \cdot 10^{-3}$ & $0.089$ & $0.902$ & $0.223$ & $0.903$ & $1.56$ & $0.89$ & $0.78$ & $0.35$ & $0.36$ & $-0.16$ & $0.29$ & $-0.82$ & $-0.06$ & $0.00$\\ 
    $45$ & $8.00 \cdot 10^{-3}$ & $0.055$ & $0.796$ & $0.217$ & $0.797$ & $1.44$ & $0.90$ & $0.80$ & $0.49$ & $0.09$ & $-0.19$ & $-0.19$ & $-0.56$ & $0.07$ & $0.00$\\ 
    $45$ & $1.30 \cdot 10^{-2}$ & $0.034$ & $0.696$ & $0.207$ & $0.696$ & $1.44$ & $0.94$ & $0.81$ & $0.47$ & $0.04$ & $-0.22$ & $-0.14$ & $-0.46$ & $0.16$ & $0.00$\\ 
    $45$ & $2.00 \cdot 10^{-2}$ & $0.022$ & $0.617$ & $0.191$ & $0.617$ & $2.15$ & $0.99$ & $0.82$ & $0.67$ & $0.23$ & $-0.28$ & $-0.14$ & $-1.53$ & $0.22$ & $0.00$\\ 
    $45$ & $3.20 \cdot 10^{-2}$ & $0.014$ & $0.543$ & $0.163$ & $0.543$ & $1.59$ & $1.08$ & $0.85$ & $0.47$ & $0.06$ & $-0.26$ & $0.10$ & $-0.31$ & $0.51$ & $0.00$\\ 
    $45$ & $6.31 \cdot 10^{-2}$ & $0.007$ & $0.457$ & $0.107$ & $0.457$ & $5.45$ & $1.05$ & $2.17$ & $0.81$ & $0.18$ & $-0.40$ & $-1.26$ & $4.60$ & $0.50$ & $0.00$\\ [1ex]

\hline\hline
\end{tabular}
\end{center}
\end{tiny}
\begin{small}
  \caption{Reduced cross section $\sigma_{r}$ for each $Q^2,x$ bin
    from the 2000 data. $F_2$ is extracted with $R$ taken from the QCD
    fit described in this paper using (\ref{e:rflf2}). The uncertainties on $\sigma_r$ are
    quoted in $\%$. $\delta_{\rm tot}$ is the total uncertainty
    determined as the quadratic sum of all uncertainties. $\delta_{\rm
      stat}$ is the statistical uncertainty. $\delta_{\rm uncor}$
    represents the uncorrelated systematic uncertainty.
    $\gamma_{E_e^{\prime}}$, $\gamma_{E_{e, lin}^{\prime}}$,
    $\gamma_{\theta_e}$, $\gamma_{E_{\rm had}}$, $\gamma_{\rm noise}$,
    $\gamma_{E_{\rm hadS}}$, and $\gamma_{\gamma p}$ are the bin-to-bin
    correlated uncertainties on the cross section measurement due to
    uncertainties on the SpaCal electromagnetic energy scale and
    linearity, positron scattering angle, LAr calorimeter hadronic
    energy scale, LAr calorimeter noise, and the
    photoproduction background, respectively. The overall
    normalisation uncertainty of $1.2\%$ is not included.
    \label{tab:table00}}
\end{small} 
\end{table}

\begin{table}
\begin{tiny}
\begin{center}
\begin{tabular}{cccccccccrrrrrrr}
\hline\hline\\[-1.5ex]
$Q^2/\gevsq$ & $x$ & $y$ & $\sigma_{r}$ & $R$ & $F_2$ & $\delta_{\rm tot}$ &
 $\delta_{\rm stat}$ & $\delta_{\rm uncor}$ 
 & \multicolumn{1}{c}{$\gamma_{E_e^{\prime}}$}
 & \multicolumn{1}{c}{$\gamma_{E_{e, lin}^{\prime}}$}
 & \multicolumn{1}{c}{$\gamma_{\theta_e}$}
 & \multicolumn{1}{c}{$\gamma_{E_{\rm had}}$ }
 & \multicolumn{1}{c}{$\gamma_{\rm noise}$}
 & \multicolumn{1}{c}{$\gamma_{E_{\rm hadS}}$}
 & \multicolumn{1}{c}{$\gamma_{\gamma p}$} \\[1.5ex]
\hline
    $60$ & $1.30 \cdot 10^{-3}$ & $0.454$ & $1.410$ & $0.227$ & $1.453$ & $1.63$ & $1.29$ & $0.89$ & $-0.26$ & $-0.10$ & $-0.17$ & $0.20$ & $-0.18$ & $-0.11$ & $-0.09$\\ 
    $60$ & $2.00 \cdot 10^{-3}$ & $0.295$ & $1.256$ & $0.225$ & $1.270$ & $1.33$ & $0.95$ & $0.80$ & $-0.45$ & $0.10$ & $-0.15$ & $0.08$ & $-0.02$ & $-0.01$ & $0.00$\\ 
    $60$ & $3.20 \cdot 10^{-3}$ & $0.185$ & $1.100$ & $0.221$ & $1.104$ & $1.43$ & $0.97$ & $0.81$ & $-0.58$ & $0.30$ & $-0.21$ & $0.06$ & $-0.02$ & $0.00$ & $0.00$\\ 
    $60$ & $5.00 \cdot 10^{-3}$ & $0.118$ & $0.988$ & $0.217$ & $0.989$ & $1.70$ & $1.01$ & $0.83$ & $-0.99$ & $0.41$ & $-0.15$ & $0.07$ & $-0.02$ & $0.00$ & $0.00$\\ 
    $60$ & $8.00 \cdot 10^{-3}$ & $0.074$ & $0.830$ & $0.210$ & $0.831$ & $1.63$ & $1.05$ & $0.84$ & $0.55$ & $0.25$ & $-0.24$ & $0.14$ & $-0.61$ & $0.16$ & $0.00$\\ 
    $60$ & $1.30 \cdot 10^{-2}$ & $0.045$ & $0.718$ & $0.198$ & $0.718$ & $1.61$ & $1.09$ & $0.85$ & $0.39$ & $0.10$ & $-0.21$ & $-0.50$ & $-0.45$ & $0.05$ & $0.00$\\ 
    $60$ & $2.00 \cdot 10^{-2}$ & $0.030$ & $0.653$ & $0.181$ & $0.653$ & $1.83$ & $1.14$ & $0.87$ & $0.51$ & $0.23$ & $-0.10$ & $0.20$ & $-0.92$ & $0.29$ & $0.00$\\ 
    $60$ & $3.20 \cdot 10^{-2}$ & $0.018$ & $0.560$ & $0.154$ & $0.560$ & $2.00$ & $1.23$ & $0.90$ & $0.22$ & $-0.11$ & $-0.21$ & $-0.41$ & $-1.12$ & $0.43$ & $0.00$\\ 
    $60$ & $6.31 \cdot 10^{-2}$ & $0.009$ & $0.464$ & $0.102$ & $0.464$ & $4.49$ & $1.15$ & $2.18$ & $1.09$ & $0.08$ & $-0.33$ & $-1.24$ & $3.31$ & $0.58$ & $0.00$\\ [1ex]
    $90$ & $2.00 \cdot 10^{-3}$ & $0.443$ & $1.321$ & $0.219$ & $1.358$ & $2.07$ & $1.74$ & $1.07$ & $-0.12$ & $0.20$ & $-0.20$ & $0.09$ & $-0.04$ & $-0.03$ & $0.00$\\ 
    $90$ & $3.20 \cdot 10^{-3}$ & $0.277$ & $1.185$ & $0.214$ & $1.196$ & $1.60$ & $1.20$ & $0.88$ & $-0.57$ & $0.08$ & $-0.12$ & $0.05$ & $-0.01$ & $-0.02$ & $0.00$\\ 
    $90$ & $5.00 \cdot 10^{-3}$ & $0.177$ & $1.046$ & $0.208$ & $1.049$ & $1.70$ & $1.15$ & $0.87$ & $-0.78$ & $0.36$ & $-0.21$ & $0.06$ & $-0.02$ & $0.00$ & $0.00$\\ 
    $90$ & $8.00 \cdot 10^{-3}$ & $0.111$ & $0.894$ & $0.200$ & $0.895$ & $1.94$ & $1.22$ & $0.90$ & $-1.05$ & $0.57$ & $-0.12$ & $0.05$ & $-0.01$ & $-0.01$ & $0.00$\\ 
    $90$ & $1.30 \cdot 10^{-2}$ & $0.068$ & $0.763$ & $0.187$ & $0.763$ & $1.78$ & $1.26$ & $0.91$ & $0.61$ & $0.49$ & $-0.20$ & $-0.11$ & $-0.14$ & $-0.32$ & $0.00$\\ 
    $90$ & $2.00 \cdot 10^{-2}$ & $0.044$ & $0.668$ & $0.169$ & $0.668$ & $2.05$ & $1.30$ & $0.93$ & $0.33$ & $-0.08$ & $-0.25$ & $-0.23$ & $-1.19$ & $0.05$ & $0.00$\\ 
    $90$ & $3.20 \cdot 10^{-2}$ & $0.028$ & $0.568$ & $0.143$ & $0.568$ & $1.98$ & $1.39$ & $0.95$ & $0.67$ & $0.11$ & $-0.18$ & $-0.41$ & $-0.51$ & $0.39$ & $0.00$\\ 
    $90$ & $5.00 \cdot 10^{-2}$ & $0.018$ & $0.492$ & $0.111$ & $0.492$ & $2.44$ & $1.58$ & $1.01$ & $0.95$ & $0.08$ & $-0.15$ & $-0.81$ & $-0.07$ & $0.92$ & $0.00$\\ 
    $90$ & $1.00 \cdot 10^{-1}$ & $0.009$ & $0.410$ & $0.065$ & $0.410$ & $7.14$ & $1.71$ & $2.26$ & $0.52$ & $0.00$ & $-0.36$ & $-2.36$ & $6.03$ & $0.78$ & $0.00$\\ [1ex]
    $120$ & $5.00 \cdot 10^{-3}$ & $0.236$ & $1.030$ & $0.202$ & $1.036$ & $2.57$ & $2.13$ & $1.23$ & $-0.66$ & $0.29$ & $-0.17$ & $0.05$ & $0.00$ & $-0.01$ & $0.00$\\ 
    $120$ & $8.00 \cdot 10^{-3}$ & $0.148$ & $0.926$ & $0.192$ & $0.928$ & $2.38$ & $1.77$ & $1.10$ & $-1.10$ & $0.29$ & $-0.06$ & $0.04$ & $0.00$ & $0.00$ & $0.00$\\ 
    $120$ & $1.30 \cdot 10^{-2}$ & $0.091$ & $0.808$ & $0.179$ & $0.808$ & $2.44$ & $1.70$ & $1.09$ & $-1.11$ & $0.75$ & $-0.23$ & $0.04$ & $-0.01$ & $-0.01$ & $0.00$\\ 
    $120$ & $2.00 \cdot 10^{-2}$ & $0.059$ & $0.675$ & $0.161$ & $0.675$ & $2.46$ & $1.67$ & $1.06$ & $0.80$ & $0.50$ & $-0.05$ & $0.43$ & $-0.93$ & $-0.46$ & $0.00$\\ 
    $120$ & $3.20 \cdot 10^{-2}$ & $0.037$ & $0.573$ & $0.135$ & $0.573$ & $2.62$ & $1.74$ & $1.08$ & $1.11$ & $0.07$ & $-0.13$ & $-0.09$ & $-1.16$ & $0.30$ & $0.00$\\ 
    $120$ & $5.00 \cdot 10^{-2}$ & $0.024$ & $0.489$ & $0.105$ & $0.489$ & $2.75$ & $1.87$ & $1.11$ & $0.94$ & $0.29$ & $-0.22$ & $-0.46$ & $-0.93$ & $0.87$ & $0.00$\\ 
    $120$ & $1.00 \cdot 10^{-1}$ & $0.012$ & $0.402$ & $0.062$ & $0.402$ & $5.52$ & $1.92$ & $1.12$ & $1.44$ & $-0.14$ & $-0.12$ & $-2.45$ & $4.17$ & $0.22$ & $0.00$\\ [1ex]
    $150$ & $1.30 \cdot 10^{-2}$ & $0.114$ & $0.764$ & $0.173$ & $0.765$ & $6.39$ & $5.42$ & $2.74$ & $-1.98$ & $0.13$ & $-0.23$ & $0.00$ & $-0.03$ & $0.00$ & $0.00$\\ 
    $150$ & $2.00 \cdot 10^{-2}$ & $0.074$ & $0.725$ & $0.155$ & $0.725$ & $5.34$ & $4.46$ & $2.41$ & $-1.23$ & $1.12$ & $-0.09$ & $0.07$ & $-0.07$ & $-0.02$ & $0.00$\\ 
    $150$ & $3.20 \cdot 10^{-2}$ & $0.046$ & $0.611$ & $0.130$ & $0.611$ & $6.60$ & $3.76$ & $2.02$ & $3.05$ & $0.76$ & $-0.07$ & $2.05$ & $-3.20$ & $-1.05$ & $0.00$\\ 
    $150$ & $5.00 \cdot 10^{-2}$ & $0.030$ & $0.500$ & $0.101$ & $0.500$ & $5.60$ & $3.76$ & $1.94$ & $3.30$ & $-0.21$ & $0.03$ & $0.54$ & $-1.45$ & $0.31$ & $0.00$\\ 
    $150$ & $1.00 \cdot 10^{-1}$ & $0.015$ & $0.422$ & $0.060$ & $0.422$ & $5.91$ & $3.45$ & $1.81$ & $3.40$ & $-0.43$ & $-0.10$ & $-1.73$ & $2.18$ & $0.41$ & $0.00$\\ 

\hline\hline
\end{tabular}
\end{center}
\end{tiny}
\begin{small}
  \caption{Reduced cross section $\sigma_{r}$
    and $F_2$ for each $Q^2,x$
    bin from the 2000 data, continued from \Tab~\ref{tab:table00}. 
    \label{tab:table00_2}}
\end{small}
\end{table}

\begin{table}
\begin{tiny}
\begin{center}
\begin{tabular}{cccccccccrrrrr}
\hline\hline\\[-1.5ex]
$Q^2/\gevsq$ & $x$ & $y$ & $\sigma_{r}$ & $R$ & $F_2$ & $\delta_{\rm tot}$ &
 $\delta_{\rm stat}$ & $\delta_{\rm uncor}$ 
 & \multicolumn{1}{c}{$\gamma_{E_e^{\prime}}$}
 & \multicolumn{1}{c}{$\gamma_{\theta_e}$}
 & \multicolumn{1}{c}{$\gamma_{E_{\rm had}}$ }
 & \multicolumn{1}{c}{$\gamma_{\rm noise}$}
 & \multicolumn{1}{c}{$\gamma_{\gamma p}$} \\[1.5ex]
\hline
    $12$ & $1.61 \cdot 10^{-4}$ & $0.825$ & $1.238$ &   ---   &   ---   & $5.85$ & $4.10$ & $3.78$ & $-1.55$ & $-0.70$ & $0.51$ & $0.00$ & $0.00$\\ 
    $12$ & $1.97 \cdot 10^{-4}$ & $0.675$ & $1.282$ &   ---   &   ---   & $3.50$ & $0.87$ & $2.15$ & $-1.06$ & $-0.43$ & $0.58$ & $0.03$ & $-2.33$\\ 
    $12$ & $3.20 \cdot 10^{-4}$ & $0.415$ & $1.229$ & $0.205$ & $1.257$ & $1.99$ & $0.57$ & $1.73$ & $-0.80$ & $-0.13$ & $0.09$ & $0.04$ & $-0.51$\\ 
    $12$ & $5.00 \cdot 10^{-4}$ & $0.266$ & $1.157$ & $0.208$ & $1.167$ & $1.84$ & $0.67$ & $1.51$ & $-0.96$ & $0.09$ & $0.00$ & $0.00$ & $-0.06$\\ [1ex]
    $15$ & $2.01 \cdot 10^{-4}$ & $0.825$ & $1.272$ &   ---   &   ---   & $5.15$ & $3.19$ & $3.61$ & $-1.61$ & $-0.65$ & $0.48$ & $0.00$ & $0.00$\\ 
    $15$ & $2.46 \cdot 10^{-4}$ & $0.675$ & $1.379$ &   ---   &   ---   & $3.35$ & $0.91$ & $2.17$ & $-1.99$ & $-0.49$ & $0.55$ & $0.04$ & $-1.19$\\ 
    $15$ & $3.20 \cdot 10^{-4}$ & $0.519$ & $1.300$ & $0.214$ & $1.352$ & $2.43$ & $0.68$ & $1.97$ & $-0.82$ & $-0.67$ & $0.19$ & $0.02$ & $-0.83$\\ 
    $15$ & $5.00 \cdot 10^{-4}$ & $0.332$ & $1.244$ & $0.217$ & $1.262$ & $1.86$ & $0.58$ & $1.70$ & $-0.53$ & $-0.44$ & $0.02$ & $0.01$ & $-0.08$\\ 
    $15$ & $8.00 \cdot 10^{-4}$ & $0.208$ & $1.130$ & $0.219$ & $1.135$ & $1.68$ & $0.60$ & $1.47$ & $-0.71$ & $-0.26$ & $0.00$ & $0.00$ & $-0.03$\\ 
    $15$ & $1.30 \cdot 10^{-3}$ & $0.127$ & $0.982$ & $0.227$ & $0.984$ & $2.86$ & $0.62$ & $1.51$ & $1.46$ & $0.29$ & $0.36$ & $1.79$ & $0.00$\\ 
    $15$ & $2.00 \cdot 10^{-3}$ & $0.083$ & $0.877$ & $0.228$ & $0.877$ & $2.54$ & $0.65$ & $1.53$ & $1.33$ & $-0.51$ & $-0.14$ & $1.27$ & $0.00$\\ 
    $15$ & $3.20 \cdot 10^{-3}$ & $0.052$ & $0.784$ & $0.231$ & $0.785$ & $2.51$ & $0.70$ & $1.54$ & $1.32$ & $-0.56$ & $-0.22$ & $1.16$ & $0.00$\\ 
    $15$ & $5.00 \cdot 10^{-3}$ & $0.033$ & $0.718$ & $0.241$ & $0.718$ & $2.42$ & $0.75$ & $1.59$ & $1.25$ & $-0.49$ & $-0.63$ & $0.74$ & $0.00$\\ 
    $15$ & $1.00 \cdot 10^{-2}$ & $0.017$ & $0.583$ & $0.228$ & $0.583$ & $2.74$ & $0.67$ & $1.52$ & $1.16$ & $-0.47$ & $-1.73$ & $-0.45$ & $0.00$\\ 
    $15$ & $3.20 \cdot 10^{-2}$ & $0.005$ & $0.459$ & $0.231$ & $0.459$ & $6.45$ & $1.03$ & $3.31$ & $1.06$ & $-0.70$ & $-2.69$ & $-4.56$ & $0.00$\\ [1ex]
    $20$ & $2.68 \cdot 10^{-4}$ & $0.825$ & $1.337$ &   ---   &   ---   & $5.19$ & $3.24$ & $3.67$ & $-1.50$ & $-0.67$ & $0.52$ & $0.00$ & $0.00$\\ 
    $20$ & $3.28 \cdot 10^{-4}$ & $0.675$ & $1.408$ &   ---   &   ---   & $2.71$ & $1.04$ & $2.10$ & $-0.84$ & $-0.39$ & $0.53$ & $0.06$ & $-1.00$\\ 
    $20$ & $5.00 \cdot 10^{-4}$ & $0.443$ & $1.308$ & $0.224$ & $1.345$ & $1.98$ & $0.64$ & $1.71$ & $-0.33$ & $-0.77$ & $0.12$ & $0.04$ & $-0.33$\\ 
    $20$ & $8.00 \cdot 10^{-4}$ & $0.277$ & $1.199$ & $0.226$ & $1.210$ & $1.81$ & $0.60$ & $1.45$ & $-0.80$ & $-0.64$ & $0.00$ & $0.00$ & $-0.01$\\ 
    $20$ & $1.30 \cdot 10^{-3}$ & $0.170$ & $1.078$ & $0.230$ & $1.081$ & $1.80$ & $0.64$ & $1.45$ & $-0.83$ & $-0.51$ & $0.00$ & $0.00$ & $0.00$\\ 
    $20$ & $2.00 \cdot 10^{-3}$ & $0.111$ & $0.956$ & $0.229$ & $0.957$ & $2.78$ & $0.69$ & $1.51$ & $1.47$ & $-0.24$ & $0.25$ & $1.64$ & $0.00$\\ 
    $20$ & $3.20 \cdot 10^{-3}$ & $0.069$ & $0.834$ & $0.235$ & $0.834$ & $2.34$ & $0.73$ & $1.54$ & $1.32$ & $-0.32$ & $-0.38$ & $0.77$ & $0.00$\\ 
    $20$ & $5.00 \cdot 10^{-3}$ & $0.044$ & $0.760$ & $0.239$ & $0.760$ & $2.16$ & $0.78$ & $1.58$ & $1.02$ & $-0.22$ & $-0.42$ & $0.50$ & $0.00$\\ 
    $20$ & $1.00 \cdot 10^{-2}$ & $0.022$ & $0.621$ & $0.238$ & $0.621$ & $2.41$ & $0.67$ & $1.52$ & $0.94$ & $-0.09$ & $-1.45$ & $-0.27$ & $0.00$\\ 
    $20$ & $3.20 \cdot 10^{-2}$ & $0.007$ & $0.463$ & $0.187$ & $0.463$ & $6.81$ & $0.97$ & $3.30$ & $1.21$ & $-0.22$ & $-3.05$ & $-4.88$ & $0.00$\\ [1ex]

\hline\hline
\end{tabular}
\end{center}
\end{tiny}
\begin{small}
  \caption{Corrected reduced cross section $\sigma_{r}$
    from H1 1996/97 data, presented in the same format as originally
    in~\cite{h1alphas}, with uncertainties quoted in $\%$.
    The correction procedure is described in the text. 
    $F_2$ is extracted for $y<0.6$ with $R$ given by the
    QCD fit described in this paper.
    The overall normalisation uncertainty of $1.7\%$ is
    not included.
    \label{tab:table97}
  }
\end{small} 
\end{table}

\begin{table}
\begin{tiny}
\begin{center}
\begin{tabular}{cccccccccrrrrr}
\hline\hline\\[-1.5ex]
$Q^2/\gevsq$ & $x$ & $y$ & $\sigma_{r}$ & $R$ & $F_2$ & $\delta_{\rm tot}$ &
 $\delta_{\rm stat}$ & $\delta_{\rm uncor}$ 
 & \multicolumn{1}{c}{$\gamma_{E_e^{\prime}}$}
 & \multicolumn{1}{c}{$\gamma_{\theta_e}$}
 & \multicolumn{1}{c}{$\gamma_{E_{\rm had}}$ }
 & \multicolumn{1}{c}{$\gamma_{\rm noise}$}
 & \multicolumn{1}{c}{$\gamma_{\gamma p}$} \\[1.5ex]
\hline
    $25$ & $3.35 \cdot 10^{-4}$ & $0.825$ & $1.408$ &   ---   &   ---   & $5.91$ & $4.09$ & $3.92$ & $-1.49$ & $-0.65$ & $0.46$ & $0.00$ & $0.00$\\ 
    $25$ & $4.10 \cdot 10^{-4}$ & $0.675$ & $1.400$ &   ---   &   ---   & $2.64$ & $1.16$ & $2.10$ & $-0.17$ & $-0.48$ & $0.59$ & $0.04$ & $-0.92$\\ 
    $25$ & $5.00 \cdot 10^{-4}$ & $0.553$ & $1.374$ & $0.229$ & $1.442$ & $2.41$ & $1.04$ & $1.88$ & $-1.04$ & $-0.37$ & $0.25$ & $0.04$ & $-0.41$\\ 
    $25$ & $8.00 \cdot 10^{-4}$ & $0.346$ & $1.268$ & $0.229$ & $1.289$ & $1.94$ & $0.67$ & $1.70$ & $-0.60$ & $-0.60$ & $0.04$ & $0.02$ & $-0.07$\\ 
    $25$ & $1.30 \cdot 10^{-3}$ & $0.213$ & $1.114$ & $0.230$ & $1.120$ & $1.78$ & $0.66$ & $1.45$ & $-0.64$ & $-0.69$ & $0.00$ & $0.00$ & $0.00$\\ 
    $25$ & $2.00 \cdot 10^{-3}$ & $0.138$ & $1.006$ & $0.233$ & $1.008$ & $2.89$ & $0.76$ & $1.51$ & $1.78$ & $-0.70$ & $0.17$ & $1.34$ & $0.00$\\ 
    $25$ & $3.20 \cdot 10^{-3}$ & $0.086$ & $0.898$ & $0.235$ & $0.898$ & $2.78$ & $0.79$ & $1.54$ & $1.80$ & $-0.77$ & $-0.23$ & $0.92$ & $0.00$\\ 
    $25$ & $5.00 \cdot 10^{-3}$ & $0.055$ & $0.770$ & $0.236$ & $0.770$ & $2.38$ & $0.85$ & $1.57$ & $1.01$ & $-0.58$ & $0.16$ & $1.03$ & $0.00$\\ 
    $25$ & $8.00 \cdot 10^{-3}$ & $0.034$ & $0.677$ & $0.241$ & $0.677$ & $2.52$ & $0.92$ & $1.62$ & $1.11$ & $-0.68$ & $-0.72$ & $0.84$ & $0.00$\\ 
    $25$ & $1.58 \cdot 10^{-2}$ & $0.018$ & $0.559$ & $0.205$ & $0.559$ & $3.71$ & $0.85$ & $1.57$ & $1.36$ & $-0.88$ & $-2.44$ & $-1.42$ & $0.00$\\ 
    $25$ & $5.00 \cdot 10^{-2}$ & $0.005$ & $0.457$ & $0.180$ & $0.457$ & $7.54$ & $1.28$ & $3.39$ & $0.99$ & $-0.68$ & $-3.28$ & $-5.62$ & $0.00$\\ [1ex]
    $35$ & $5.74 \cdot 10^{-4}$ & $0.675$ & $1.512$ &   ---   &   ---   & $2.65$ & $1.36$ & $2.01$ & $-0.37$ & $-0.70$ & $0.64$ & $0.08$ & $-0.55$\\ 
    $35$ & $8.00 \cdot 10^{-4}$ & $0.484$ & $1.390$ & $0.232$ & $1.440$ & $2.18$ & $0.88$ & $1.75$ & $-0.87$ & $-0.49$ & $0.19$ & $0.05$ & $-0.28$\\ 
    $35$ & $1.30 \cdot 10^{-3}$ & $0.298$ & $1.212$ & $0.231$ & $1.226$ & $1.76$ & $0.77$ & $1.45$ & $-0.21$ & $-0.77$ & $0.01$ & $0.01$ & $-0.02$\\ 
    $35$ & $2.00 \cdot 10^{-3}$ & $0.194$ & $1.058$ & $0.229$ & $1.063$ & $1.76$ & $0.77$ & $1.47$ & $-0.55$ & $-0.55$ & $0.00$ & $0.00$ & $0.00$\\ 
    $35$ & $3.20 \cdot 10^{-3}$ & $0.121$ & $0.960$ & $0.229$ & $0.961$ & $3.07$ & $0.90$ & $1.54$ & $2.07$ & $-0.77$ & $0.27$ & $1.14$ & $0.00$\\ 
    $35$ & $5.00 \cdot 10^{-3}$ & $0.077$ & $0.843$ & $0.231$ & $0.843$ & $2.67$ & $0.94$ & $1.57$ & $1.47$ & $-0.69$ & $-0.18$ & $1.06$ & $0.00$\\ 
    $35$ & $8.00 \cdot 10^{-3}$ & $0.048$ & $0.738$ & $0.228$ & $0.738$ & $2.40$ & $1.00$ & $1.61$ & $1.02$ & $-0.68$ & $-0.15$ & $0.82$ & $0.00$\\ 
    $35$ & $1.30 \cdot 10^{-2}$ & $0.030$ & $0.642$ & $0.211$ & $0.642$ & $3.24$ & $1.15$ & $1.69$ & $1.31$ & $-0.68$ & $-1.84$ & $-0.87$ & $0.00$\\ 
    $35$ & $2.51 \cdot 10^{-2}$ & $0.015$ & $0.538$ & $0.200$ & $0.538$ & $4.14$ & $1.14$ & $1.67$ & $1.46$ & $-0.88$ & $-2.80$ & $-1.50$ & $0.00$\\ 
    $35$ & $8.00 \cdot 10^{-2}$ & $0.005$ & $0.424$ & $0.089$ & $0.424$ & $9.21$ & $1.84$ & $3.55$ & $0.69$ & $-0.81$ & $-3.16$ & $-7.59$ & $0.00$\\ [1ex]
    $45$ & $1.30 \cdot 10^{-3}$ & $0.383$ & $1.321$ & $0.230$ & $1.348$ & $1.94$ & $0.92$ & $1.75$ & $-0.24$ & $-0.23$ & $0.11$ & $0.03$ & $-0.05$\\ 
    $45$ & $2.00 \cdot 10^{-3}$ & $0.249$ & $1.141$ & $0.228$ & $1.149$ & $1.75$ & $0.88$ & $1.47$ & $-0.25$ & $-0.54$ & $0.00$ & $0.00$ & $0.00$\\ 
    $45$ & $3.20 \cdot 10^{-3}$ & $0.156$ & $1.009$ & $0.225$ & $1.011$ & $1.81$ & $0.94$ & $1.48$ & $-0.17$ & $-0.66$ & $0.00$ & $0.00$ & $0.00$\\ 
    $45$ & $5.00 \cdot 10^{-3}$ & $0.099$ & $0.898$ & $0.226$ & $0.899$ & $2.81$ & $1.08$ & $1.58$ & $1.64$ & $-0.53$ & $0.26$ & $1.08$ & $0.00$\\ 
    $45$ & $8.00 \cdot 10^{-3}$ & $0.062$ & $0.766$ & $0.219$ & $0.766$ & $2.50$ & $1.15$ & $1.61$ & $1.24$ & $-0.54$ & $0.07$ & $0.72$ & $0.00$\\ 
    $45$ & $1.30 \cdot 10^{-2}$ & $0.038$ & $0.669$ & $0.211$ & $0.669$ & $2.85$ & $1.28$ & $1.69$ & $1.49$ & $-0.81$ & $-0.85$ & $-0.51$ & $0.00$\\ 
    $45$ & $2.51 \cdot 10^{-2}$ & $0.020$ & $0.541$ & $0.175$ & $0.541$ & $4.28$ & $1.25$ & $1.66$ & $1.36$ & $-0.54$ & $-2.99$ & $-1.70$ & $0.00$\\ 
    $45$ & $8.00 \cdot 10^{-2}$ & $0.006$ & $0.408$ & $0.099$ & $0.408$ & $7.60$ & $2.06$ & $3.52$ & $1.27$ & $-0.86$ & $-2.98$ & $-5.47$ & $0.00$\\ [1ex]
    $60$ & $2.00 \cdot 10^{-3}$ & $0.332$ & $1.285$ & $0.225$ & $1.303$ & $2.10$ & $1.03$ & $1.76$ & $0.29$ & $-0.62$ & $0.04$ & $0.03$ & $0.00$\\ 
    $60$ & $3.20 \cdot 10^{-3}$ & $0.208$ & $1.086$ & $0.220$ & $1.091$ & $1.93$ & $1.07$ & $1.51$ & $-0.65$ & $-0.32$ & $0.00$ & $0.00$ & $0.00$\\ 
    $60$ & $5.00 \cdot 10^{-3}$ & $0.133$ & $0.929$ & $0.216$ & $0.930$ & $3.08$ & $1.25$ & $1.65$ & $1.96$ & $-0.56$ & $0.34$ & $0.96$ & $0.00$\\ 
    $60$ & $8.00 \cdot 10^{-3}$ & $0.083$ & $0.829$ & $0.210$ & $0.829$ & $2.77$ & $1.31$ & $1.69$ & $1.51$ & $-0.47$ & $0.24$ & $0.78$ & $0.00$\\ 
    $60$ & $1.30 \cdot 10^{-2}$ & $0.051$ & $0.705$ & $0.199$ & $0.705$ & $2.94$ & $1.43$ & $1.74$ & $1.49$ & $-0.76$ & $-0.87$ & $-0.40$ & $0.00$\\ 
    $60$ & $2.00 \cdot 10^{-2}$ & $0.033$ & $0.616$ & $0.184$ & $0.616$ & $4.07$ & $1.70$ & $1.88$ & $1.84$ & $-0.88$ & $-2.34$ & $-0.71$ & $0.00$\\ 
    $60$ & $3.98 \cdot 10^{-2}$ & $0.017$ & $0.522$ & $0.132$ & $0.522$ & $4.71$ & $1.82$ & $1.95$ & $0.88$ & $-0.32$ & $-3.03$ & $-2.25$ & $0.00$\\ 
    $60$ & $1.30 \cdot 10^{-1}$ & $0.005$ & $0.371$ & $0.057$ & $0.371$ & $9.42$ & $3.03$ & $3.88$ & $1.38$ & $-0.56$ & $-3.98$ & $-6.82$ & $0.00$\\ [1ex]
    $90$ & $3.20 \cdot 10^{-3}$ & $0.311$ & $1.142$ & $0.214$ & $1.156$ & $2.57$ & $1.35$ & $1.89$ & $0.64$ & $-1.03$ & $0.01$ & $0.01$ & $0.00$\\ 
    $90$ & $5.00 \cdot 10^{-3}$ & $0.199$ & $1.031$ & $0.208$ & $1.035$ & $2.20$ & $1.28$ & $1.61$ & $-0.67$ & $-0.67$ & $0.00$ & $0.00$ & $0.00$\\ 
    $90$ & $8.00 \cdot 10^{-3}$ & $0.124$ & $0.872$ & $0.202$ & $0.873$ & $3.35$ & $1.52$ & $1.76$ & $1.97$ & $-0.51$ & $0.83$ & $0.94$ & $0.00$\\ 
    $90$ & $1.30 \cdot 10^{-2}$ & $0.076$ & $0.751$ & $0.190$ & $0.752$ & $2.75$ & $1.61$ & $1.84$ & $1.06$ & $-0.27$ & $-0.23$ & $0.58$ & $0.00$\\ 
    $90$ & $2.00 \cdot 10^{-2}$ & $0.050$ & $0.638$ & $0.168$ & $0.638$ & $3.87$ & $1.83$ & $1.94$ & $1.74$ & $-0.82$ & $-2.02$ & $-0.33$ & $0.00$\\ 
    $90$ & $3.98 \cdot 10^{-2}$ & $0.025$ & $0.522$ & $0.128$ & $0.522$ & $3.83$ & $1.97$ & $1.95$ & $1.23$ & $-0.53$ & $-2.28$ & $-0.08$ & $0.00$\\ 
    $90$ & $1.30 \cdot 10^{-1}$ & $0.008$ & $0.350$ & $0.049$ & $0.350$ & $4.84$ & $3.35$ & $2.06$ & $0.89$ & $0.16$ & $-1.69$ & $-2.08$ & $0.00$\\ [1ex]
    $120$ & $5.00 \cdot 10^{-3}$ & $0.266$ & $1.043$ & $0.201$ & $1.051$ & $3.69$ & $2.36$ & $2.20$ & $0.39$ & $-1.80$ & $0.00$ & $0.00$ & $0.00$\\ 
    $120$ & $8.00 \cdot 10^{-3}$ & $0.166$ & $0.866$ & $0.192$ & $0.868$ & $3.00$ & $1.94$ & $1.90$ & $-0.41$ & $-1.31$ & $0.00$ & $0.00$ & $0.00$\\ 
    $120$ & $1.30 \cdot 10^{-2}$ & $0.102$ & $0.768$ & $0.179$ & $0.768$ & $4.61$ & $2.10$ & $2.09$ & $2.77$ & $0.49$ & $1.44$ & $1.68$ & $0.00$\\ 
    $120$ & $2.00 \cdot 10^{-2}$ & $0.066$ & $0.623$ & $0.164$ & $0.624$ & $3.95$ & $2.26$ & $2.15$ & $2.33$ & $-0.36$ & $-0.64$ & $-0.43$ & $0.00$\\ 
    $120$ & $3.20 \cdot 10^{-2}$ & $0.041$ & $0.576$ & $0.139$ & $0.576$ & $5.80$ & $2.77$ & $2.51$ & $2.65$ & $-0.49$ & $-3.25$ & $-1.34$ & $0.00$\\ 
    $120$ & $6.31 \cdot 10^{-2}$ & $0.021$ & $0.477$ & $0.090$ & $0.477$ & $4.95$ & $3.05$ & $2.71$ & $1.57$ & $-0.41$ & $-2.25$ & $-0.50$ & $0.00$\\ 
    $120$ & $2.00 \cdot 10^{-1}$ & $0.007$ & $0.322$ & $0.034$ & $0.322$ & $10.59$ & $4.76$ & $4.82$ & $1.22$ & $0.37$ & $-3.21$ & $-7.37$ & $0.00$\\ [1ex]
    $150$ & $2.00 \cdot 10^{-2}$ & $0.083$ & $0.732$ & $0.155$ & $0.732$ & $8.85$ & $4.45$ & $3.86$ & $5.11$ & $0.85$ & $2.89$ & $2.95$ & $0.00$\\ 
    $150$ & $3.20 \cdot 10^{-2}$ & $0.052$ & $0.568$ & $0.129$ & $0.568$ & $7.99$ & $5.05$ & $4.10$ & $4.56$ & $-0.53$ & $-0.66$ & $0.68$ & $0.00$\\ 
    $150$ & $6.31 \cdot 10^{-2}$ & $0.026$ & $0.431$ & $0.088$ & $0.431$ & $8.97$ & $5.68$ & $4.44$ & $4.69$ & $-1.27$ & $-2.19$ & $-0.52$ & $0.00$\\ 
    $150$ & $2.00 \cdot 10^{-1}$ & $0.008$ & $0.305$ & $0.038$ & $0.305$ & $12.86$ & $8.39$ & $7.12$ & $3.05$ & $-0.72$ & $-3.74$ & $-4.50$ & $0.00$\\ 

\hline\hline
\end{tabular}
\end{center}
\end{tiny}
\begin{small}
  \caption{Corrected reduced cross section $\sigma_{r}$
    and $F_2$, for $y < 0.6$,  from H1 1996/97 data,
    continued from \Tab~\ref{tab:table97}.
    \label{tab:table97_2}
}
\end{small} 
\end{table}

\begin{sidewaystable}
\begin{tiny}
\begin{center}
\begin{tabular}{cccccccccrrrrrrrrrrrrrr}
  \hline\hline\\[-1.5ex]
  $Q^2/\gevsq$ & $x$ & $y$ & $\sigma_{r}$ & $R$ & $F_2$ & $\delta_{\rm ave, tot}$
  & $\delta_{\rm ave, stat}$
  & $\delta_{\rm ave, uncor}$
  & \multicolumn{1}{c}{$\gamma^{\rm ave}_{1}$}
  & \multicolumn{1}{c}{$\gamma^{\rm ave}_{2}$}
  & \multicolumn{1}{c}{$\gamma^{\rm ave}_{3}$}
  & \multicolumn{1}{c}{$\gamma^{\rm ave}_{4}$}
  & \multicolumn{1}{c}{$\gamma^{\rm ave}_{5}$}
  & \multicolumn{1}{c}{$\gamma^{\rm ave}_{6}$}
  & \multicolumn{1}{c}{$\gamma^{\rm ave}_{7}$}
  & \multicolumn{1}{c}{$\gamma^{\rm ave}_{8}$}
  & \multicolumn{1}{c}{$\gamma^{\rm ave}_{9}$}
  & \multicolumn{1}{c}{$\gamma^{\rm ave}_{10}$}
  & \multicolumn{1}{c}{$\gamma^{\rm ave}_{11}$}
  & \multicolumn{1}{c}{$\gamma^{\rm ave}_{12}$}
  & \multicolumn{1}{c}{$\gamma^{\rm ave}_{13}$}
  & \multicolumn{1}{c}{$\gamma^{\rm ave}_{14}$}\\[1.5ex]
  \hline
    $12$ & $1.61 \cdot 10^{-4}$ & $0.826$ & $1.236$ &   ---   &   ---   & $5.72$ & $4.11$ & $3.78$ & $0.97$ & $0.00$ & $-0.06$ & $-0.15$ & $-0.04$ & $-0.02$ & $0.02$ & $-0.14$ & $0.23$ & $-0.47$ & $0.33$ & $-0.46$ & $-0.08$ & $-0.06$\\ 
    $12$ & $1.97 \cdot 10^{-4}$ & $0.675$ & $1.278$ &   ---   &   ---   & $3.48$ & $0.88$ & $2.15$ & $1.02$ & $1.82$ & $1.37$ & $-0.17$ & $0.16$ & $-0.02$ & $0.07$ & $0.07$ & $0.22$ & $-0.40$ & $0.22$ & $-0.38$ & $-0.10$ & $-0.07$\\ 
    $12$ & $2.00 \cdot 10^{-4}$ & $0.591$ & $1.318$ & $0.202$ & $1.387$ & $2.92$ & $1.21$ & $0.71$ & $0.92$ & $1.44$ & $-1.80$ & $-0.21$ & $-0.48$ & $0.14$ & $0.15$ & $-0.19$ & $-0.03$ & $-0.10$ & $-0.09$ & $0.11$ & $0.04$ & $0.08$\\ 
    $12$ & $3.20 \cdot 10^{-4}$ & $0.369$ & $1.238$ & $0.205$ & $1.259$ & $1.57$ & $0.86$ & $0.68$ & $0.96$ & $0.24$ & $-0.31$ & $-0.19$ & $-0.23$ & $-0.06$ & $0.07$ & $-0.15$ & $0.03$ & $-0.15$ & $-0.05$ & $0.12$ & $0.04$ & $0.08$\\ 
    $12$ & $3.20 \cdot 10^{-4}$ & $0.415$ & $1.234$ & $0.205$ & $1.262$ & $2.19$ & $0.57$ & $1.73$ & $0.98$ & $0.39$ & $0.26$ & $-0.20$ & $0.01$ & $-0.09$ & $0.00$ & $0.17$ & $-0.03$ & $-0.29$ & $0.19$ & $-0.30$ & $-0.04$ & $-0.10$\\ 
    $12$ & $5.00 \cdot 10^{-4}$ & $0.236$ & $1.164$ & $0.209$ & $1.171$ & $1.40$ & $0.65$ & $0.67$ & $0.96$ & $0.03$ & $-0.03$ & $-0.22$ & $-0.10$ & $-0.10$ & $0.00$ & $0.02$ & $0.01$ & $-0.29$ & $0.05$ & $0.01$ & $0.03$ & $0.03$\\ 
    $12$ & $8.00 \cdot 10^{-4}$ & $0.148$ & $1.064$ & $0.213$ & $1.067$ & $1.62$ & $0.94$ & $0.69$ & $0.96$ & $0.00$ & $-0.03$ & $-0.30$ & $0.01$ & $-0.13$ & $-0.06$ & $-0.09$ & $0.10$ & $-0.41$ & $0.00$ & $0.19$ & $0.07$ & $0.09$\\ [1ex]
    $15$ & $2.01 \cdot 10^{-4}$ & $0.827$ & $1.269$ &   ---   &   ---   & $4.98$ & $3.20$ & $3.61$ & $0.97$ & $0.00$ & $-0.06$ & $-0.15$ & $-0.05$ & $-0.03$ & $0.01$ & $-0.11$ & $0.20$ & $-0.47$ & $0.34$ & $-0.47$ & $-0.07$ & $-0.06$\\ 
    $15$ & $2.46 \cdot 10^{-4}$ & $0.675$ & $1.369$ &   ---   &   ---   & $2.94$ & $0.92$ & $2.17$ & $0.99$ & $0.93$ & $0.67$ & $-0.18$ & $-0.02$ & $-0.04$ & $0.01$ & $0.00$ & $0.13$ & $-0.52$ & $0.42$ & $-0.54$ & $-0.09$ & $-0.03$\\ 
    $15$ & $3.20 \cdot 10^{-4}$ & $0.462$ & $1.329$ & $0.214$ & $1.369$ & $1.91$ & $0.84$ & $0.70$ & $0.94$ & $0.70$ & $-0.87$ & $-0.06$ & $-0.44$ & $0.07$ & $0.11$ & $-0.15$ & $0.00$ & $-0.16$ & $-0.07$ & $0.11$ & $0.04$ & $0.08$\\ 
    $15$ & $3.20 \cdot 10^{-4}$ & $0.519$ & $1.307$ & $0.214$ & $1.359$ & $2.51$ & $0.67$ & $1.97$ & $0.98$ & $0.65$ & $0.46$ & $-0.15$ & $0.03$ & $-0.05$ & $-0.03$ & $-0.15$ & $0.17$ & $-0.33$ & $0.20$ & $-0.32$ & $-0.04$ & $-0.11$\\ 
    $15$ & $5.00 \cdot 10^{-4}$ & $0.295$ & $1.225$ & $0.217$ & $1.238$ & $1.34$ & $0.55$ & $0.66$ & $0.96$ & $0.07$ & $-0.07$ & $-0.16$ & $-0.14$ & $-0.08$ & $0.04$ & $-0.12$ & $0.04$ & $-0.20$ & $0.00$ & $0.04$ & $0.03$ & $0.04$\\ 
    $15$ & $8.00 \cdot 10^{-4}$ & $0.185$ & $1.119$ & $0.220$ & $1.123$ & $1.35$ & $0.55$ & $0.64$ & $0.96$ & $0.01$ & $-0.03$ & $-0.23$ & $-0.02$ & $-0.10$ & $-0.01$ & $-0.06$ & $0.05$ & $-0.30$ & $0.04$ & $0.04$ & $0.04$ & $0.03$\\ 
    $15$ & $1.30 \cdot 10^{-3}$ & $0.114$ & $0.992$ & $0.224$ & $0.993$ & $1.43$ & $0.59$ & $0.67$ & $0.96$ & $-0.01$ & $-0.03$ & $-0.29$ & $0.04$ & $-0.14$ & $-0.13$ & $0.04$ & $-0.01$ & $-0.39$ & $-0.07$ & $0.21$ & $-0.05$ & $0.00$\\ 
    $15$ & $2.00 \cdot 10^{-3}$ & $0.074$ & $0.883$ & $0.228$ & $0.883$ & $1.49$ & $0.61$ & $0.67$ & $0.97$ & $0.00$ & $-0.03$ & $-0.24$ & $0.06$ & $-0.39$ & $-0.34$ & $-0.24$ & $-0.07$ & $0.10$ & $-0.27$ & $0.01$ & $-0.05$ & $0.00$\\ 
    $15$ & $3.20 \cdot 10^{-3}$ & $0.046$ & $0.797$ & $0.233$ & $0.797$ & $1.46$ & $0.66$ & $0.69$ & $0.97$ & $0.00$ & $-0.02$ & $-0.25$ & $-0.01$ & $-0.22$ & $-0.21$ & $-0.27$ & $-0.10$ & $0.09$ & $-0.24$ & $0.02$ & $-0.05$ & $-0.01$\\ 
    $15$ & $5.00 \cdot 10^{-3}$ & $0.030$ & $0.728$ & $0.237$ & $0.728$ & $1.49$ & $0.70$ & $0.70$ & $0.97$ & $0.00$ & $-0.02$ & $-0.23$ & $-0.04$ & $-0.28$ & $-0.26$ & $-0.21$ & $-0.06$ & $0.14$ & $-0.20$ & $0.03$ & $-0.01$ & $-0.01$\\ 
    $15$ & $1.00 \cdot 10^{-2}$ & $0.015$ & $0.609$ & $0.242$ & $0.609$ & $1.38$ & $0.59$ & $0.65$ & $0.96$ & $0.00$ & $-0.02$ & $-0.20$ & $0.00$ & $-0.12$ & $-0.17$ & $-0.21$ & $-0.10$ & $0.17$ & $-0.13$ & $0.04$ & $0.08$ & $-0.01$\\ 
    $15$ & $2.51 \cdot 10^{-2}$ & $0.006$ & $0.505$ & $0.224$ & $0.505$ & $2.63$ & $0.72$ & $1.79$ & $0.96$ & $-0.01$ & $-0.03$ & $0.20$ & $0.29$ & $-0.12$ & $0.98$ & $-0.02$ & $0.43$ & $0.61$ & $0.67$ & $0.26$ & $0.26$ & $-0.08$\\ [1ex]
    $20$ & $2.68 \cdot 10^{-4}$ & $0.826$ & $1.334$ &   ---   &   ---   & $5.05$ & $3.25$ & $3.67$ & $0.97$ & $0.00$ & $-0.06$ & $-0.15$ & $-0.03$ & $-0.02$ & $0.03$ & $-0.12$ & $0.23$ & $-0.46$ & $0.32$ & $-0.46$ & $-0.08$ & $-0.06$\\ 
    $20$ & $3.28 \cdot 10^{-4}$ & $0.675$ & $1.408$ &   ---   &   ---   & $2.79$ & $1.04$ & $2.10$ & $0.99$ & $0.78$ & $0.56$ & $-0.17$ & $0.10$ & $-0.03$ & $0.08$ & $0.07$ & $0.20$ & $-0.36$ & $0.18$ & $-0.33$ & $-0.09$ & $-0.09$\\ 
    $20$ & $5.00 \cdot 10^{-4}$ & $0.394$ & $1.319$ & $0.225$ & $1.348$ & $1.50$ & $0.74$ & $0.70$ & $0.96$ & $0.26$ & $-0.33$ & $-0.12$ & $-0.20$ & $-0.01$ & $0.06$ & $-0.16$ & $0.00$ & $-0.12$ & $-0.05$ & $0.11$ & $0.04$ & $0.08$\\ 
    $20$ & $5.00 \cdot 10^{-4}$ & $0.443$ & $1.321$ & $0.225$ & $1.359$ & $2.16$ & $0.64$ & $1.71$ & $0.97$ & $0.26$ & $0.16$ & $-0.13$ & $0.04$ & $-0.05$ & $-0.03$ & $-0.22$ & $0.21$ & $-0.27$ & $0.10$ & $-0.23$ & $-0.04$ & $-0.14$\\ 
    $20$ & $8.00 \cdot 10^{-4}$ & $0.246$ & $1.209$ & $0.227$ & $1.218$ & $1.30$ & $0.49$ & $0.64$ & $0.96$ & $0.02$ & $-0.05$ & $-0.15$ & $-0.05$ & $-0.07$ & $-0.02$ & $-0.13$ & $0.06$ & $-0.23$ & $0.04$ & $0.02$ & $0.03$ & $0.03$\\ 
    $20$ & $1.30 \cdot 10^{-3}$ & $0.151$ & $1.087$ & $0.229$ & $1.089$ & $1.34$ & $0.51$ & $0.64$ & $0.96$ & $0.00$ & $-0.03$ & $-0.17$ & $-0.01$ & $-0.09$ & $-0.06$ & $-0.08$ & $0.08$ & $-0.34$ & $0.06$ & $0.05$ & $0.04$ & $0.03$\\ 
    $20$ & $2.00 \cdot 10^{-3}$ & $0.098$ & $0.979$ & $0.231$ & $0.980$ & $1.44$ & $0.55$ & $0.66$ & $0.96$ & $-0.01$ & $-0.03$ & $-0.22$ & $0.03$ & $-0.13$ & $-0.16$ & $-0.04$ & $0.06$ & $-0.49$ & $-0.04$ & $0.23$ & $-0.02$ & $0.00$\\ 
    $20$ & $3.20 \cdot 10^{-3}$ & $0.062$ & $0.863$ & $0.233$ & $0.864$ & $1.38$ & $0.57$ & $0.66$ & $0.96$ & $0.00$ & $-0.02$ & $-0.12$ & $0.00$ & $-0.23$ & $-0.23$ & $-0.17$ & $-0.06$ & $0.12$ & $-0.21$ & $0.02$ & $-0.02$ & $0.01$\\ 
    $20$ & $5.00 \cdot 10^{-3}$ & $0.039$ & $0.774$ & $0.235$ & $0.774$ & $1.39$ & $0.61$ & $0.68$ & $0.96$ & $0.00$ & $-0.02$ & $-0.22$ & $-0.01$ & $-0.11$ & $-0.22$ & $-0.14$ & $-0.04$ & $0.13$ & $-0.16$ & $0.02$ & $-0.01$ & $0.01$\\ 
    $20$ & $1.00 \cdot 10^{-2}$ & $0.020$ & $0.653$ & $0.235$ & $0.653$ & $1.37$ & $0.49$ & $0.63$ & $0.96$ & $-0.01$ & $-0.01$ & $-0.19$ & $-0.03$ & $-0.23$ & $-0.17$ & $-0.21$ & $-0.20$ & $0.12$ & $-0.28$ & $-0.03$ & $0.06$ & $0.02$\\ 
    $20$ & $2.51 \cdot 10^{-2}$ & $0.008$ & $0.524$ & $0.207$ & $0.524$ & $2.37$ & $0.56$ & $1.78$ & $0.96$ & $-0.01$ & $-0.02$ & $0.00$ & $0.23$ & $-0.39$ & $0.82$ & $-0.16$ & $0.04$ & $0.40$ & $0.08$ & $0.06$ & $0.31$ & $-0.04$\\ [1ex]
    $25$ & $3.35 \cdot 10^{-4}$ & $0.827$ & $1.407$ &   ---   &   ---   & $5.80$ & $4.09$ & $3.92$ & $0.97$ & $0.00$ & $-0.06$ & $-0.15$ & $-0.04$ & $-0.03$ & $0.02$ & $-0.12$ & $0.20$ & $-0.45$ & $0.32$ & $-0.45$ & $-0.07$ & $-0.06$\\ 
    $25$ & $4.10 \cdot 10^{-4}$ & $0.675$ & $1.408$ &   ---   &   ---   & $2.79$ & $1.15$ & $2.10$ & $0.99$ & $0.72$ & $0.51$ & $-0.14$ & $0.18$ & $0.00$ & $0.14$ & $0.03$ & $0.33$ & $-0.28$ & $0.03$ & $-0.22$ & $-0.10$ & $-0.12$\\ 
    $25$ & $5.00 \cdot 10^{-4}$ & $0.492$ & $1.405$ & $0.229$ & $1.457$ & $1.79$ & $0.94$ & $0.75$ & $0.95$ & $0.52$ & $-0.65$ & $-0.12$ & $-0.25$ & $0.11$ & $0.11$ & $-0.19$ & $-0.05$ & $-0.08$ & $-0.08$ & $0.09$ & $0.03$ & $0.08$\\ 
    $25$ & $5.00 \cdot 10^{-4}$ & $0.553$ & $1.376$ & $0.229$ & $1.445$ & $2.46$ & $1.04$ & $1.88$ & $0.97$ & $0.32$ & $0.20$ & $-0.18$ & $0.00$ & $-0.06$ & $0.00$ & $0.04$ & $0.08$ & $-0.35$ & $0.23$ & $-0.35$ & $-0.06$ & $-0.09$\\ 
    $25$ & $8.00 \cdot 10^{-4}$ & $0.308$ & $1.279$ & $0.230$ & $1.294$ & $1.33$ & $0.54$ & $0.67$ & $0.96$ & $0.04$ & $-0.05$ & $-0.13$ & $-0.09$ & $-0.05$ & $0.00$ & $-0.13$ & $0.05$ & $-0.20$ & $0.01$ & $0.04$ & $0.03$ & $0.04$\\ 
    $25$ & $1.30 \cdot 10^{-3}$ & $0.189$ & $1.143$ & $0.231$ & $1.148$ & $1.33$ & $0.52$ & $0.65$ & $0.96$ & $0.00$ & $-0.04$ & $-0.15$ & $-0.02$ & $-0.07$ & $-0.04$ & $-0.13$ & $0.09$ & $-0.27$ & $0.05$ & $0.04$ & $0.03$ & $0.03$\\ 
    $25$ & $2.00 \cdot 10^{-3}$ & $0.123$ & $1.034$ & $0.232$ & $1.036$ & $1.38$ & $0.55$ & $0.66$ & $0.96$ & $0.00$ & $-0.03$ & $-0.14$ & $0.06$ & $-0.09$ & $-0.13$ & $-0.13$ & $0.09$ & $-0.33$ & $-0.06$ & $0.20$ & $-0.02$ & $0.00$\\ 
    $25$ & $3.20 \cdot 10^{-3}$ & $0.077$ & $0.913$ & $0.232$ & $0.914$ & $1.40$ & $0.56$ & $0.67$ & $0.97$ & $0.00$ & $-0.03$ & $-0.18$ & $0.09$ & $-0.17$ & $-0.21$ & $-0.27$ & $-0.04$ & $0.12$ & $-0.23$ & $0.04$ & $-0.03$ & $0.00$\\ 
    $25$ & $5.00 \cdot 10^{-3}$ & $0.049$ & $0.800$ & $0.232$ & $0.800$ & $1.40$ & $0.57$ & $0.67$ & $0.97$ & $0.00$ & $-0.03$ & $-0.16$ & $0.10$ & $-0.20$ & $-0.23$ & $-0.23$ & $-0.06$ & $0.06$ & $-0.23$ & $0.00$ & $-0.04$ & $0.02$\\ 
    $25$ & $8.00 \cdot 10^{-3}$ & $0.031$ & $0.706$ & $0.231$ & $0.706$ & $1.40$ & $0.61$ & $0.69$ & $0.96$ & $0.00$ & $-0.01$ & $-0.15$ & $-0.12$ & $-0.10$ & $-0.25$ & $-0.22$ & $-0.05$ & $0.13$ & $-0.15$ & $0.03$ & $-0.01$ & $0.01$\\ 
    $25$ & $1.30 \cdot 10^{-2}$ & $0.019$ & $0.639$ & $0.224$ & $0.639$ & $1.48$ & $0.61$ & $0.70$ & $0.96$ & $0.00$ & $-0.01$ & $-0.15$ & $-0.07$ & $-0.22$ & $-0.22$ & $-0.32$ & $-0.13$ & $0.23$ & $-0.29$ & $-0.04$ & $0.15$ & $-0.01$\\ 
    $25$ & $2.00 \cdot 10^{-2}$ & $0.012$ & $0.584$ & $0.211$ & $0.584$ & $1.75$ & $0.84$ & $0.77$ & $0.97$ & $-0.01$ & $-0.01$ & $-0.19$ & $0.08$ & $-0.48$ & $-0.05$ & $-0.40$ & $-0.31$ & $-0.10$ & $-0.50$ & $-0.10$ & $0.04$ & $0.12$\\ 
    $25$ & $3.98 \cdot 10^{-2}$ & $0.006$ & $0.500$ & $0.167$ & $0.500$ & $2.68$ & $0.62$ & $1.81$ & $0.97$ & $-0.01$ & $-0.03$ & $0.00$ & $0.29$ & $-0.09$ & $0.98$ & $-0.05$ & $0.43$ & $0.89$ & $0.67$ & $0.21$ & $0.27$ & $-0.07$\\ [1ex]

  \hline\hline
\end{tabular}
\end{center}
\end{tiny}
\begin{small}
  \caption{Combined reduced cross section measurement from H1.
    The uncertainties are
    quoted in $\%$.  $\delta_{\rm ave, tot}$ is the total uncertainty
    determined as the quadratic sum of all uncertainties.
    $\delta_{\rm ave, stat}$ is the statistical uncertainty. $\delta_{\rm
      ave, uncor}$ represents the uncorrelated systematic uncertainty.
    $\gamma^{\rm ave}_{i}$ are the bin-to-bin
    correlated systematic uncertainties on the cross section
    measurement after diagonalisation.
    All normalisation uncertainties are included, except for the
    correlated $0.5\%$ common to all H1 cross section measurements.
    \label{tab:table9700comb}
}
\end{small} 
\end{sidewaystable}

\begin{sidewaystable}
\begin{tiny}
\begin{center}
\begin{tabular}{cccccccccrrrrrrrrrrrrrr}
  \hline\hline\\[-1.5ex]
  $Q^2\gevsq$ & $x$ & $y$ & $\sigma_{r}$ & $R$ & $F_2$ & $\delta_{\rm ave, tot}$
  & $\delta_{\rm ave, stat}$
  & $\delta_{\rm ave, uncor}$
  & \multicolumn{1}{c}{$\gamma^{\rm ave}_{1}$}
  & \multicolumn{1}{c}{$\gamma^{\rm ave}_{2}$}
  & \multicolumn{1}{c}{$\gamma^{\rm ave}_{3}$}
  & \multicolumn{1}{c}{$\gamma^{\rm ave}_{4}$}
  & \multicolumn{1}{c}{$\gamma^{\rm ave}_{5}$}
  & \multicolumn{1}{c}{$\gamma^{\rm ave}_{6}$}
  & \multicolumn{1}{c}{$\gamma^{\rm ave}_{7}$}
  & \multicolumn{1}{c}{$\gamma^{\rm ave}_{8}$}
  & \multicolumn{1}{c}{$\gamma^{\rm ave}_{9}$}
  & \multicolumn{1}{c}{$\gamma^{\rm ave}_{10}$}
  & \multicolumn{1}{c}{$\gamma^{\rm ave}_{11}$}
  & \multicolumn{1}{c}{$\gamma^{\rm ave}_{12}$}
  & \multicolumn{1}{c}{$\gamma^{\rm ave}_{13}$}
  & \multicolumn{1}{c}{$\gamma^{\rm ave}_{14}$}\\[1.5ex]
  \hline
    $35$ & $5.74 \cdot 10^{-4}$ & $0.675$ & $1.519$ &   ---   &   ---   & $2.74$ & $1.36$ & $2.01$ & $0.98$ & $0.43$ & $0.28$ & $-0.12$ & $0.14$ & $0.01$ & $0.11$ & $-0.11$ & $0.37$ & $-0.33$ & $0.08$ & $-0.26$ & $-0.11$ & $-0.12$\\ 
    $35$ & $8.00 \cdot 10^{-4}$ & $0.431$ & $1.402$ & $0.231$ & $1.440$ & $1.55$ & $0.83$ & $0.74$ & $0.96$ & $0.18$ & $-0.21$ & $-0.04$ & $-0.31$ & $0.05$ & $0.08$ & $-0.14$ & $0.00$ & $-0.14$ & $-0.06$ & $0.11$ & $0.04$ & $0.08$\\ 
    $35$ & $8.00 \cdot 10^{-4}$ & $0.484$ & $1.396$ & $0.231$ & $1.446$ & $2.27$ & $0.87$ & $1.75$ & $0.97$ & $0.22$ & $0.12$ & $-0.17$ & $0.00$ & $-0.06$ & $-0.02$ & $-0.04$ & $0.10$ & $-0.33$ & $0.20$ & $-0.32$ & $-0.05$ & $-0.10$\\ 
    $35$ & $1.30 \cdot 10^{-3}$ & $0.265$ & $1.234$ & $0.231$ & $1.244$ & $1.34$ & $0.58$ & $0.67$ & $0.96$ & $0.01$ & $-0.04$ & $-0.13$ & $-0.02$ & $-0.04$ & $-0.02$ & $-0.16$ & $0.09$ & $-0.18$ & $0.02$ & $0.04$ & $0.02$ & $0.01$\\ 
    $35$ & $2.00 \cdot 10^{-3}$ & $0.172$ & $1.103$ & $0.230$ & $1.107$ & $1.37$ & $0.59$ & $0.68$ & $0.96$ & $0.00$ & $-0.05$ & $-0.19$ & $0.08$ & $-0.07$ & $-0.07$ & $-0.11$ & $0.07$ & $-0.26$ & $0.05$ & $0.04$ & $0.03$ & $0.02$\\ 
    $35$ & $3.20 \cdot 10^{-3}$ & $0.108$ & $0.980$ & $0.229$ & $0.981$ & $1.46$ & $0.63$ & $0.69$ & $0.96$ & $0.00$ & $-0.04$ & $-0.20$ & $0.20$ & $-0.08$ & $-0.15$ & $-0.13$ & $0.14$ & $-0.35$ & $-0.07$ & $0.23$ & $-0.01$ & $-0.01$\\ 
    $35$ & $5.00 \cdot 10^{-3}$ & $0.069$ & $0.858$ & $0.227$ & $0.858$ & $1.43$ & $0.62$ & $0.69$ & $0.97$ & $0.00$ & $-0.03$ & $-0.21$ & $0.10$ & $-0.12$ & $-0.16$ & $-0.28$ & $-0.09$ & $0.12$ & $-0.21$ & $0.03$ & $-0.04$ & $0.00$\\ 
    $35$ & $8.00 \cdot 10^{-3}$ & $0.043$ & $0.761$ & $0.223$ & $0.762$ & $1.44$ & $0.65$ & $0.70$ & $0.96$ & $0.00$ & $-0.02$ & $-0.14$ & $0.01$ & $-0.21$ & $-0.24$ & $-0.22$ & $-0.01$ & $0.16$ & $-0.17$ & $0.00$ & $-0.03$ & $0.01$\\ 
    $35$ & $1.30 \cdot 10^{-2}$ & $0.027$ & $0.673$ & $0.214$ & $0.673$ & $1.48$ & $0.69$ & $0.72$ & $0.96$ & $0.00$ & $0.00$ & $-0.20$ & $-0.14$ & $-0.01$ & $-0.20$ & $-0.24$ & $-0.09$ & $0.19$ & $-0.23$ & $0.00$ & $0.09$ & $0.00$\\ 
    $35$ & $2.00 \cdot 10^{-2}$ & $0.017$ & $0.609$ & $0.199$ & $0.609$ & $1.62$ & $0.71$ & $0.72$ & $0.96$ & $0.00$ & $-0.02$ & $-0.13$ & $0.04$ & $-0.45$ & $-0.28$ & $-0.35$ & $-0.20$ & $0.15$ & $-0.37$ & $-0.05$ & $0.16$ & $0.00$\\ 
    $35$ & $3.98 \cdot 10^{-2}$ & $0.009$ & $0.512$ & $0.156$ & $0.512$ & $2.69$ & $0.76$ & $2.14$ & $0.96$ & $-0.01$ & $-0.02$ & $0.00$ & $0.14$ & $-0.21$ & $0.61$ & $-0.16$ & $0.17$ & $0.59$ & $0.51$ & $0.23$ & $-0.06$ & $0.04$\\ 
    $35$ & $8.00 \cdot 10^{-2}$ & $0.004$ & $0.452$ & $0.097$ & $0.452$ & $4.89$ & $1.72$ & $3.54$ & $0.97$ & $-0.01$ & $-0.06$ & $0.19$ & $0.57$ & $0.15$ & $1.29$ & $0.34$ & $1.36$ & $0.96$ & $-0.09$ & $-0.40$ & $1.52$ & $-0.24$\\ [1ex]
    $45$ & $8.00 \cdot 10^{-4}$ & $0.554$ & $1.473$ & $0.231$ & $1.548$ & $1.97$ & $1.38$ & $0.89$ & $0.96$ & $0.25$ & $-0.33$ & $-0.11$ & $-0.09$ & $0.14$ & $0.10$ & $-0.20$ & $-0.08$ & $-0.07$ & $-0.07$ & $0.09$ & $0.03$ & $0.08$\\ 
    $45$ & $1.30 \cdot 10^{-3}$ & $0.341$ & $1.321$ & $0.230$ & $1.342$ & $1.53$ & $0.87$ & $0.76$ & $0.96$ & $0.04$ & $-0.07$ & $-0.06$ & $-0.08$ & $0.00$ & $-0.01$ & $-0.13$ & $0.02$ & $-0.12$ & $-0.03$ & $0.12$ & $0.04$ & $0.08$\\ 
    $45$ & $1.30 \cdot 10^{-3}$ & $0.383$ & $1.333$ & $0.230$ & $1.359$ & $2.24$ & $0.92$ & $1.75$ & $0.97$ & $0.03$ & $-0.02$ & $-0.18$ & $0.05$ & $-0.07$ & $0.04$ & $0.12$ & $0.07$ & $-0.22$ & $0.07$ & $-0.20$ & $-0.04$ & $-0.13$\\ 
    $45$ & $2.00 \cdot 10^{-3}$ & $0.222$ & $1.167$ & $0.228$ & $1.173$ & $1.39$ & $0.65$ & $0.70$ & $0.96$ & $0.00$ & $-0.03$ & $-0.15$ & $0.04$ & $-0.05$ & $-0.05$ & $-0.11$ & $0.08$ & $-0.19$ & $0.03$ & $0.04$ & $0.02$ & $0.01$\\ 
    $45$ & $3.20 \cdot 10^{-3}$ & $0.138$ & $1.021$ & $0.226$ & $1.023$ & $1.43$ & $0.68$ & $0.71$ & $0.96$ & $0.00$ & $-0.04$ & $-0.15$ & $0.12$ & $-0.06$ & $-0.09$ & $-0.11$ & $0.11$ & $-0.30$ & $0.04$ & $0.06$ & $0.04$ & $0.01$\\ 
    $45$ & $5.00 \cdot 10^{-3}$ & $0.089$ & $0.902$ & $0.223$ & $0.903$ & $1.48$ & $0.71$ & $0.71$ & $0.96$ & $0.00$ & $-0.03$ & $-0.12$ & $0.21$ & $-0.03$ & $-0.09$ & $-0.28$ & $-0.14$ & $0.01$ & $-0.28$ & $0.03$ & $-0.04$ & $0.00$\\ 
    $45$ & $8.00 \cdot 10^{-3}$ & $0.055$ & $0.790$ & $0.217$ & $0.790$ & $1.47$ & $0.73$ & $0.73$ & $0.96$ & $0.00$ & $-0.02$ & $-0.11$ & $0.02$ & $-0.03$ & $-0.23$ & $-0.19$ & $-0.01$ & $0.16$ & $-0.20$ & $0.01$ & $-0.04$ & $0.00$\\ 
    $45$ & $1.30 \cdot 10^{-2}$ & $0.034$ & $0.694$ & $0.207$ & $0.694$ & $1.50$ & $0.77$ & $0.74$ & $0.96$ & $0.00$ & $-0.02$ & $-0.09$ & $0.00$ & $-0.09$ & $-0.16$ & $-0.22$ & $0.03$ & $0.23$ & $-0.18$ & $0.01$ & $0.04$ & $0.00$\\ 
    $45$ & $2.00 \cdot 10^{-2}$ & $0.022$ & $0.613$ & $0.191$ & $0.613$ & $1.66$ & $0.79$ & $0.76$ & $0.96$ & $0.00$ & $-0.02$ & $-0.17$ & $-0.01$ & $-0.21$ & $-0.44$ & $-0.27$ & $-0.21$ & $0.27$ & $-0.38$ & $-0.08$ & $0.18$ & $-0.01$\\ 
    $45$ & $3.20 \cdot 10^{-2}$ & $0.014$ & $0.542$ & $0.163$ & $0.542$ & $1.79$ & $1.09$ & $0.85$ & $0.96$ & $-0.01$ & $-0.01$ & $-0.04$ & $0.00$ & $-0.45$ & $-0.04$ & $-0.23$ & $-0.03$ & $0.28$ & $-0.13$ & $0.00$ & $0.00$ & $0.09$\\ 
    $45$ & $6.31 \cdot 10^{-2}$ & $0.007$ & $0.459$ & $0.107$ & $0.459$ & $3.00$ & $0.92$ & $1.85$ & $0.97$ & $-0.01$ & $-0.06$ & $-0.09$ & $0.43$ & $0.00$ & $0.51$ & $0.14$ & $0.81$ & $1.29$ & $0.95$ & $0.28$ & $0.19$ & $-0.09$\\ [1ex]
    $60$ & $1.30 \cdot 10^{-3}$ & $0.454$ & $1.400$ & $0.227$ & $1.442$ & $1.87$ & $1.30$ & $0.89$ & $0.96$ & $0.05$ & $-0.07$ & $-0.04$ & $-0.17$ & $0.05$ & $0.07$ & $-0.17$ & $-0.03$ & $-0.10$ & $-0.06$ & $0.10$ & $0.03$ & $0.08$\\ 
    $60$ & $2.00 \cdot 10^{-3}$ & $0.295$ & $1.262$ & $0.225$ & $1.276$ & $1.45$ & $0.75$ & $0.75$ & $0.96$ & $0.00$ & $-0.03$ & $-0.06$ & $0.03$ & $-0.02$ & $-0.02$ & $-0.12$ & $0.09$ & $-0.15$ & $0.00$ & $0.07$ & $0.02$ & $0.01$\\ 
    $60$ & $3.20 \cdot 10^{-3}$ & $0.185$ & $1.091$ & $0.221$ & $1.095$ & $1.46$ & $0.75$ & $0.73$ & $0.96$ & $-0.01$ & $-0.04$ & $-0.14$ & $0.10$ & $-0.05$ & $-0.07$ & $-0.06$ & $0.04$ & $-0.22$ & $0.06$ & $0.01$ & $0.03$ & $0.02$\\ 
    $60$ & $5.00 \cdot 10^{-3}$ & $0.118$ & $0.968$ & $0.217$ & $0.970$ & $1.55$ & $0.81$ & $0.76$ & $0.96$ & $-0.01$ & $-0.04$ & $-0.10$ & $0.23$ & $-0.06$ & $-0.13$ & $-0.09$ & $0.12$ & $-0.30$ & $-0.08$ & $0.21$ & $-0.02$ & $-0.01$\\ 
    $60$ & $8.00 \cdot 10^{-3}$ & $0.074$ & $0.832$ & $0.210$ & $0.833$ & $1.56$ & $0.84$ & $0.77$ & $0.96$ & $-0.01$ & $-0.03$ & $-0.13$ & $0.15$ & $-0.14$ & $-0.08$ & $-0.24$ & $-0.07$ & $0.14$ & $-0.24$ & $0.01$ & $-0.05$ & $0.00$\\ 
    $60$ & $1.30 \cdot 10^{-2}$ & $0.045$ & $0.719$ & $0.198$ & $0.719$ & $1.60$ & $0.88$ & $0.78$ & $0.96$ & $0.00$ & $-0.03$ & $-0.13$ & $0.01$ & $0.02$ & $-0.35$ & $-0.16$ & $0.09$ & $0.24$ & $-0.17$ & $0.02$ & $0.04$ & $-0.01$\\ 
    $60$ & $2.00 \cdot 10^{-2}$ & $0.030$ & $0.650$ & $0.181$ & $0.650$ & $1.70$ & $0.94$ & $0.81$ & $0.96$ & $0.00$ & $-0.02$ & $0.00$ & $0.07$ & $-0.27$ & $-0.21$ & $-0.34$ & $-0.20$ & $0.20$ & $-0.29$ & $0.00$ & $0.10$ & $-0.01$\\ 
    $60$ & $3.20 \cdot 10^{-2}$ & $0.018$ & $0.565$ & $0.154$ & $0.565$ & $1.78$ & $1.02$ & $0.83$ & $0.96$ & $0.00$ & $-0.01$ & $-0.07$ & $-0.23$ & $-0.29$ & $-0.48$ & $-0.09$ & $-0.03$ & $0.20$ & $-0.26$ & $-0.05$ & $0.19$ & $0.01$\\ 
    $60$ & $6.31 \cdot 10^{-2}$ & $0.009$ & $0.466$ & $0.102$ & $0.466$ & $3.22$ & $1.14$ & $2.18$ & $0.97$ & $-0.01$ & $-0.05$ & $-0.10$ & $0.27$ & $-0.09$ & $0.11$ & $0.10$ & $0.63$ & $1.38$ & $0.95$ & $0.28$ & $-0.14$ & $0.01$\\ 
    $60$ & $1.30 \cdot 10^{-1}$ & $0.005$ & $0.400$ & $0.054$ & $0.400$ & $5.41$ & $2.82$ & $3.88$ & $0.97$ & $-0.01$ & $-0.04$ & $0.11$ & $0.41$ & $-0.01$ & $0.92$ & $0.29$ & $0.83$ & $1.08$ & $-0.16$ & $-0.18$ & $1.50$ & $-0.30$\\ [1ex]
    $90$ & $2.00 \cdot 10^{-3}$ & $0.443$ & $1.312$ & $0.219$ & $1.348$ & $2.28$ & $1.75$ & $1.07$ & $0.96$ & $0.00$ & $-0.03$ & $-0.10$ & $0.10$ & $0.00$ & $-0.01$ & $-0.17$ & $-0.02$ & $0.00$ & $-0.02$ & $0.10$ & $0.03$ & $0.08$\\ 
    $90$ & $3.20 \cdot 10^{-3}$ & $0.277$ & $1.173$ & $0.214$ & $1.183$ & $1.61$ & $0.94$ & $0.83$ & $0.96$ & $0.00$ & $-0.02$ & $-0.04$ & $0.02$ & $-0.01$ & $-0.05$ & $-0.18$ & $0.15$ & $-0.17$ & $-0.01$ & $0.09$ & $0.02$ & $0.00$\\ 
    $90$ & $5.00 \cdot 10^{-3}$ & $0.177$ & $1.037$ & $0.208$ & $1.040$ & $1.58$ & $0.88$ & $0.79$ & $0.96$ & $0.00$ & $-0.04$ & $-0.14$ & $0.12$ & $-0.06$ & $-0.11$ & $-0.12$ & $0.09$ & $-0.30$ & $0.08$ & $0.02$ & $0.03$ & $0.01$\\ 
    $90$ & $8.00 \cdot 10^{-3}$ & $0.111$ & $0.884$ & $0.200$ & $0.885$ & $1.70$ & $0.97$ & $0.82$ & $0.96$ & $-0.01$ & $-0.05$ & $-0.09$ & $0.35$ & $-0.03$ & $-0.12$ & $-0.06$ & $0.17$ & $-0.33$ & $-0.09$ & $0.21$ & $-0.03$ & $-0.01$\\ 
    $90$ & $1.30 \cdot 10^{-2}$ & $0.068$ & $0.761$ & $0.187$ & $0.761$ & $1.71$ & $1.01$ & $0.83$ & $0.97$ & $-0.01$ & $-0.05$ & $-0.21$ & $0.29$ & $0.19$ & $-0.13$ & $-0.19$ & $-0.12$ & $0.21$ & $-0.10$ & $0.03$ & $-0.03$ & $0.00$\\ 
    $90$ & $2.00 \cdot 10^{-2}$ & $0.044$ & $0.666$ & $0.169$ & $0.666$ & $1.81$ & $1.06$ & $0.86$ & $0.96$ & $0.00$ & $0.00$ & $-0.15$ & $-0.19$ & $-0.08$ & $-0.43$ & $-0.26$ & $-0.11$ & $0.14$ & $-0.33$ & $0.00$ & $0.09$ & $-0.01$\\ 
    $90$ & $3.20 \cdot 10^{-2}$ & $0.028$ & $0.571$ & $0.143$ & $0.571$ & $1.86$ & $1.13$ & $0.87$ & $0.96$ & $0.00$ & $-0.02$ & $-0.08$ & $-0.05$ & $-0.25$ & $-0.45$ & $-0.21$ & $-0.13$ & $0.36$ & $-0.15$ & $0.00$ & $0.07$ & $-0.01$\\ 
    $90$ & $5.00 \cdot 10^{-2}$ & $0.018$ & $0.493$ & $0.111$ & $0.493$ & $2.39$ & $1.58$ & $1.02$ & $0.96$ & $0.00$ & $-0.03$ & $0.08$ & $0.02$ & $-0.56$ & $-0.59$ & $-0.06$ & $0.23$ & $0.71$ & $-0.05$ & $-0.04$ & $-0.04$ & $0.08$\\ 
    $90$ & $1.00 \cdot 10^{-1}$ & $0.009$ & $0.404$ & $0.065$ & $0.404$ & $3.19$ & $1.59$ & $1.63$ & $0.97$ & $-0.01$ & $-0.06$ & $-0.11$ & $0.25$ & $-0.03$ & $0.12$ & $0.49$ & $0.86$ & $1.25$ & $1.12$ & $0.37$ & $0.06$ & $-0.11$\\ [1ex]
    $120$ & $5.00 \cdot 10^{-3}$ & $0.236$ & $1.037$ & $0.202$ & $1.044$ & $2.24$ & $1.60$ & $1.13$ & $0.96$ & $0.00$ & $-0.03$ & $-0.06$ & $0.12$ & $-0.02$ & $-0.11$ & $-0.39$ & $0.24$ & $-0.23$ & $0.01$ & $0.05$ & $0.02$ & $-0.03$\\ 
    $120$ & $8.00 \cdot 10^{-3}$ & $0.148$ & $0.901$ & $0.192$ & $0.903$ & $1.99$ & $1.33$ & $1.00$ & $0.96$ & $0.00$ & $-0.04$ & $-0.02$ & $0.08$ & $-0.04$ & $-0.15$ & $-0.24$ & $0.20$ & $-0.39$ & $0.09$ & $0.03$ & $0.04$ & $-0.01$\\ 
    $120$ & $1.30 \cdot 10^{-2}$ & $0.091$ & $0.790$ & $0.179$ & $0.791$ & $2.07$ & $1.34$ & $1.00$ & $0.96$ & $-0.01$ & $-0.06$ & $-0.21$ & $0.49$ & $-0.05$ & $-0.10$ & $0.14$ & $0.12$ & $-0.33$ & $-0.16$ & $0.28$ & $-0.10$ & $-0.02$\\ 
    $120$ & $2.00 \cdot 10^{-2}$ & $0.059$ & $0.664$ & $0.161$ & $0.664$ & $2.06$ & $1.35$ & $0.99$ & $0.96$ & $-0.01$ & $-0.04$ & $-0.10$ & $0.35$ & $0.21$ & $0.06$ & $-0.29$ & $-0.23$ & $0.21$ & $-0.36$ & $0.01$ & $0.04$ & $-0.02$\\ 
    $120$ & $3.20 \cdot 10^{-2}$ & $0.037$ & $0.582$ & $0.135$ & $0.582$ & $2.20$ & $1.46$ & $1.02$ & $0.96$ & $0.00$ & $-0.01$ & $-0.03$ & $-0.03$ & $-0.23$ & $-0.33$ & $-0.30$ & $-0.24$ & $0.49$ & $-0.39$ & $-0.04$ & $0.12$ & $-0.01$\\ 
    $120$ & $5.00 \cdot 10^{-2}$ & $0.024$ & $0.499$ & $0.105$ & $0.499$ & $2.35$ & $1.58$ & $1.05$ & $0.97$ & $-0.01$ & $-0.04$ & $-0.05$ & $0.09$ & $-0.56$ & $-0.56$ & $-0.19$ & $-0.07$ & $0.46$ & $-0.28$ & $-0.06$ & $0.06$ & $0.01$\\ 
    $120$ & $1.00 \cdot 10^{-1}$ & $0.012$ & $0.405$ & $0.062$ & $0.405$ & $3.52$ & $1.90$ & $1.12$ & $0.97$ & $0.00$ & $-0.06$ & $-0.02$ & $0.10$ & $0.56$ & $-0.37$ & $0.37$ & $0.95$ & $1.85$ & $1.23$ & $0.32$ & $-0.18$ & $-0.02$\\ 
    $120$ & $2.00 \cdot 10^{-1}$ & $0.006$ & $0.343$ & $0.037$ & $0.343$ & $7.21$ & $4.46$ & $4.81$ & $0.97$ & $-0.02$ & $-0.07$ & $0.08$ & $0.62$ & $0.09$ & $1.38$ & $1.06$ & $1.00$ & $1.09$ & $-0.22$ & $-0.25$ & $1.48$ & $-0.24$\\ [1ex]
    $150$ & $1.30 \cdot 10^{-2}$ & $0.114$ & $0.752$ & $0.173$ & $0.753$ & $6.30$ & $5.50$ & $2.73$ & $0.96$ & $0.00$ & $-0.02$ & $-0.12$ & $-0.09$ & $-0.09$ & $-0.25$ & $0.10$ & $0.27$ & $-0.84$ & $0.09$ & $0.31$ & $0.11$ & $0.08$\\ 
    $150$ & $2.00 \cdot 10^{-2}$ & $0.074$ & $0.725$ & $0.155$ & $0.725$ & $4.16$ & $3.20$ & $2.15$ & $0.96$ & $-0.01$ & $-0.08$ & $-0.17$ & $0.82$ & $0.01$ & $-0.01$ & $0.32$ & $0.23$ & $-0.31$ & $-0.47$ & $0.49$ & $-0.32$ & $-0.08$\\ 
    $150$ & $3.20 \cdot 10^{-2}$ & $0.046$ & $0.606$ & $0.130$ & $0.607$ & $4.18$ & $2.99$ & $1.89$ & $0.96$ & $-0.01$ & $-0.02$ & $-0.17$ & $0.59$ & $0.32$ & $0.49$ & $-0.90$ & $-0.98$ & $0.52$ & $-1.09$ & $-0.16$ & $-0.03$ & $-0.04$\\ 
    $150$ & $5.00 \cdot 10^{-2}$ & $0.030$ & $0.504$ & $0.101$ & $0.504$ & $4.05$ & $3.06$ & $1.86$ & $0.96$ & $0.00$ & $0.01$ & $0.16$ & $0.01$ & $-0.20$ & $0.15$ & $-0.72$ & $-0.45$ & $1.15$ & $-0.69$ & $-0.13$ & $0.00$ & $-0.06$\\ 
    $150$ & $1.00 \cdot 10^{-1}$ & $0.015$ & $0.429$ & $0.060$ & $0.429$ & $4.67$ & $3.40$ & $1.80$ & $0.96$ & $0.00$ & $-0.01$ & $0.08$ & $-0.14$ & $0.28$ & $-0.19$ & $-0.10$ & $0.34$ & $2.35$ & $0.50$ & $-0.10$ & $-0.23$ & $0.02$\\ 
    $150$ & $2.00 \cdot 10^{-1}$ & $0.007$ & $0.331$ & $0.034$ & $0.331$ & $10.72$ & $7.73$ & $7.14$ & $0.96$ & $-0.01$ & $-0.01$ & $0.04$ & $0.33$ & $-0.12$ & $0.41$ & $-0.06$ & $0.45$ & $1.05$ & $-0.47$ & $0.24$ & $1.11$ & $-0.40$\\ 

  \hline\hline
\end{tabular}
\end{center}
\end{tiny}
\begin{small}
  \caption{Combined cross section measurement from H1, continued
    from \Tab~\ref{tab:table9700comb}.
    \label{tab:table9700comb2}
}
\end{small} 
\end{sidewaystable}

\clearpage

\end{document}